\documentclass[conference,compsoc]{IEEEtran}
\ifCLASSOPTIONcompsoc
  \usepackage[nocompress]{cite}
\else
  \usepackage{cite}
\fi

\ifCLASSINFOpdf
\else
\fi
\usepackage{graphicx}
\usepackage{comment}
\usepackage{nicefrac}
\usepackage{siunitx}
\usepackage{array,framed}
\usepackage{booktabs}
\usepackage{hyperref}
\usepackage{upgreek}

\usepackage{textcomp}
\usepackage{setspace}
\usepackage{amsfonts}
\usepackage{latexsym,fancyhdr,url}
\usepackage{enumerate}
\usepackage{algorithm2e}
\usepackage{algpseudocode}
\usepackage{graphics}
\usepackage{xparse} 
\usepackage{xspace}
\usepackage{multirow}
\usepackage{csvsimple}
\usepackage{amsmath}
\usepackage{balance}
\usepackage{enumerate}
\usepackage{enumitem}
\usepackage{subcaption}
\usepackage{threeparttable}

\usepackage{
  tikz,
  pgfplots,
  pgfplotstable
}

\usetikzlibrary{
  shapes.geometric,
  arrows,
  external,
  pgfplots.groupplots,
  matrix
}

\usepackage{pifont}
\newcommand{\cmark}{\ding{51}}%
\newcommand{\xmark}{\ding{55}}%
\usepackage{color, colortbl}
\definecolor{LightCyan}{rgb}{0.88,1,1}
\definecolor{Gray}{gray}{0.9}
\newcommand{\midsepremove}{\aboverulesep = 0mm \belowrulesep = 0mm}
\midsepremove
\newcommand{\midsepdefault}{\aboverulesep = 0.605mm \belowrulesep = 0.984mm}
\midsepdefault

\pgfplotsset{compat=1.9}

\usepackage{mathtools}

\DeclarePairedDelimiter\norm{\lVert}{\rVert}

\DeclareMathAlphabet{\mathcal}{OMS}{cmsy}{m}{n}

\def\P(#1){\Phelper#1|\relax\Pchoice(#1)}
\def\Phelper#1|#2\relax{\ifx\relax#2\relax\def\Pchoice{\Pone}\else\def\Pchoice{\Ptwo}\fi}
\def\Pone(#1){\Pr\left( #1 \right)}
\def\Ptwo(#1|#2){\Pr( #1 | #2 )}
\def\Pr{\text{Pr}}

\def\PP(#1){\Phelper#1|\relax\Pchoice(#1)}
\def\Phelper#1|#2\relax{\ifx\relax#2\relax\def\Pchoice{\Pone}\else\def\Pchoice{\PtwoP}\fi}
\def\Pone(#1){\PrP\left( #1 \right)}
\def\PtwoP(#1|#2){\PrP( #1 | #2 )}
\def\PrP{\text{Pr}^{k}_{s}}

\usepackage{cleveref}

\crefformat{section}{\S#2#1#3} 
\crefformat{subsection}{\S#2#1#3}
\crefformat{subsubsection}{\S#2#1#3}

\DeclareGraphicsExtensions{%
    .png,.PNG,%
    .pdf,.PDF,%
    .jpg,.mps,.jpeg,.jbig2,.jb2,.JPG,.JPEG,.JBIG2,.JB2}

\newcommand{\sys}{EveGuard\xspace}
\newcommand{\gan}{Eve-GAN\xspace}
\newcommand{\pgm}{PGM\xspace}
\newcommand{\presec}{\vspace{-0.05in}}
\def\jungwoo{\textcolor{black}}
\def\todo{\textcolor{black}}
\def\revision{\textcolor{black}}

\begin{document}

\title{\LARGE EveGuard: Defeating Vibration-based Side-Channel Eavesdropping with Audio Adversarial Perturbations}

\author{
\parbox{\linewidth}{\centering
Jung-Woo Chang\IEEEauthorrefmark{1}\textsuperscript{\textsection},
Ke Sun\IEEEauthorrefmark{1}\IEEEauthorrefmark{2}\textsuperscript{\textsection},
David Xia\IEEEauthorrefmark{3},
Xinyu Zhang\IEEEauthorrefmark{1},
Farinaz Koushanfar\IEEEauthorrefmark{1} \\
\IEEEauthorrefmark{1}University of California San Diego \quad \IEEEauthorrefmark{2}University of Michigan \quad \IEEEauthorrefmark{3}University of Illinois Urbana-Champaign \\
}
\vspace{+0.6cm}}

\maketitle
\begingroup\renewcommand\thefootnote{\textsection}
\footnotetext{The first two authors contributed equally to this work.}
\endgroup
\begin{abstract}
Vibrometry-based side channels pose a significant privacy risk, exploiting sensors like mmWave radars, light sensors, and accelerometers to detect vibrations from sound sources or proximate objects, enabling speech eavesdropping. 
%
Despite various proposed defenses, these involve costly hardware solutions with inherent physical limitations.
This paper presents \sys, a software-driven defense framework that creates adversarial audio, protecting voice privacy from side channels without compromising human perception.
%
We leverage the distinct sensing capabilities of side channels and traditional microphones—where side channels capture vibrations and microphones record changes in air pressure, resulting in different frequency responses.
\sys first proposes a perturbation generator model (PGM) that effectively suppresses sensor-based eavesdropping while maintaining high audio quality. Second, to enable end-to-end training of PGM, we introduce a new domain translation task called \gan for inferring an eavesdropped signal from a given audio. \todo{We further apply few-shot learning to mitigate the data collection overhead for \gan training.} 
Our extensive experiments show that \sys achieves a protection rate of more than 97$\%$ from audio classifiers and significantly hinders eavesdropped audio reconstruction. We further validate the performance of \sys across three adaptive attack mechanisms. We have conducted a user study to verify the perceptual quality of our perturbed audio.



\end{abstract}

\IEEEpeerreviewmaketitle

\section{Introduction}
\label{sec:intro}
Loudspeakers are omnipresent in today's technology-based society. Their use extends beyond facilitating phone calls and video conferencing for the exchange of private information. They have been widely integrated into intelligent mobile and IoT devices, enhancing human-machine interaction through speech recognition. 
The associated use cases are anticipated to reach a market size of $\$150.68$ billion by 2032~\cite{market_vc, market_ai}.  
As people increasingly rely on loudspeaker-equipped devices, voice privacy is becoming increasingly important.

Unfortunately, the diverse sensors in intelligent devices are imposing an alarming risk to voice privacy. Although these sensors are not originally designed for voice recording, they can be repurposed by adversaries to serve as side channels to capture voice-induced vibrations, thereby facilitating unauthorized eavesdropping.
For example, the prevalent accelerometers on smartphones have been exploited to eavesdrop on voice playout \cite{sun2023stealthyimu, hu2022accear}. Millimeter-wave (mmWave) radars can remotely detect vibrations from sound sources and recover speech signals through walls \cite{hu2023mmecho, hu2022milliear, shi2023privacy, wangvibspeech, wang2022mmeve, zhao2023radio2text}.
Such side-channel speech eavesdropping attacks (SSEAs) lead to severe individual privacy breaches \cite{news_accelerometer} and may compromise sensitive organizational intellectual property~\cite{news_alexa}.

\begin{figure}[t]
\centering
\includegraphics[width=0.94\columnwidth]{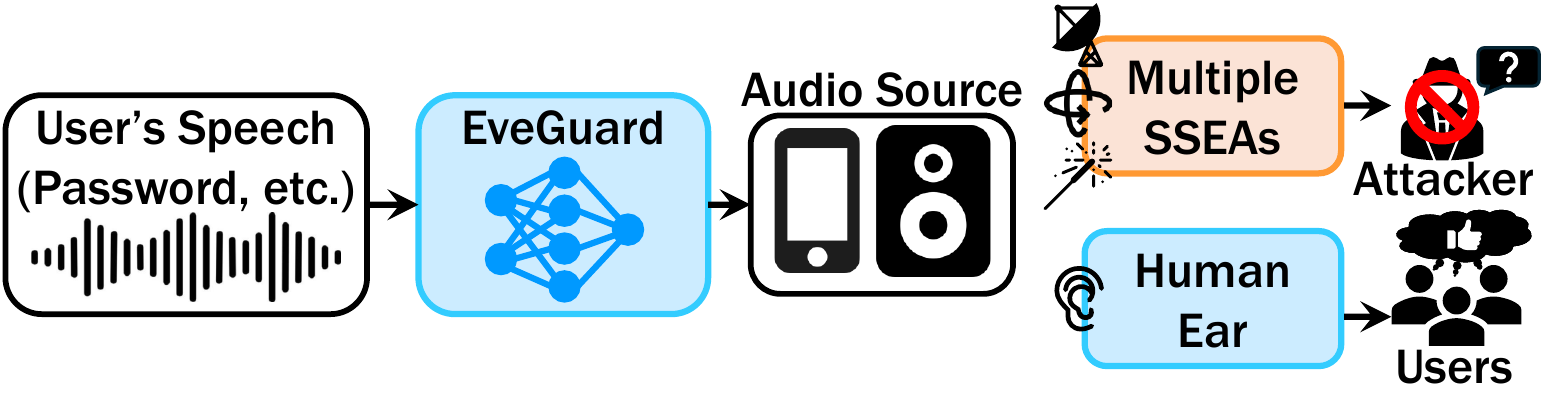}
\caption{Overview of \sys, inserting imperceptible adversarial perturbations to the target speech to protect users' voice communication from multi-sensor eavesdropping attacks.}
\label{fig:intro}
\end{figure}

Existing research has devised hardware-based defenses against SSEAs.
For instance, jamming-based methods \cite{jiao2021openwifi, sun2024sok, yang2013design} can block adversarial mmWave SSEAs. However, they may degrade the sensing function of legitimate mmWave devices. Moreover, jamming is generally prohibited in non-military applications \cite{fc_jammer}. Intelligent reflecting surface (IRS) has also been used as a security shield~\cite{shenoy2022rf, staat2022irshield}, yet it can only protect its immediate vicinity. 
As for defending against accelerometer-based SSEAs, vibration motors have been used to generate low-amplitude vibrations that disrupt eavesdropping~\cite{zhang2023spy}. However, this method may cause user discomfort and hasten the depletion of smartphone batteries.

\todo{We propose \sys, an innovative software-driven defense mechanism to protect against privacy leakage from the loudspeaker-generated voice in SSEAs.} As shown in Figure~\ref{fig:intro}, \sys mitigates SSEA threats by introducing audio adversarial examples to the original audio signals prior to playback. \sys ensures that (i) the perturbed speech signals remain natural to human ears and microphones, and (ii) any attempt by SSEAs to capture and reconstruct the perturbed speech will produce content that is difficult to interpret, both for humans and automated speech recognition systems. \todo{Note that \sys cannot protect voice from a human speaker when SSEAs target eavesdropping on throat vibrations, a challenge also for state-of-the-art (SOTA) attacks \cite{walker2021sok}.}

To attain these salient properties, \sys must address four main design challenges. First, it is crucial to ensure the effectiveness of perturbations against SSEAs while maintaining the quality of legitimate voice communication. Existing adversarial speech generation methods commonly rely on additive perturbations, which can introduce noticeable, conspicuous noise~\cite{yu2023smack}. 
In contrast, \sys \textit{leverages the distinct sensing mechanisms of the side channels versus traditional microphones or human hearing}, i.e., the former only captures low-frequency vibrations whereas the latter senses the subtle changes in air pressure.
\sys devises a two-stage \textit{Perturbation Generator Model} (PGM) that integrates: (1) finite impulse response (FIR) convolution to perturb the low-frequency attributes of speech while preserving the speech quality and (2) inaudible low-frequency adversarial perturbations (LFAPs) to corrupt the eavesdropped signals.

\todo{Second, to \todo{automate and optimize} the perturbation signal generation, \sys requires a new differentiable computational model to represent the SSEA.  
To tackle this challenge, we propose \gan, a deep generative network aiming at learning an audio-to-SSEA translation that can map the source audio to the targeted eavesdropped audio. Once trained, \gan serves as a differentiable layer, enabling end-to-end training of our PGM.
Yet training \gan requires collecting sufficient SSEA samples across various attack scenarios.
%
Additionally, obtaining paired training data is tedious as it requires input-output pairs with the same speaker, speech attributes (e.g., prosody), and utterance content.
To address this, we leverage advancements in few-shot unsupervised learning~\cite{liu2019few}. We propose a few-shot, unpaired audio-to-SSEA translation, which learns to convert source audio into eavesdropped audio by referencing an unpaired SSEA sample. By extracting domain features from the few-shot real-world SSEA samples, \gan facilitates a generalizable conversion applicable to unseen samples during training.}
Third, the rapid growth of ML empowers attackers to devise sophisticated SSEAs~\cite{hu2022milliear, hu2022accear, shi2023privacy, sun2023stealthyimu}. For instance, the attacker can transcribe private conversations using speech recognition, and identify digits with audio classifiers. However, the defender has no prior knowledge of the SSEA model deployed by the eavesdropper. To overcome this hindrance, we utilize the transferability of adversarial examples, which means perturbations learned to fool an ensemble of diverse surrogate models can also be effective against unknown black-box models~\cite{liu2016delving, cai2022blackbox}. 
To this end, we first build a set of surrogate ML models based on multiple hypothetical SSEAs. We then concatenate the PGM with \gan and ensemble surrogate models to encourage the PGM to learn robust perturbations in an end-to-end manner. 


Finally, an adaptive attacker who knows the existence of \sys may attempt to mitigate the effects of the perturbation. 
Thus, we apply three preventive techniques to PGM as follows: (1) the use of a discriminator inside the PGM to enforce undetectable constraints, (2) style diversification by integrating VAE-GAN~\cite{gur2020hierarchical} into FIR perturbation generator, and \todo{(3) ensuring the LFAP generator uses a random latent vector as input to produce diverse LFAP samples.}




We implement \sys by integrating the above solutions. To evaluate \sys, we reproduce white-box SSEAs based on representative eavesdropping sensors, mmWave radar, accelerometer, and optical sensor. 
Built upon these, we extensively validate \sys under various attack settings including distance, orientation, materials, hardware configurations, etc. Our experimental results show that \sys achieves a protection rate of more than 97$\%$ from SSEA's digit classifiers and hinders the recovery of eavesdropped audio with an MCD (Mel-Cepstral Distortion) of over 13.4, and a WER (Word Error Rate) of over 68.2$\%$. 
To show that our adversarial audio generated by PGM is imperceptible to humans, we verify the indistinguishability through a user study involving 24 participants.

\todo {The main contributions of \sys are as follows.}

\begin{itemize}[leftmargin=*]
\item We introduce \sys, a novel software-driven defense framework that leverages black-box adversarial examples to protect loudspeaker-generated voice from SSEAs. 

\item We design a PGM that leverages the unique features of eavesdropping devices to ensure robust perturbations across diverse attack scenarios, including variations in distance, orientation, materials, and hardware configurations.

\item We develop \gan, a differential framework that enables our PGM to learn the distribution of adversarial examples end-to-end, incorporating a few-shot, unpaired audio-to-SSEA translation framework to reduce data collection overhead for training \gan.

\item We perform extensive experiments to verify the effectiveness of the \sys defense, using both objective metrics and subjective user studies. Audio samples are available at \textcolor{magenta}{\href{https://eveguard.github.io/demo/} {https://eveguard.github.io/demo/}}.
\end{itemize}    
\section{{Background and Related Work}}
\label{sec:background_ke}
In this section, we provide an overview of vibration-based SSEAs and related work.

\subsection{Vibration-based Speech Eavesdropping}
%
%
%
Speech can be recovered by measuring sound-induced vibrations on objects using sensors or vibration-sensitive devices.
Table~\ref{tab:summary} provides an overview of conventional SSEA methods, summarizing three representative types of SSEA techniques and associated defense mechanisms.
%
%

\begin{table*}[t]
\caption{Comparison with conventional vibration-based SSEAs. Existing defenders must rely on inefficient hardware-based techniques to disable each SSEA. (\cmark: the item is supported; \xmark: the item is not supported.)} 
\centering
\scriptsize
\label{tab:summary}
\resizebox{0.96\textwidth}{!}{
\midsepremove
\begin{tabular}{ c|c|c|c|c|c|c|c|c|c }
\toprule
\begin{tabular}[c]{@{}c@{}}Previous Work\end{tabular} & 
\begin{tabular}[c]{@{}c@{}}Sensor\end{tabular} & 
\begin{tabular}[c]{@{}c@{}}Sensing Target\end{tabular} & \begin{tabular}[c]{@{}c@{}}Sampling\\ Rate\end{tabular} &  \begin{tabular}[c]{@{}c@{}}Sensing Distance\end{tabular} &
\begin{tabular}[c]{@{}c@{}}Non- \\ Invasive \end{tabular} &
\begin{tabular}[c]{@{}c@{}}Through-Wall\\ (Opaque)\end{tabular} & 
\begin{tabular}[c]{@{}c@{}}Aided by \\ ML\end{tabular} &
\begin{tabular}[c]{@{}c@{}}Existing Defense \\Solution\end{tabular} &
\begin{tabular}[c]{@{}c@{}}Our Defense \\ (\sys) \end{tabular}\\
\hline
\begin{tabular}[c]{@{}c@{}}  Lamphone~\cite{nassi2020lamphone}\\ LidarPhone~\cite{sami2020lidarphone}\end{tabular}   & 
\begin{tabular}[c]{@{}c@{}}Optical\\ Sensor\end{tabular} & \begin{tabular}[c]{@{}c@{}} Loudspeaker or \\ Vibrating Object \end{tabular} & $\leq$ 2kHz  & \begin{tabular}[c]{@{}c@{}} Far ($\simeq$ 35m) \\ Moderate ($\simeq$ 1.5m) \end{tabular} & \begin{tabular}[c]{@{}c@{}} \cmark \\ \xmark \end{tabular} & \begin{tabular}[c]{@{}c@{}} \xmark \\ \xmark \end{tabular}& \begin{tabular}[c]{@{}c@{}} \xmark \\ \cmark \end{tabular} & \begin{tabular}[c]{@{}c@{}} Blocking Visible Channel \\(e.g., Wood, Curtains, etc.) \end{tabular} & \cmark \\ 
\hline
\begin{tabular}[c]{@{}c@{}} StealthyIMU~\cite{sun2023stealthyimu}  \\
Accear~\cite{hu2022accear}\end{tabular} &  
\begin{tabular}[c]{@{}c@{}}Motion \\ Sensor\end{tabular} &
Loudspeaker & $\leq$ 0.5kHz  & Close ($\leq 0.1$m) & \begin{tabular}[c]{@{}c@{}} \xmark \\ \xmark \end{tabular} & \begin{tabular}[c]{@{}c@{}} \xmark \\ \xmark \end{tabular} & \begin{tabular}[c]{@{}c@{}} \cmark \\ \cmark \end{tabular} &\begin{tabular}[c]{@{}c@{}}Smartphone's \\ Vibration Motor \end{tabular} &  \cmark \\
\hline
\begin{tabular}[c]{@{}c@{}}  mmSpy~\cite{basak2022mmspy} \\ mmEcho~\cite{hu2023mmecho} \\ Shi \textit{et al.}~\cite{shi2023privacy} \\  VibSpeech~\cite{wangvibspeech} \end{tabular} &
\begin{tabular}[c]{@{}c@{}}mmWave\\ Radar\end{tabular}&  \begin{tabular}[c]{@{}c@{}}Loudspeaker or \\ Vibrating Object\end{tabular} & $\leq$16kHz  & \begin{tabular}[c]{@{}c@{}} Moderate ($\simeq$ 5m) \end{tabular}& \begin{tabular}[c]{@{}c@{}} \cmark \\ \cmark \\ \cmark \\ \cmark \end{tabular} & \begin{tabular}[c]{@{}c@{}} \cmark \\ \cmark \\ \cmark \\ \cmark \end{tabular} & \begin{tabular}[c]{@{}c@{}} \cmark \\ \xmark \\ \cmark \\ \cmark \end{tabular} &\begin{tabular}[c]{@{}c@{}}Jamming~\cite{jiao2021openwifi,  yang2013design} or \\ IRS~\cite{shenoy2022rf, staat2022irshield} \end{tabular} & \cmark \\  
\bottomrule
\end{tabular}}
\end{table*}

\noindent\textbf{Motion Sensor-based SSEA.} Recent research~\cite{sun2023stealthyimu, hu2022accear} shows that malicious apps can capture sound-induced motion signals using a smartphone's accelerometer.
Using pre-trained ML models~\cite{sun2023stealthyimu, hu2022accear}, 
the attacker can recover speech from the vibration motion despite the IMU's low sampling rates ($\leq$ 500~Hz). 
While a smartphone's vibration motor~\cite{zhang2023spy} can obfuscate the sound-induced motion, it may cause battery drain and user discomfort.

\noindent\textbf{Optical Sensor-based SSEA.}
Sound-induced vibration can also be captured through optical sensors.
Lamphone~\cite{nassi2020lamphone} measures the vibration of a light bulb near a loudspeaker, while LidarPhone~\cite{sami2020lidarphone} uses a vacuum cleaner's lidar sensor to sample vibrations. However, these attacks leave visual clues (e.g., laser dots, bulky cameras) and are easily prevented by blocking the line-of-sight~\cite{walker2021sok, wang2022mmeve, shi2023privacy}. Therefore, a recent SoK deems these attacks impractical~\cite{walker2021sok}.

\noindent\textbf{mmWave Radar-based SSEA.} 
mmWave radars are commonly used for object ranging or imaging.  
Yet, slight object displacements produce subtle phase changes in the reflected radar signals, enabling the reconstruction of audio signals from these changes \cite{hu2023mmecho, hu2022milliear, shi2023privacy, wangvibspeech}.
%
Although some hardware-enabled defense techniques~\cite{jiao2021openwifi,  yang2013design, shenoy2022rf, staat2022irshield} exist, like jamming and IRS shield, these approaches typically provide only a limited area of defense and often fall short in effectiveness. For instance, IRS-based approaches~\cite{shenoy2022rf, staat2022irshield} generally require individuals targeted by SSEAs to procure and deploy extra hardware. The IRS device needs to be oriented toward the attacker and placed close (e.g., 1.2m~\cite{shenoy2022rf}) to the attacker's mmWave radar. 
Additionally, these defense mechanisms may unintentionally compromise the standard sensing capabilities of radars.

Among the aforementioned side channels, mmWave-based attacks pose the most serious threat to the user's privacy because: (1) radar can cover the human-voice frequency spectrum with a high sampling rate \cite{hu2023mmecho}, (2) radar allows attacks at a distance \cite{shi2023privacy}, and (3) radar can penetrate soundproofing materials \cite{hu2023mmecho,shi2023privacy}.
\emph{Therefore, \sys prioritizes a defense against mmWave radar-based attacks, yet we will also show that \sys remains adaptable to other side-channel threats, i.e., optical-based and accelerometer-based SSEA (Sec.~\ref{sec:eval_motion}).} 
Unlike existing defense mechanisms \cite{shenoy2022rf, staat2022irshield, yao2023interference, yang2013design}, \sys is a software-driven framework designed to generate adversarial speech before playback, 
without affecting either microphone recordings or human hearing.

\presec
\subsection{Adversarial Examples in Audio Domain}
Adversarial examples pose a significant threat to ML systems, affecting audio-based systems like automatic speech recognition (ASR). Carlini \textit{et al.}~\cite{carlini2018audio} introduced an end-to-end white-box attack where 
the ML model is manipulated to translate speech signals into the attacker's desired phrase. 
Recent studies~\cite{guo2022specpatch, chen2020devil, zheng2021black, chen2020metamorph} focus on achieving over-the-air delivery of adversarial audio. 
%
Semantic perturbation approaches~\cite{yu2023smack,o2022voiceblock} have also emerged that deviate from additive perturbations. 
SMACK~\cite{yu2023smack} devised semantic audio attacks targeting speech transcription and speaker recognition systems.
Voiceblock~\cite{o2022voiceblock} applies a time-varying FIR filter to outgoing audio, enabling effective and inconspicuous perturbations. While existing work addressed speech and speaker recognition systems, \sys focuses on the distinct vibrometry-based side-channel attacks.

\presec
\subsection{Adversarial Examples for Privacy Protection}
The rapid progress in ML has enabled attackers to misuse ML models for malicious purposes, such as device fingerprinting attacks, DeepFake audio generation, and unauthorized face and speaker recognition. In response, researchers have developed defensive strategies utilizing adversarial examples~\cite{yu2023antifake, kreuk2018fooling, cherepanova2021lowkey, shan2020fawkes, shenoi2023ipet}. For instance, 
iPET~\cite{shenoi2023ipet} protects the privacy of IoT users by perturbing network traffic to prevent fingerprinting attacks. 
Antifake~\cite{yu2023antifake} specifically targets DeepFake by adding perturbations to the user's speech, disrupting the speech synthesis process. 
Fawkes~\cite{shan2020fawkes} inserts imperceptible pixel-level perturbations into the user's photo to thwart unauthorized facial recognition from learning the user's identity.
\sys closely aligns with the goal of these studies. However, protecting users' voice privacy from side-channel eavesdroppers through audio perturbation is a new research field.

\presec
\section{Threat Model, and Defense Goals}
\label{sec:threatmodel_ke}

In this section, we introduce the threat model and defender's objectives and capability.

\begin{figure*}[t]
    \centering
    \begin{tabular}{@{}ccccccc@{}}
        \includegraphics[height=0.076\textheight]{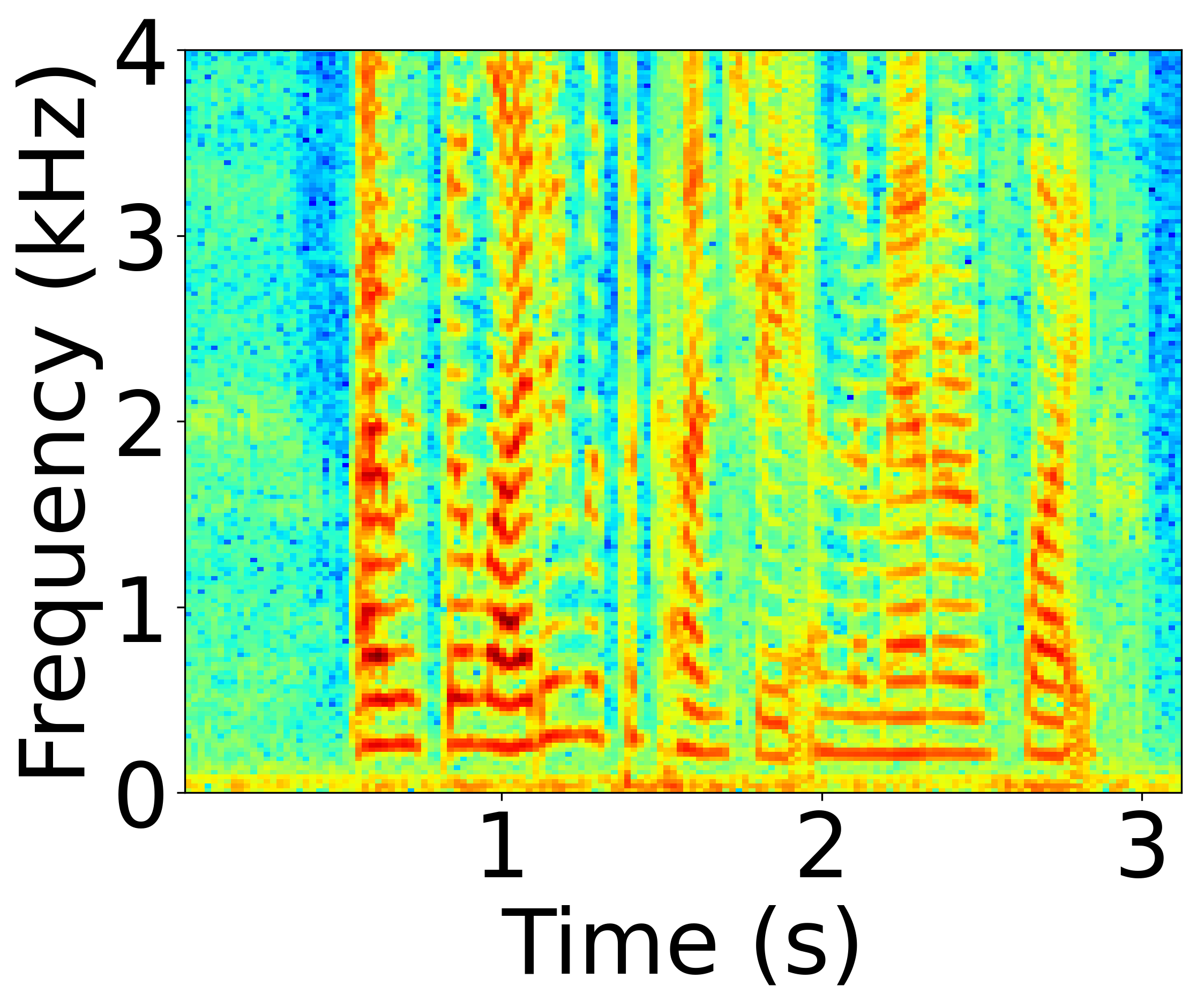} &
        \includegraphics[height=0.076\textheight]{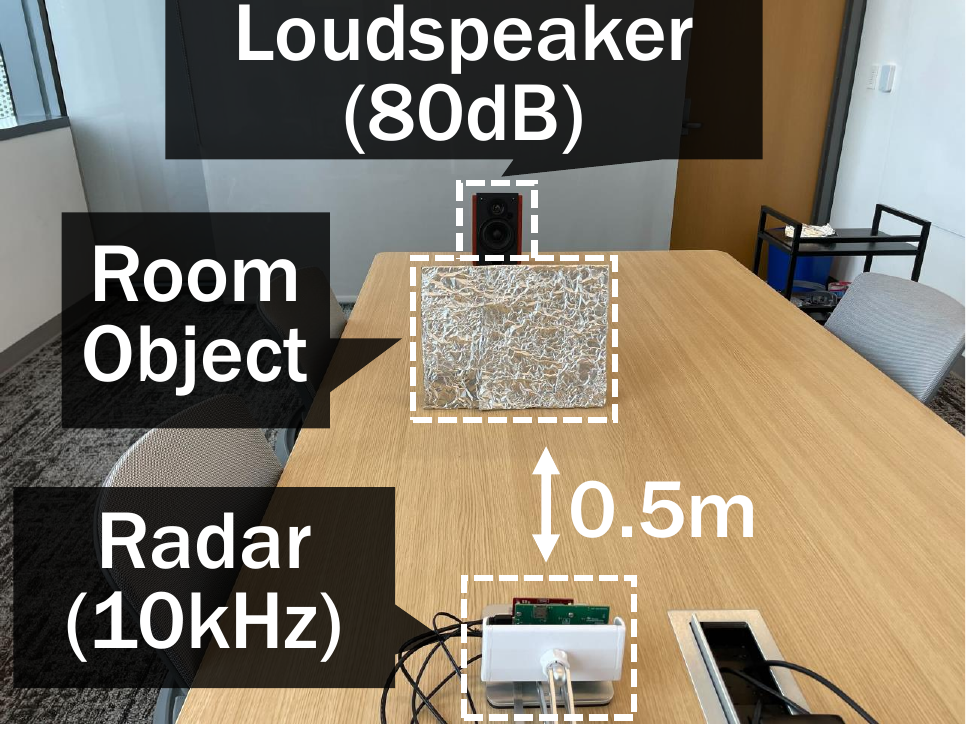} &
        \includegraphics[height=0.076\textheight]{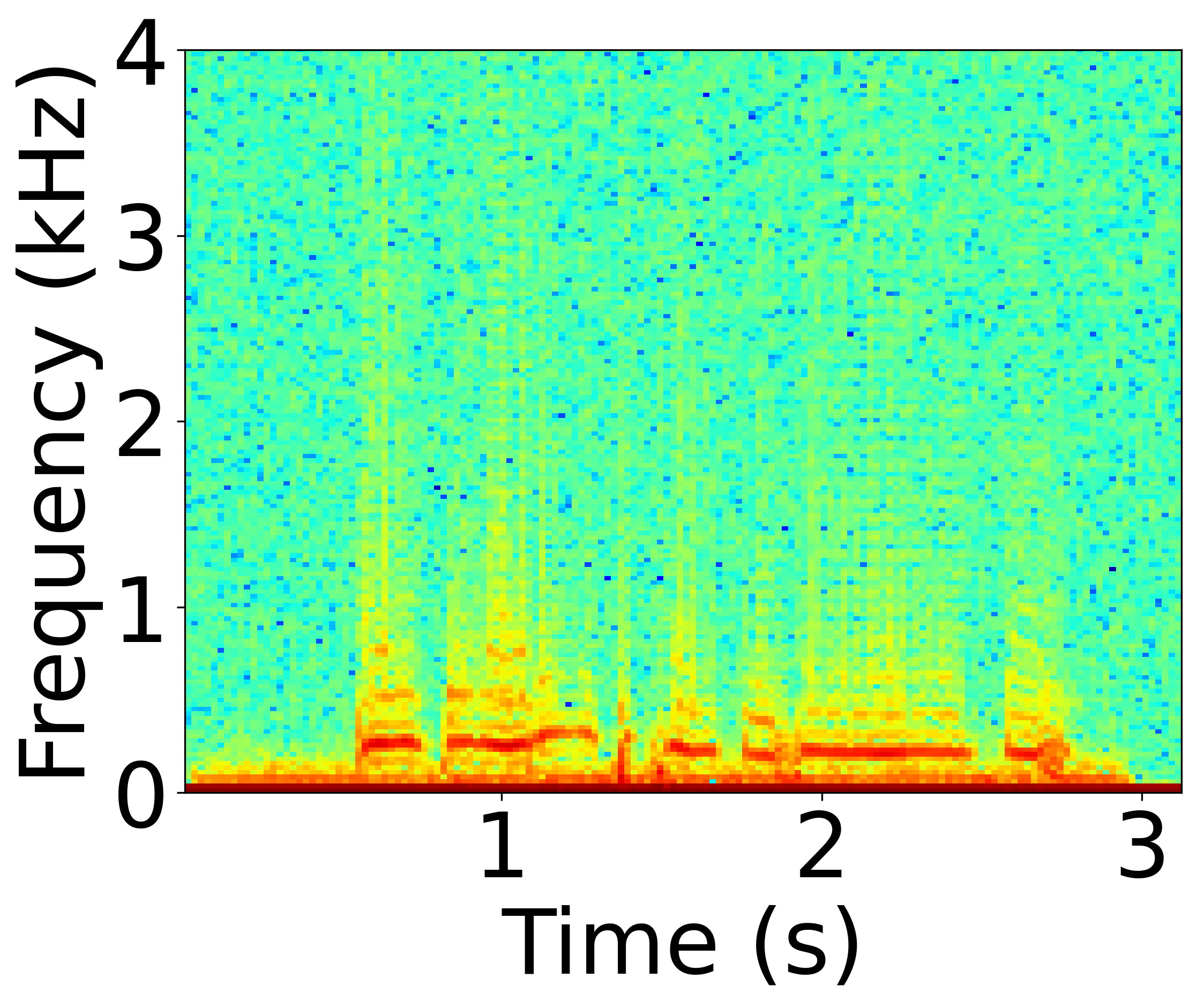} &
         \includegraphics[height=0.076\textheight]{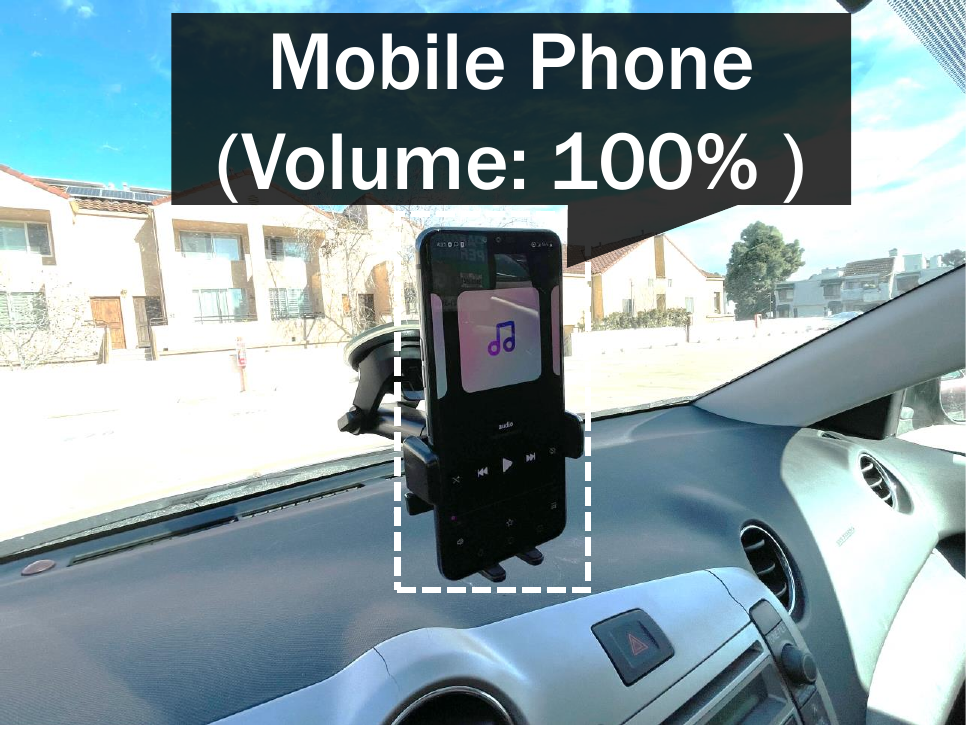} &
         \includegraphics[height=0.076\textheight]{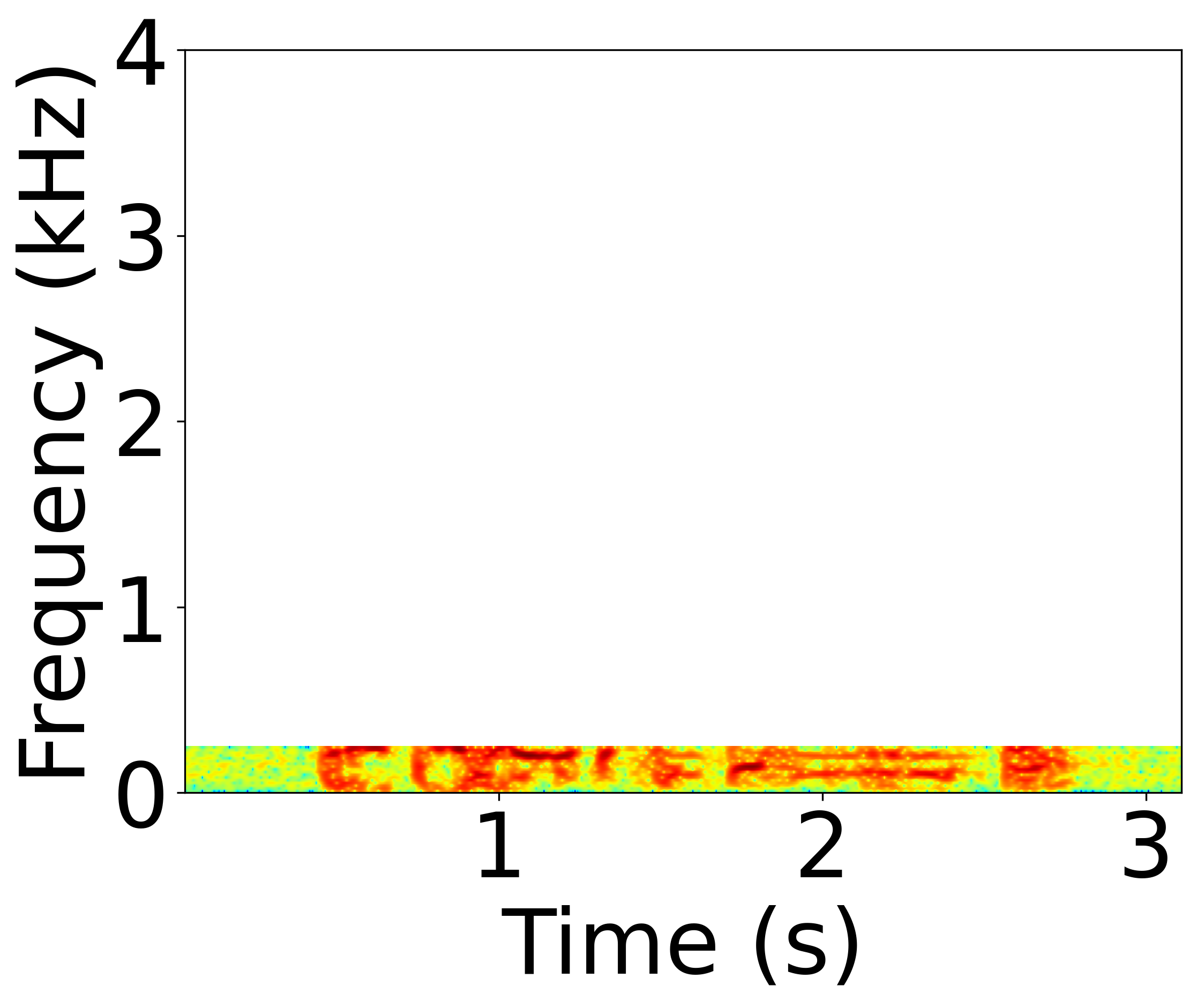} &
         \includegraphics[height=0.076\textheight]{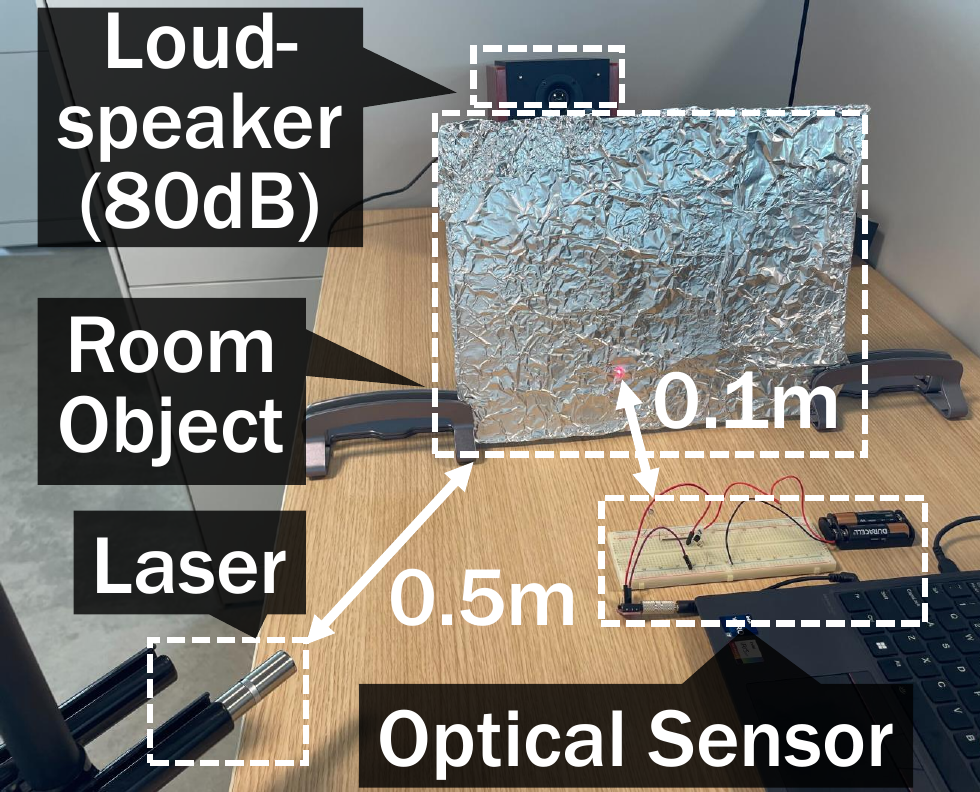} &
        \includegraphics[height=0.076\textheight]{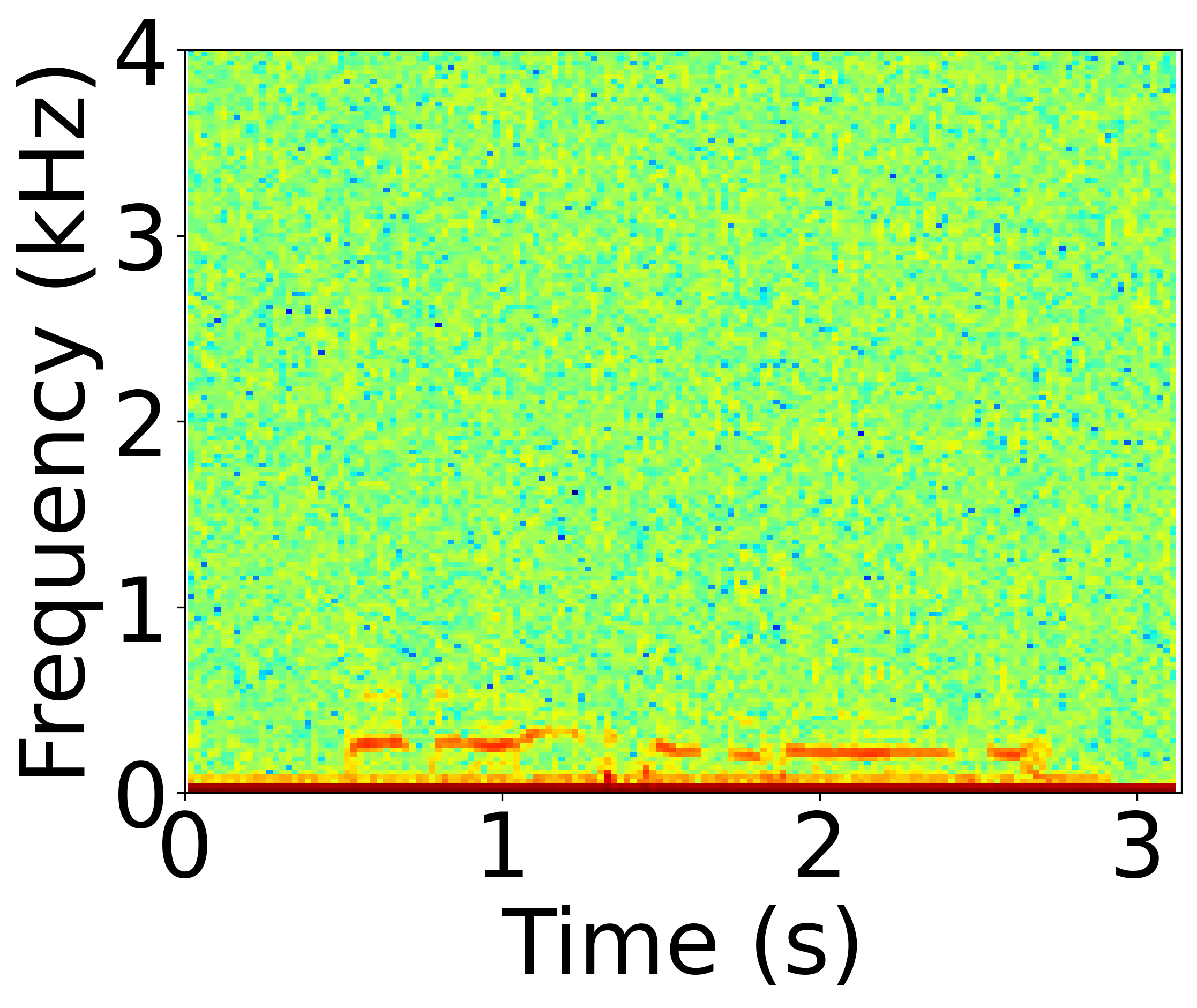} \\
        (a) Original &
        \multicolumn{2}{c}{(b) mmWave Radar} &
        \multicolumn{2}{c}{(c) Smartphone Accelerometer} &
        \multicolumn{2}{c}{(d) Optical Sensor} \\
    \end{tabular}
    \caption{Feasibility study settings and results (The sampling rate of smartphone's accelerometer is $500$ Hz).}
    \label{fig:feasibility}
\end{figure*}

\subsection{\sys Threat Model}
\label{sec:threat}
\revision{The attacker's goal is to eavesdrop on the audio content played by a loudspeaker.
However, direct microphone deployment inside or outside the victim's room is infeasible due to the following reasons: (a) hidden microphones inside the room can be easily detected by victims using spy detectors~\cite{mitev2020leakypick, zhou2023dehirec}, making covert placement ineffective and (b) if victims use loudspeakers in a soundproof room, external eavesdropping microphones cannot capture the audio.
As a result, the attacker seeks to exploit SSEAs, leveraging side-channel sensors to capture speech-induced vibrations from the loudspeakers. Specifically, mmWave radar-based techniques~\cite{hu2023mmecho, shi2023privacy, wei2015acoustic} enable the attacker to penetrate soundproofing materials and eavesdrop on speech. Additionally, zero-permission motion sensors\cite{sun2023stealthyimu} can be exploited to infer private speech information from a smartphone. The attacker may also detect speech-induced vibrations by sensing reflected ambient light~\cite{nassi2020lamphone}.}


%
\revision{In this work, rather than limiting our focus to constrained SSEAs, we aim to develop defenses capable of mitigating even the worst-case scenarios, ensuring robustness against highly capable adversaries.}
We assume that the attacker has strong capabilities to i). compromise the side-channel sensors and their sensing systems; ii). employ advanced ML and signal processing techniques to derive private information from the captured signals; iii). gain access to the victims' speech samples or even create training datasets to facilitate the eavesdropping. 

Specifically, for mmWave radar-based SSEA, the attacker can access the radar's raw Analog-to-Digital Converter (ADC) data, by either employing their radar or planting malware into a radar-equipped IoT device near the victim~\cite{wei2018status}.
Moreover, the attacker possesses knowledge of the victim's room layout, enabling them to isolate sound vibrations not just from the speaker but also from everyday objects (e.g., chip bags, etc.) near the loudspeaker.
Additionally, the attacker can acquire speech samples of the victim from publicly available sources. Prior to initiating the attack, they might physically access the intended attack environment, play the victim's audio samples through a loudspeaker, and gather a dataset of eavesdropping signals with mmWave radar.

\todo{For optical sensor-based SSEA, consistent with recently proposed attacks \cite{sami2020lidarphone, nassi2020lamphone}, we assume the attacker can deploy their laser transmitter towards the reverberator without line-of-sight obstruction. The optical sensor captures the reflected laser, and the attacker accesses the ADC data, following similar assumptions as in mmWave-based SSEAs.}

For the motion sensor-based SSEA, we assume that the attacker can trick the victim into installing a malicious app, which collects motion signals in the background and can even stream the data to the adversaries' server. \revision{Note that detecting the malicious app is a challenging problem~\cite{yang2015appcontext}, as attackers can effectively disguise it among legitimate apps, allowing them to bypass security measures.}

%

\presec
\subsection{Defense Objectives}
\label{sec:design_objective}
\noindent\textbf{Design Goal.} 
\todo{\sys introduces adversarial examples to the original audio signals prior to playback, aiming to protect loudspeaker-generated voice from SSEAs with minimal impact on the intelligible voice quality.
Note that \sys cannot protect voice from a human speaker when SSEAs target eavesdropping on throat vibrations, a challenge also for existing SOTA attacks~\cite{walker2021sok}.}
To this end, \sys must meet five criteria. First, the adversarial voice audio generated by \sys should prevent the attacker from restoring audible and intelligent speech. Neither humans nor ML models should be able to transcribe the recovered audio into words and sentences. Second, the perturbation should be imperceptible to humans, and the perturbed speech audio must remain high quality. Third, since the defender does not know the attacker's ML models and attack scenario (e.g., audio volume, distance), the generated adversarial perturbations should be effective regardless of black-box knowledge. Fourth, \sys should be robust against adaptive attacks who know the presence of adversarial perturbations. \sys needs to enforce that perturbations are undetectable to attackers, leading to failure in attackers' attempts to remove perturbations from eavesdropped audio.
Finally, \sys should be applicable to both offline and online scenarios. In offline scenarios, such as intelligent speakers delivering private content to users, \sys runs without latency limitations on the user's device. For challenging online scenarios, such as VoIP,
\sys must meet low-latency requirements (e.g., $\leq$ 150ms for real-time VoIP communication~\cite{series2003transmission}).

\noindent\textbf{Defender's Capability.} To achieve the aforementioned design goals, \sys employs black-box perturbations to speech signals prior to playback, aiming to protect voice privacy from the sound source, i.e., loudspeakers, against side-channel eavesdropping.
We assume that \sys has access to the input audio of the voice communication device (i.e., loudspeaker), and can directly convert the input audio into adversarial audio before playing out. The defender follows black-box settings where he/she has no knowledge about the attack model (e.g., ML model and parameters) and scenario (e.g., distance, audio volume, etc.). 


\section{Preliminary Study} 

%
In this section, we investigate the fundamental differences between air-pressure-based sound-capturing methods (e.g., microphones and human hearing) and vibrometry-based side-channel attacks.
%
We further conduct a preliminary study to understand how to leverage these insights in the development of \sys's PGM.


\subsection{Understanding Side-Channel Attacks} \label{sec::preli:understand}
Microphones convert air pressure variations into electrical signals using a diaphragm, while vibrometry-based side channels measure the vibration displacement or acceleration of physical objects.
Due to these different mechanisms, \textit{microphones can detect sounds across the entire audible frequency range, while the SNR of vibrometry-based sensors drops sharply at higher frequencies.}
More specifically: i). The vibration displacement of a speaker diaphragm is approximately inversely proportional to the sound frequency to maintain consistent sound pressure across different frequencies, making it harder for side-channel sensors to detect high-frequency vibrations \cite{kinsler2000fundamentals, moser2009engineering}.
ii). Sound waves experience higher attenuation at higher frequencies when traveling through structural materials or air \cite{kinsler2000fundamentals}.
Although accelerometers can theoretically capture high-frequency sounds by measuring vibration acceleration, the actual frequency response is diminished due to structural propagation loss.
iii). Side-channel sensors are not designed for precise sound recording, typically having limited sampling rates and sensitivity to high frequencies. For instance, the maximum sampling rate of a smartphone accelerometer is about $500$ Hz, whereas capturing full-band speech signals requires at least $8$ kHz \cite{hu2022accear}.

\textit{The vibrometry-based side channels share consistent frequency response characteristics due to the aforementioned inherent limitations, which can be leveraged by the \sys defender.}
\todo{To validate this observation, we conduct experiments using three representative side channels: a COTS mmWave radar, an accelerometer embedded in a smartphone, and a laser microphone.}
Specifically, following the SSEA in ~\cite{hu2023mmecho,shi2023privacy}, we use the TI IWR1843-Boost radar~\cite{iwr1843} to sense speech-induced vibrations from an everyday object (i.e., tinfoil) while a loudspeaker (i.e., Edifier R1700BTs) plays acoustic signals. The chirp rate of the mmWave sensor is set to 10kHz. Figure~\ref{fig:feasibility} illustrates our basic experimental setup. Through a range-FFT operation, we isolate phase changes at the tinfoil's location in the mmWave data, and then apply a short-time Fourier transform (STFT) to these data points.
For the accelerometer SSEA, we follow \cite{sun2023stealthyimu,hu2022accear} and mount a smartphone (i.e., LG V50) in a car phone holder, as shown in Figure~\ref{fig:feasibility}(c). We then collect accelerometer data at the device's maximum sampling rate (i.e., 500 Hz).
\jungwoo{To build an optical-based SSEA, we aim a laser beam at the tinfoil vibrated by the loudspeaker, as shown in Figure~\ref{fig:feasibility}(d). Since the optical sensor converts the intensity of the laser reflected from the tinfoil into an electrical signal, we can effectively extract audio information~\cite{nassi2020lamphone}.}
For all the experiments, we employ two types of audio signals: a 3-second clip from the Librispeech corpus~\cite{panayotov2015librispeech} to visualize eavesdropping signals and sweep tones ranging from $50$ Hz to $4$ kHz to analyze frequency response.

\begin{figure}[t]
    \centering
    \begin{minipage}[b]{0.21\textwidth}
        \centering
        \resizebox{\linewidth}{!}{
        \begin{tabular}{@{}c@{}}
        \includegraphics[width=\linewidth]{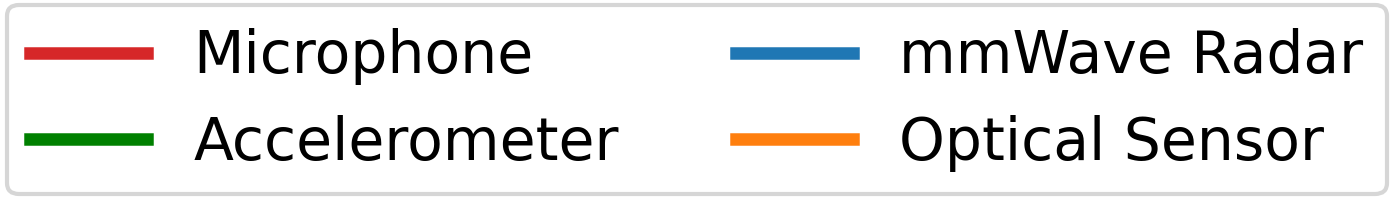} \\  
        \end{tabular}}
        \includegraphics[width=\linewidth]{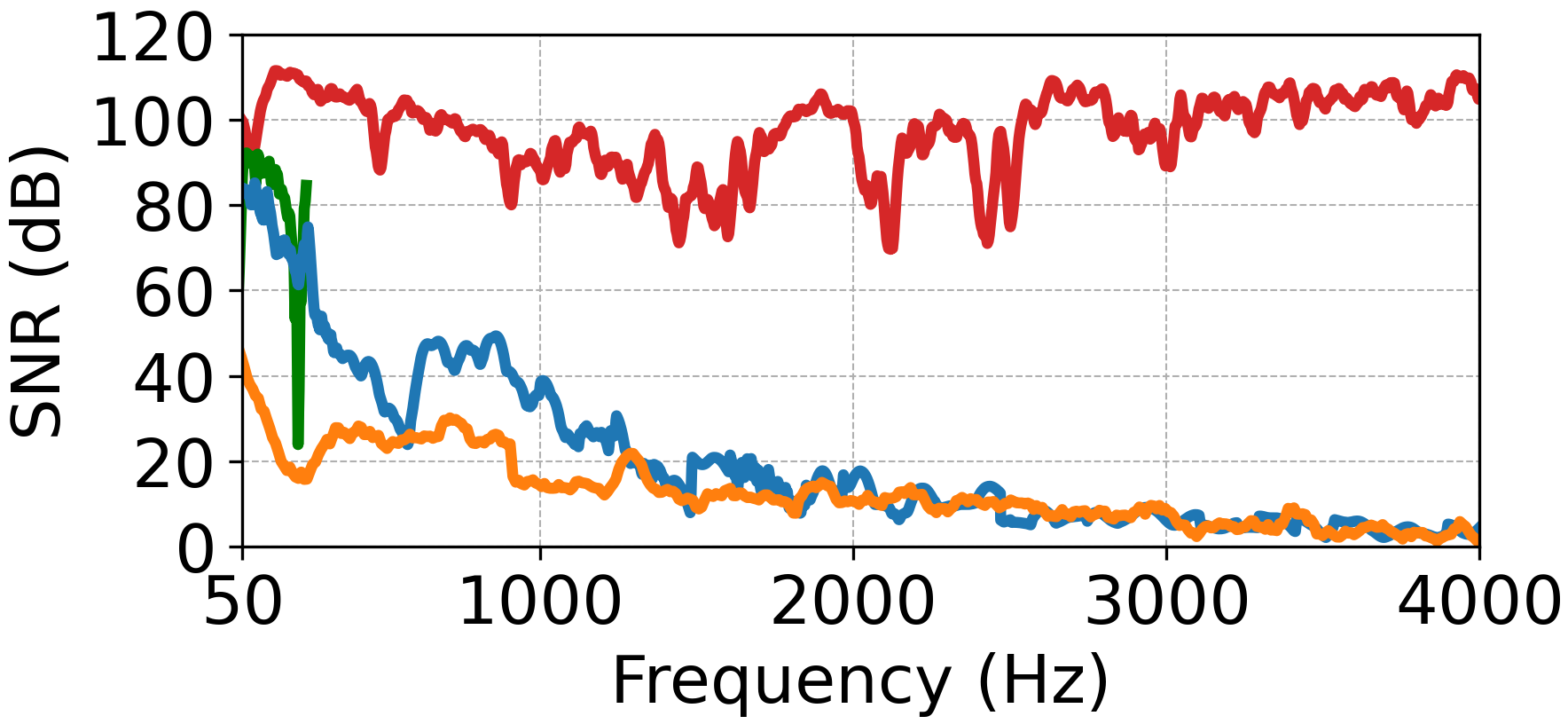} 
        \caption{The frequency responses of different side channels.}
        \label{fig:snr}
    \end{minipage}
    \hfill
    \begin{minipage}[b]{0.25\textwidth}
        \centering
        \resizebox{\linewidth}{!}{
        \begin{tabular}{@{}c@{}}
        \includegraphics[width=\linewidth]{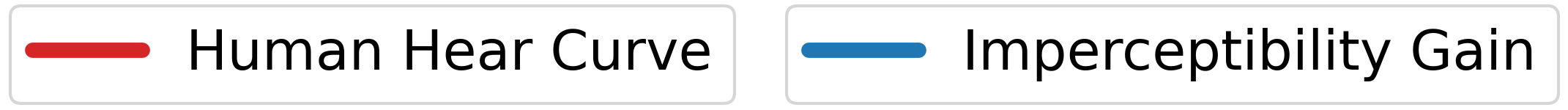} \\ 
        \end{tabular}}
        \includegraphics[width=\linewidth]{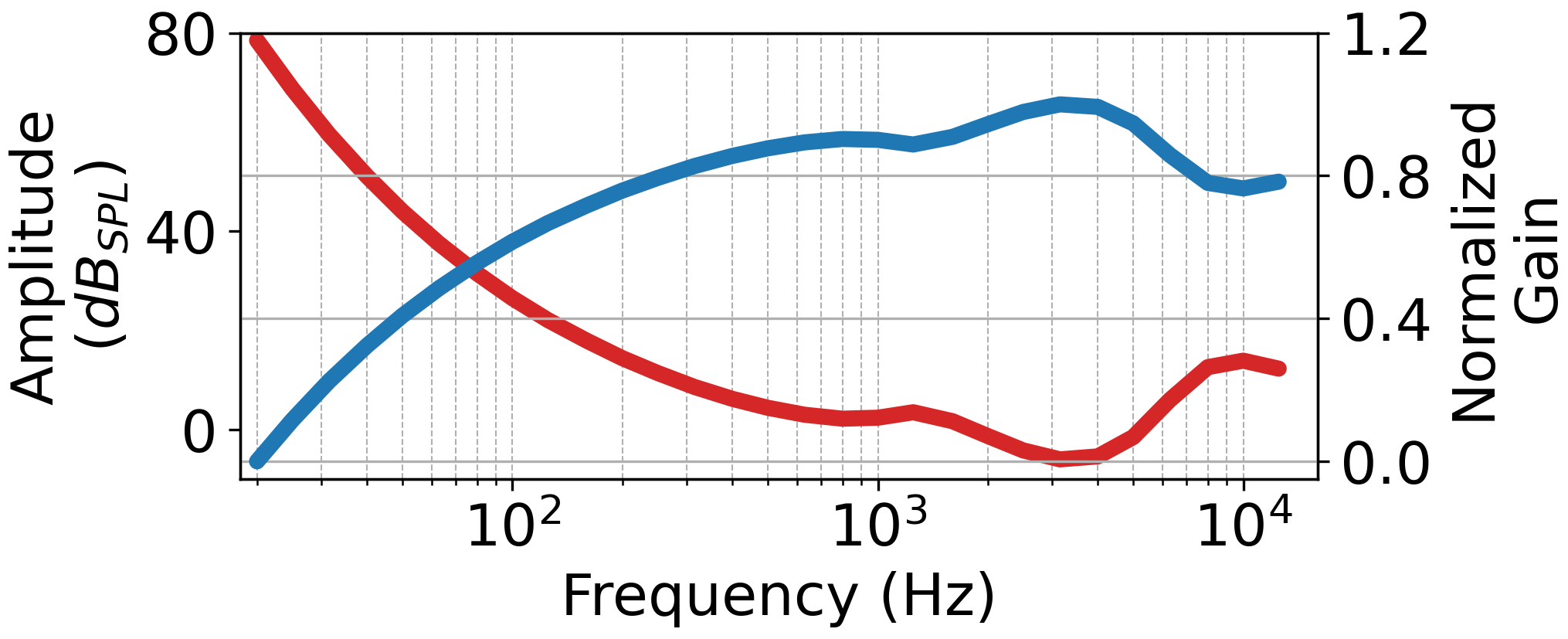}  \\
        \caption{Equal-soundness contour and imperceptibility gain curve.}
        \label{fig:human}
    \end{minipage}
\end{figure}

\begin{figure}[t]
    \centering
    \begin{tabular}{@{}cc@{}}
    \includegraphics[width=0.43\linewidth]{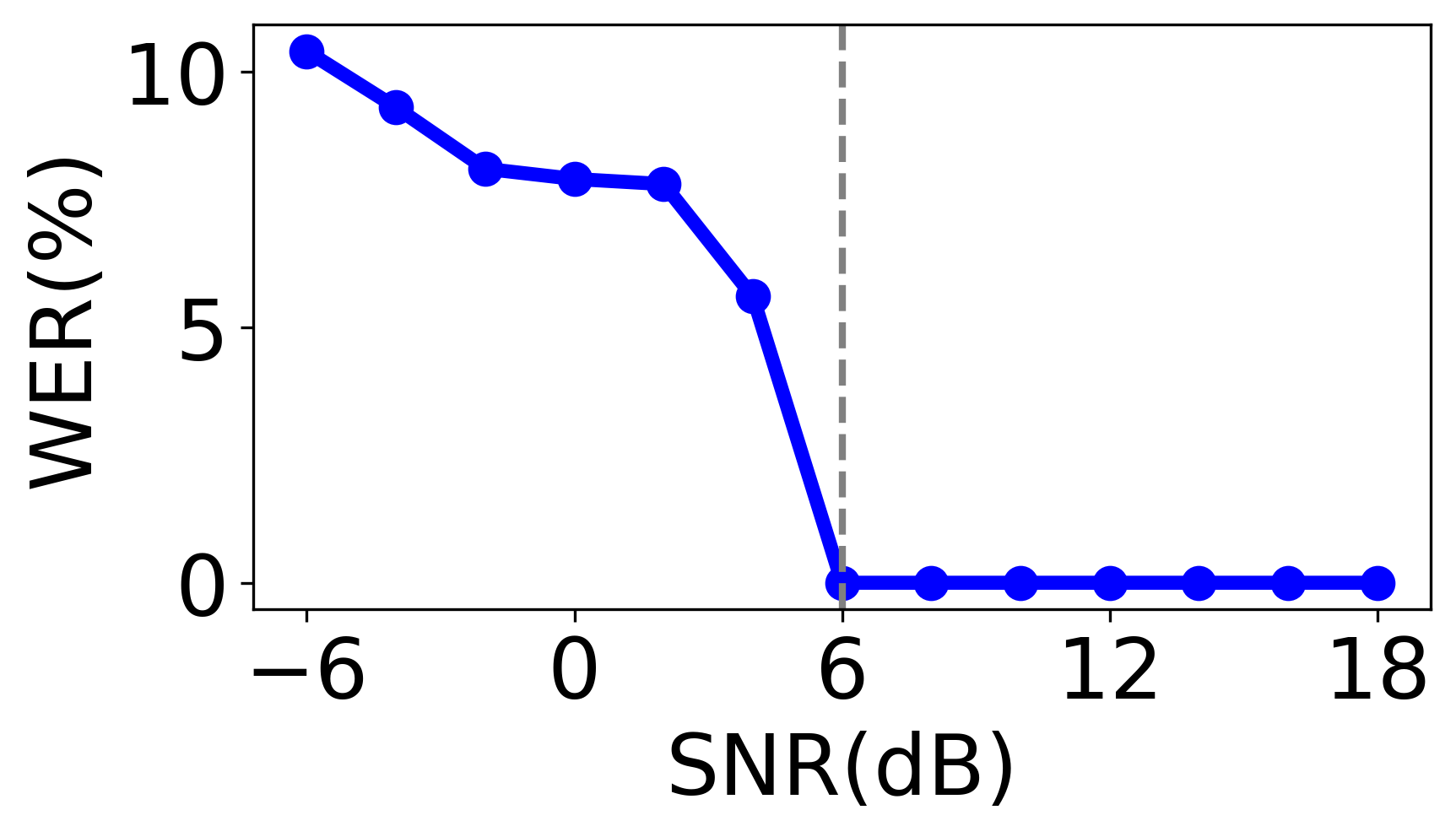} &
    \includegraphics[width=0.43\linewidth]{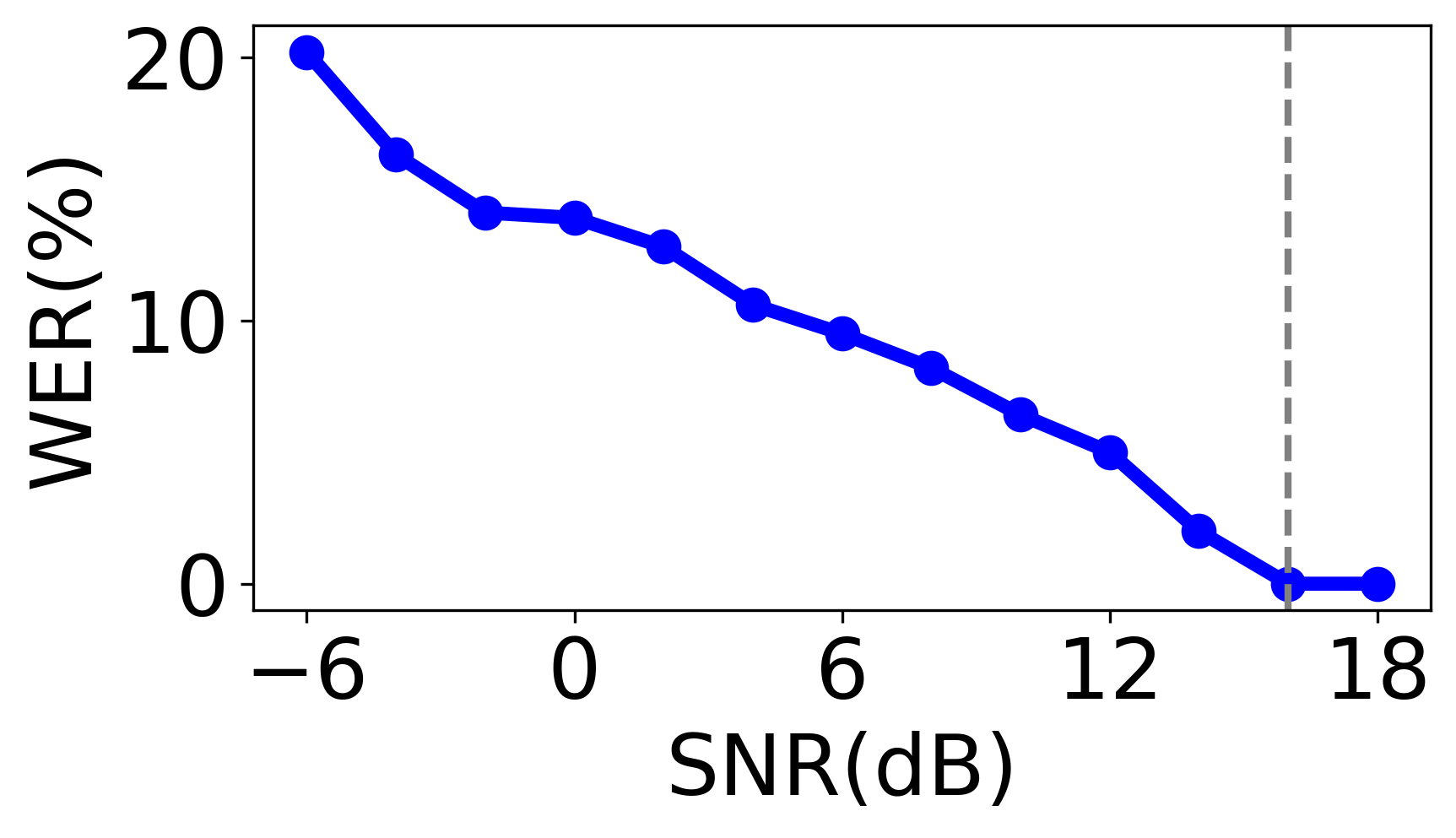} \\
    (a) AWGN (under 500Hz) &
    (b) AWGN (under 1000Hz) \\
    \end{tabular}
    \caption{Impact of AWGN with different frequency bands on human speech understanding.}
    \label{fig:feasibility_wer}
\end{figure}

\noindent\textbf{mmWave radar.}
We compare the spectrograms of the original and the radar-reconstructed speech in Figure~\ref{fig:feasibility}(a), \ref{fig:feasibility}(b).
We observe that the reconstructed speech signals maintain a high-frequency response in the low- and mid-frequency bands (i.e., $<$ 1~kHz).
However, the SNR tends to diminish beyond these frequencies, down to nearly 0~dB for frequencies above 2~kHz. 
This drop highlights the radar's limitations in detecting high-frequency vibrations. 
We further confirm that variations in the radar's sampling rate (see Figure~\ref{fig:snr}) or different attack scenarios (see Appendix~\ref{sec:append_frequency}) have little impact on its frequency response, suggesting these characteristics can be reliably utilized by \sys to defend against various SSEA radar hardware configurations and attack scenarios. 

\noindent\textbf{Accelerometer.}
According to the Nyquist sampling theorem, an accelerometer with a sampling rate of 500 Hz can capture data only up to 250 Hz.
As shown in Figure~\ref{fig:feasibility}(c) and Figure~\ref{fig:snr}, the audio reconstructed by the accelerometer is similar to the original audio in low-frequency components (i.e., $<$ 250 Hz). However, due to the accelerometer's limited sampling rates, the raw vibration signal loses mid- and high-frequency components (i.e., $>$ 250 Hz). The lost speech spectrum can be recovered by the audio enhancement, as shown in Figure~\ref{fig:result_spec}(c).
%

\noindent\textbf{Laser.} \jungwoo{As shown in Figure~\ref{fig:feasibility}(d) and Figure~\ref{fig:snr}, optical-based SSEA can recover wideband intelligible speech, but its sensing capability is worse than mmWave radar in most frequency bands. The frequency response can be enhanced with a high-end laser vibrometer (LV-FS01~\cite{wangvibspeech}), but this makes the attack device bulky and more easily identifiable.}

\presec
\subsection{Characterizing Human Hearing}

We then investigate human auditory sensitivity across different frequency ranges to devise an undetectable defense. Our study is based on equal-loudness contour~\cite{acoustics2003normal}, a well-established model that delineates the sound pressure level (SPL) perceived by the human ear across the frequency bands. We convert SPL measurements from $-20dB_{SPL}$ to $80dB_{SPL}$ into a normalized scale from $0$ to $1$ to visualize human insensitivity to different frequency bands.
Figure~\ref{fig:human} shows the hearing curve and the imperceptibility gain obtained through psychoacoustic experiments to describe the human ear's sensitivity~\cite{acoustics2003normal}. The hearing curve represents the amplitude required for a purely continuous tone of a certain frequency that humans can hear. 
Frequencies with high imperceptibility are harder to perceive by human. It shows that human ears are most sensitive to frequencies between 1.6~kHz and 4~kHz, with a marked insensitivity to frequencies below 500~Hz.
\jungwoo{Furthermore, we conducted an experiment to investigate the effects of low- and mid-frequency noise on human speech understanding. We asked 24 participants to listen to noisy audio containing either low-frequency or mid-frequency AWGN and then translate sentences. \revision{Please refer to Sec.~\ref{sec:user} for a detailed discussion of the ethical considerations in our study.} We calculated the WER based on their translations. From Fig~\ref{fig:feasibility_wer}(a), we observe that at SNRs above 6dB, low-frequency noise does not impact human understanding. However, as shown in Fig~\ref{fig:feasibility_wer}(b), for participants to accurately hear audio containing mid-frequency noise, the SNR must be over 16dB. Furthermore, SNR and defense success rate have an inverse relationship (see Sec.~\ref{sec:ablation}).}
\emph{Thus, to guarantee the sound quality to the human ear, \sys can mainly generate perturbations within the low-frequency bands (i.e., $<$ 500 Hz) while minimally affecting the mid- and high-frequency ranges (i.e., $>$ 500 Hz).} 

\begin{figure}[t]
\centering
\begin{tabular}{@{}c@{}}
    \includegraphics[width=0.94\linewidth]{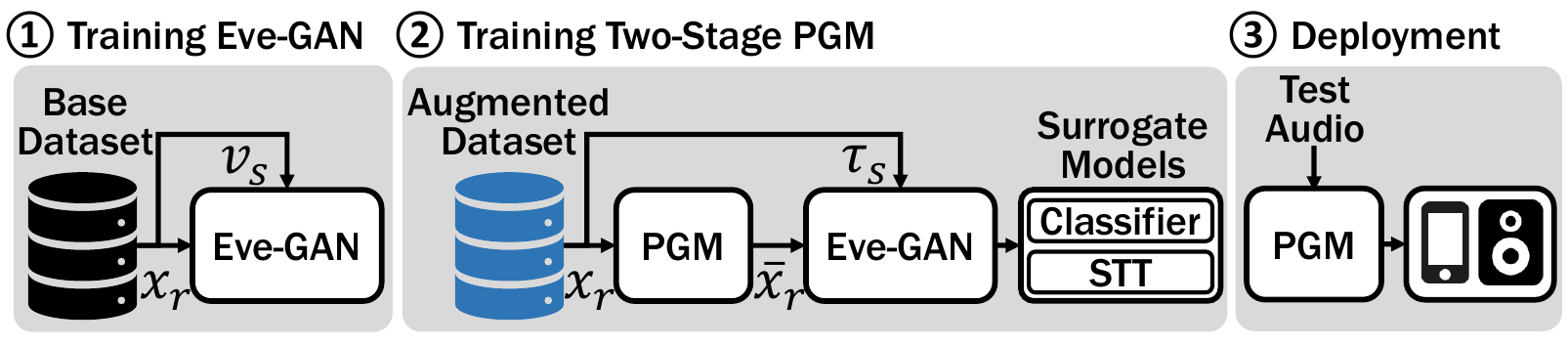} \\
\end{tabular}
\caption{\sys training and deployment process.}
\label{fig:overview}
\end{figure}

\presec
\section{Design of \sys} 
\label{sec:defense}
In this section, we present the optimization problem and methodology for designing \sys. 
An overview of the \sys system and its modules is provided in Figure~\ref{fig:overview}.

\subsection{Problem Formulation} 
\sys aims to automate and optimize robust audio adversarial perturbations to protect loudspeaker-generated voice from SSEAs while maintaining the quality of legitimate voice communication.
Suppose an attacker eavesdrops on a $n$-dimensional victim's speech $x_{r}\in[-1, 1]^{n}$ with sampling rate $r$ and reconstruct a waveform $\mathcal{A}_{s}(x_{r}, \zeta_{s})$ via his/her audio reconstruction method $\mathcal{A}_{s}$, where $\zeta_{s}$ is a vector representing the attack scenarios (e.g., distance, orientation, materials, hardware configurations, and loudspeaker's volume, etc.) for a specific sensor $s \in \{\text{side channel sensors}\}$. 
The attacker then uses the reconstructed audio as input to the following ML models: (1) a speech recognition model $M^{sr}_{s}(\cdot)$ that converts audio into text transcriptions and (2) an audio classifier $M^{ac}_{s}(\cdot)$ that identifies specific digits or keywords. Note that ML models have different model parameters depending on the sensor type $s$.
\sys applies adversarial perturbations to prevent eavesdroppers from recovering audible speech.
The objective is to find a minimal perturbation $\delta$ as follows: 
\begin{equation} \label{eq:general1}
\resizebox{0.87\columnwidth}{!}{$
\begin{aligned}
\operatorname*{arg\,max}_{\delta} \quad & \underset{x_{r}, \zeta_{s}}{\mathbb{E}} [\mathcal{L}_{qd}(\mathcal{A}_{s}(x_{r}+\delta, \zeta_{s}), {A}_{s}(x_{r}, \zeta_{s}))] - \alpha \norm{\delta}_{2}, \\
\text{subject to}
\quad & WER(M^{sr}_{s}(\mathcal{A}_{s}(x_{r}+\delta, \zeta_{s})), y_{sr}) > t_{sr}, \\
\quad & M^{ac}_{s}(\mathcal{A}_{s}(x_{r}+\delta, \zeta_{s})) \neq y_{ac},
\end{aligned}$}
\end{equation}
where $\mathcal{L}_{qd}$ is the loss that measures the quality difference between the eavesdropping results from clean and perturbed audio.
$WER(\cdot)$ (Word Error Rate) is a metric to assess speech recognition~\cite{wang2003word}. It calculates accuracy by dividing the number of errors by the total number of words in the reference $y_{sr}$. $t_{sr}$ is a threshold that determines the success of our defense. $y_{ac}$ is a label for audio classification. $\alpha$ is a hyper-parameter that controls the relative importance of imperceptibility of $\delta$ and defense performance, respectively. 

 

The key challenge of \sys is how to automate and optimize the adversarial perturbations $\delta$ to solve Eq.~\ref{eq:general1}. Specifically, \sys needs to address four challenges. Firstly, it must model the audio reconstruction $\mathcal{A}_{s}$ without laborious data collection across numerous eavesdropper hardware configurations and audio profiles. 
Secondly, side-channel eavesdropping characteristics must be considered when modeling $\delta$. Otherwise, the optimization may become stuck in local optima. 
Thirdly, the \sys defender has no knowledge of the ML model and parameters used by the eavesdropper (i.e., $M^{sr}_{s}(\cdot)$ and $M^{ac}_{s}(\cdot)$), and cannot even perform black-box queries.
Finally, \sys must be immune to adaptive attackers who attempt to learn the perturbation and denoise it from the eavesdropped speech.

\begin{figure*}[t]
\centering
\includegraphics[width=0.83\textwidth]{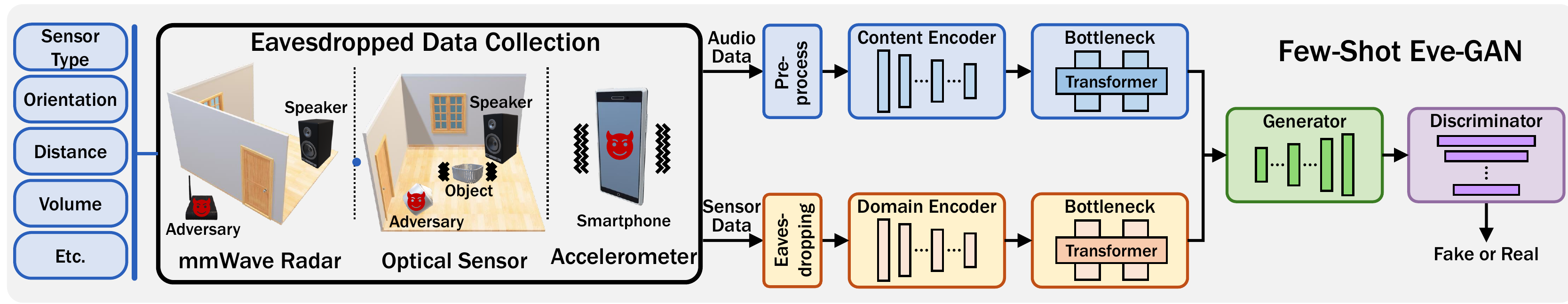}
\caption{The general workflow of few-shot \gan consists of a few-shot audio translator and discriminator.}
\label{fig:gan}
\end{figure*}

\presec
\subsection{Overview of \sys}
\label{sec:overview}
To address them, we designed \sys to train and deploy end-to-end, comprising three major phases, as illustrated in Figure \ref{fig:overview}.

\noindent\textbf{Phase $\#$1 - Training \gan (Sec. \ref{sec:proposed_gan}).} 
To achieve automatic optimization of these perturbations, we model the SSEA audio reconstruction process $\mathcal{A}_{s}(x_{r}, \zeta_{s})$ within a differentiable framework, allowing PGM to learn the distribution of adversarial examples end-to-end. Specifically, we design a deep generative network called \gan to convert audio signals into eavesdropped data. To ensure generalization and reduce data collection effort, we propose a few-shot audio-to-SSEA translator that trains with a base dataset consisting only of unpaired audio-SSEA data.

\noindent\textbf{Phase $\#$2 - Training \pgm (Sec. \ref{sec:proposed_pgm}).}
Then, we aim to train the PGM using a set of surrogate models to enhance the transferability of adversarial examples. We concatenate the PGM with the few-shot translator and surrogate models (i.e., $M^{sr}_{s}(\cdot)$ and $M^{ac}_{s}(\cdot)$), setting all ML modules except PGM as non-trainable to allow end-to-end gradient backpropagation.
To ensure scenario-agnostic perturbations resilient to variations in SSEA scenarios $\zeta_{s}$, we use the pretrained few-shot audio-to-SSEA translator in the first phase to augment the base dataset by generating eavesdropping signals in unseen SSEA domains.
Finally, we optimize the PGM to minimize the intelligibility of audio in the augmented SSEA domain.


\noindent\textbf{Phase $\#$3 - Deployment.} Once the PGM's training process is complete, we place the trained PGM prior to the audio sources to convert the original audio into perturbed audio in the audio front-end processing pipeline.
\sys operates within acceptable latencies for Voice-over-IP (VoIP) applications, making it suitable for real-time VoIP communications (see Sec.~\ref{sec:runtime}).

\presec
\subsection{Modeling of Few-Shot \gan}
\label{sec:proposed_gan}
We first devise \gan (Figure \ref{fig:gan}), a deep generative model that establishes a non-linear relationship between the original audio and the eavesdropped audio (i.e, $\mathcal{A}_{s}(x_{r}, \zeta_{s})$) within a differentiable framework. 
\gan learns to extract generalizable style patterns that can be applied to unseen SSEA samples. It leverages few-shot unpaired audio-to-SSEA translation to alleviate the data collection overhead.


\noindent\textbf{SSEA Data Collection.} Our few-shot approach addresses two major issues: (i) obtaining paired training data and (ii) collecting SSEA samples under infinite attack scenarios.
To achieve this, we construct a \textit{base dataset} that facilitates the learning of generalizable translation capabilities.
The base dataset generally requires diverse data samples~\cite{liu2019few, ding2020rf}.
Thus, we collect data by considering several crucial factors that determine the SSEA sensors' vibration sensing capability~\cite{hu2022accear, hu2023mmecho, shi2023privacy, sun2023stealthyimu, hu2022milliear}. \todo{We use mmWave-based SSEA as an example for illustration and include the details of other SSEA in Table~\ref{tab:data1} and Table~\ref{tab:data_laser}.}
As shown in Table~\ref{tab:data2}, we thoroughly consider a total of $72=2^{3}\times3^{2}$ attack scenarios for mmWave radar, respectively. Within this base dataset, we collect $120$ samples per scenario for mmWave radar, ensuring a comprehensive representation of potential attack vectors. 
Similarly, we construct base datasets from accelerometers and optical sensors, as shown in Table~\ref{tab:data1} and Table~\ref{tab:data_laser} and collect 450$\sim$500 samples per scenario.

\begin{table}[b]
\vspace{+0.1in}
\caption{Defender's dataset settings for mmWave radar-based SSEAs, where Loud$_{1}$ refers to Logitech Z313. V, O, and R denote voice source, reverberating object, and mmWave radar, respectively.}
\centering
\footnotesize
\label{tab:data2}
\resizebox{0.96\linewidth}{!}{
\midsepremove
\begin{tabular}{ ccccccc }
\toprule
Dataset & \begin{tabular}[c]{@{}c@{}}Voice\\ Source\end{tabular} & Material & \begin{tabular}[c]{@{}c@{}}V-to-O\\ Distance\end{tabular} & \begin{tabular}[c]{@{}c@{}}R-to-O\\ Distance\end{tabular} & \begin{tabular}[c]{@{}c@{}}R-to-O\\ Angle\end{tabular} &  \begin{tabular}[c]{@{}c@{}}Audio\\ Volume\end{tabular}   \\ 
\hline
Base  & \begin{tabular}[c]{@{}c@{}}Loud$_{1}$\end{tabular} & \begin{tabular}[c]{@{}c@{}} tinfoil \\ chip bag \\ carton \end{tabular} & \begin{tabular}[c]{@{}c@{}} 0.5m \\ 1.5m \end{tabular} & \begin{tabular}[c]{@{}c@{}} 0.5m \\ 1.5m \end{tabular} & \begin{tabular}[c]{@{}c@{}} $-15^{\circ}$ \\ $0^{\circ}$ \\ $15^{\circ}$ \end{tabular}  & \begin{tabular}[c]{@{}c@{}} 70dB \\ 80dB \end{tabular}  \\ 
\hline
Few-Shot & \begin{tabular}[c]{@{}c@{}}Loud$_{1}$\end{tabular} & \begin{tabular}[c]{@{}c@{}}plastic\\ cotton \\ paper\end{tabular} & \begin{tabular}[c]{@{}c@{}} 0.5m \\ 1.5m \\  \end{tabular}  & \begin{tabular}[c]{@{}c@{}} 0.5m \\ 1.5m \end{tabular} & \begin{tabular}[c]{@{}c@{}} $-15^{\circ}$ \\ $0^{\circ}$ \\ $15^{\circ}$ \end{tabular} & \begin{tabular}[c]{@{}c@{}} 70dB \\ 80dB \end{tabular}\\
\bottomrule
\end{tabular}}
\end{table}

After the few-shot \gan is trained on the base dataset, we can convert original audio into eavesdropping signals in an unseen SSEA scenario. We leverage this few-shot capability to aid the PGM in learning the scenario-invariant perturbation distribution with minimal additional SSEA samples. To this end, we create a \textit{few-shot dataset} consisting of one sample per unseen scenario that is not included in the base dataset, as shown in Table~\ref{tab:data2}. Then, we integrate base and few-shot datasets into an augmented dataset for PGM training, as depicted in Figure~\ref{fig:overview}. 






\noindent\textbf{Few-Shot Audio Translator.} 
Next, we use the \textit{base dataset} to train a few-shot translator $\hat{x}_{r,s} = T_{r,s}(x_{r}, v_{s})$ that transforms original audio $x_{r}$ into eavesdropped audio $\hat{x}_{r,s} \in [-1, 1]^{n}$ with the domain of a given SSEA example $v_{s}\in [-1, 1]^{n}$. Note that the sampling rate of $\hat{x}_{r,s}$ is not the same as that of $v_{s}$. 
The few-shot translator consists of several modules.
The pre-processing module resamples $x_{r}$ to a sampling rate of $v_{s}$ through the differentiable resampling operation~\cite{yang2022torchaudio} and applies zero-mean normalization to the data. The content encoder, comprised of 1-D convolutional (Conv1d) layers, maps the human speech to a content latent code. The domain encoder consists of a stack of Conv1d layers to produce SSEA information. We introduce the bottleneck extractor to refine the representation. Then, the decoder has several 1-D adaptive instance normalization (AdaIN) residual blocks~\cite{huang2018multimodal} followed by upscale Conv1d layers. By feeding the content and SSEA latent codes to the decoder, we ensure that the reference SSEA sample $v_{s}$ controls the output domain while the victim's speech determines the utterance content.

\noindent\textbf{Discriminator.} Our goal is to make $\hat{x}_{r,s} = T_{r,s}(x_{r}, v_{s})$ close to the real eavesdropped audio. To this end, we adopt an adversarial training method where a discriminator $D^{e}_{r,s}$ learns to distinguish between real-world eavesdropped audio and fake audio generated by $T_{r,s}$. We adopt the multi-period discriminator in~\cite{kong2020hifi}.


\begin{table}[b]
\vspace{+0.1in}
\caption{Defender's dataset settings for IMU sensor (i.e., Accelerometer)-based SSEAs.}
\centering
\footnotesize
\label{tab:data1}
\makebox[\linewidth][c]{
\midsepremove
\begin{tabular}{ ccccc }
\toprule
Dataset & \begin{tabular}[c]{@{}c@{}}Smartphone\\ Model\end{tabular} & \begin{tabular}[c]{@{}c@{}}Sampling\\ Rate\end{tabular} & \begin{tabular}[c]{@{}c@{}}Surface of\\ Placement\end{tabular} &  \begin{tabular}[c]{@{}c@{}}Audio\\ Volume\end{tabular}   \\ 
\hline
Base  & \begin{tabular}[c]{@{}c@{}}Samsung S20\end{tabular} & \begin{tabular}[c]{@{}c@{}} 200Hz \\ 500Hz  \end{tabular} & \begin{tabular}[c]{@{}c@{}} table \\ sofa \\ floor \end{tabular}  & \begin{tabular}[c]{@{}c@{}} $60\%$ \\ $80\%$ \\ $100\%$ \end{tabular}  \\ 
\hline
Few-Shot & \begin{tabular}[c]{@{}c@{}}Samsung S20\end{tabular} & \begin{tabular}[c]{@{}c@{}} 200Hz \\ 500Hz \end{tabular} & \begin{tabular}[c]{@{}c@{}} Phone holder \\ handhold \\ bed \end{tabular} & \begin{tabular}[c]{@{}c@{}} $60\%$ \\ $80\%$ \\ $100\%$ \end{tabular}\\
\bottomrule
\end{tabular}}
\end{table}

\begin{table}[b]
\vspace{+0.1in}
\caption{Defender's dataset settings for optical sensor-based SSEAs.}
\centering
\footnotesize
\label{tab:data_laser}
\makebox[\linewidth][c]{
\midsepremove
\begin{tabular}{ ccccc }
\toprule
Dataset & \begin{tabular}[c]{@{}c@{}}Laser-to-O\\ Distance\end{tabular} & \begin{tabular}[c]{@{}c@{}}Sensor-to-O\\ Distance\end{tabular} & Material &  \begin{tabular}[c]{@{}c@{}}Audio\\ Volume\end{tabular}   \\ 
\hline
Base  & \begin{tabular}[c]{@{}c@{}}0.5m \\ 1.0m \\ 1.5m \end{tabular} & \begin{tabular}[c]{@{}c@{}} 0.05m \\ 0.1m \\ 0.15m \end{tabular} & \begin{tabular}[c]{@{}c@{}} tinfoil \\ chip bag \end{tabular}  & \begin{tabular}[c]{@{}c@{}} 70dB \\ 80dB \end{tabular}  \\ 
\hline
Few-Shot & \begin{tabular}[c]{@{}c@{}}0.5m \\ 1.0m \\ 1.5m \end{tabular} & \begin{tabular}[c]{@{}c@{}} 0.05m \\ 0.1m \\ 0.15m \end{tabular} & \begin{tabular}[c]{@{}c@{}} plastic \\ cotton \\ paper \end{tabular} & \begin{tabular}[c]{@{}c@{}} 70dB \\ 80dB \end{tabular}\\
\bottomrule
\end{tabular}}
\end{table}

\begin{figure*}[t]
\centering
\includegraphics[width=0.83\textwidth]{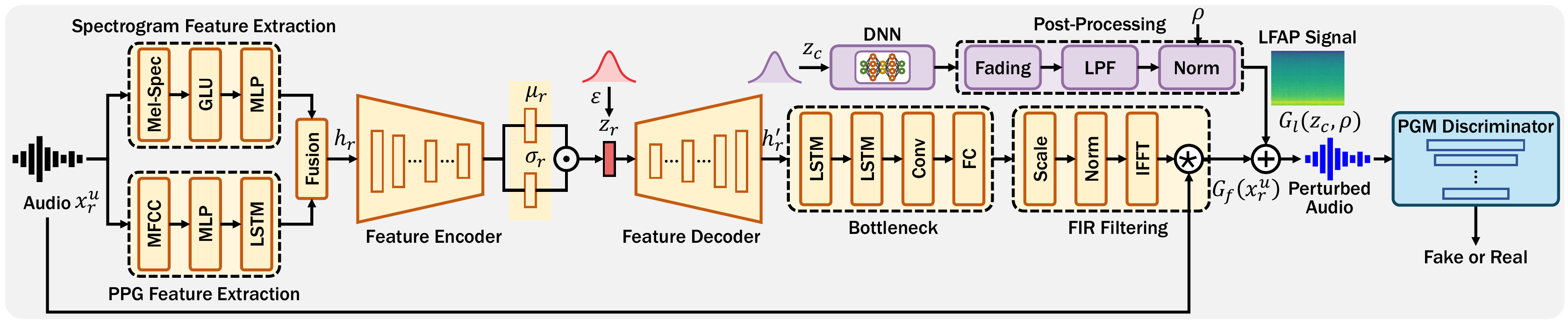}
\caption{Visualization of \pgm. Yellow modules form the FIR generator; purple modules form the LFAP generator.}
\label{fig:pgm}
\end{figure*}

\noindent\textbf{Training Loss.} We train the proposed few-shot \gan by solving a minimax optimization problem given by:
\begin{equation} \label{eq:general2}
\resizebox{0.87\columnwidth}{!}{$
\begin{aligned}
\operatorname*{max}_{D^{e}_{r,s}} \operatorname*{min}_{T_{r,s}} \ \mathcal{L}_{gan}(T_{r,s},D^{e}_{r,s}) + \beta_{con} \mathcal{L}_{con}(T_{r,s}) + \beta_{fm} \mathcal{L}_{fm}(T_{r,s}),
\end{aligned}$}
\end{equation}
where $\beta_{con}$ and $\beta_{fm}$ are hyper-parameters for each term. $\mathcal{L}_{gan}$, $\mathcal{L}_{con}$, and $\mathcal{L}_{fm}$ are the GAN loss~\cite{goodfellow2014generative}, the consistency loss~\cite{zhu2017unpaired}, and the feature matching loss~\cite{larsen2016autoencoding}. We define each loss function as:

\noindent $\bullet$ \textbf{GAN Loss.} We obtain the GAN loss as:
    \begin{equation} \label{eq:general3}
    \resizebox{0.87\columnwidth}{!}{$
    \begin{aligned}
    \mathcal{L}_{gan} = \underset{v_{s}}{\mathbb{E}}[\log D^{e}_{r,s}(v_{s})] + \underset{x_{r},v_{s}}{\mathbb{E}}[\log (1-D^{e}_{r,s}(T_{r,s}(x_{r}, v_{s})))].
    \end{aligned}$}
    \end{equation}

\noindent $\bullet$ \textbf{Consistency Loss.} The consistency loss encourages the model to preserve the properties of the original audio, ensuring that the content of the utterance is preserved. When the original audio is used on both inputs of $T_{r,s}$, the result should be identical to the input. We calculate the consistency loss as follows:
    \begin{equation} \label{eq:general4}
    \begin{aligned}
    \mathcal{L}_{con} = \underset{x_{r}}{\mathbb{E}}[\norm{x_{r} - T_{r,s}(x_{r}, x_{r})}^{1}_{1}].
    \end{aligned}
    \end{equation}

\noindent $\bullet$ \textbf{Feature Matching Loss.} The feature matching loss improves the stability of the training and the quality of the translation outputs. To this end, we design a feature extractor $D^{m}_{r,s}$, which is a model excluding the last (prediction) layer of $D^{e}_{r,s}$. We compute the loss by extracting features from the translation output and reference SSEA example as:
    \begin{equation} \label{eq:general5}
    \begin{aligned}
    \mathcal{L}_{fm} = \underset{x_{r},v_{s}}{\mathbb{E}}[\norm{D^{m}_{r,s}(T_{r,s}(x_{r}, v_{s})) - D^{m}_{r,s}(v_{s})}^{1}_{1}].
    \end{aligned}
    \end{equation}

\noindent\textbf{Inference Stage.} After the training is completed, the few-shot translator is used as a non-trainable differentiable layer, which helps the PGM to find robust perturbations. As shown in Figure~\ref{fig:overview}, the PGM is located in front of the few-shot audio translator. By feeding a perturbed audio $\bar{x}_{r}$ to the few-shot translator, we enable cross-domain conversion as:
\begin{equation} \label{eq:general_perturbed}
\begin{aligned}
\centering
\tilde{x}_{r,s} = T_{r,s}(\bar{x}_{r}, \tau_{s}),
\end{aligned}
\end{equation}
where $\tau_{s}$ is an SSEA example sampled from augmented datasets consisting of the base and few-shot datasets and $\tilde{x}_{r,s}$ is the domain conversion result of the perturbed audio $\bar{x}_{r}$. Note that the sampling rate of $\tilde{x}_{r,s}$ follows that of $\tau_{s}$.


\presec
\subsection{Modeling of PGM}
\label{sec:proposed_pgm}
With pretrained \gan, we can train the \pgm end-to-end. Our optimization goal is to generate robust adversarial examples against SSEAs with minimal impact on human auditory perception. Guided by the unique characteristics of SSEA, we transform the original audio $x_{r}$ into adversarial audio $\bar{x}_{r}$ using two principles: adversarial FIR filtering, which perturbs the audio in the frequency domain, and 
low-frequency adversarial perturbations (LFAPs), which are negligible to the human ear. 
We also build a set of surrogate models for multiple SSEAs to ensure that the adversarial examples have strong transferability through ensemble learning. This approach helps to generate generalizable perturbations despite the lack of knowledge about the attackers' speech-processing models.
Additionally, we apply a robustness constraint to the optimization problem, making \sys unlearnable by adaptive eavesdroppers.

Figure~\ref{fig:pgm} shows PGM's architecture, comprising three main modules: (1) an FIR generator $G_{f}$ learns the perturbation distribution in the frequency domain to avoid noisy artifacts common in additive attacks; (2) an LFAP generator $G_{l}$ uses low-frequency adversarial perturbations ($< 500$ Hz) to prevent the attacker from restoring speech, leveraging the unique frequency response characteristics of the sensors, and (3) a discriminator $D^{p}_{r}$ is employed to distinguish whether the PGM generates real or fake audio, which helps to make $\bar{x}_{r}$ close to $x_{r}$ to prevent an adaptive attacker from learning our perturbation patterns. \jungwoo{To make \sys suitable for the audio streaming, PGM continuously processes $x_{r}^{u}$, which is a segment of $x_{r}$, where $u$ denotes the segment index.}


\noindent\textbf{FIR Generator.} 
Conventional FIR generators~\cite{o2022voiceblock, o2022effective} are primarily used to disrupt speaker recognition systems. However, these generators are designed for microphone recordings and are equally sensitive across the audible frequency range. Their formulations do not consider domain constraints for eavesdropping side channels, thus making them ineffective for SSEA side channels.  
Furthermore, their perturbation lacks variability as they do not enforce a robustness constraint. Thus, an adaptive attacker aware of the existence of defense can easily identify perturbations and then build a robust SSEA with adversarial training. 

To overcome these deficiencies, we incorporate the VAE-GAN architecture~\cite{gur2020hierarchical} into the conventional FIR generator~\cite{o2022voiceblock}.
Instead of the FIR filter being fixed to the input, VAE-GAN enables $G_{f}$ to produce diverse FIR filters. To this end, we first extract representative acoustic features, including spectrogram features~\cite{kaneko2020cyclegan}, phonetic posteriorgrams (PPG)~\cite{ronssin2021ac}. Then, the feature encoder maps acoustic features $h_{r}$ extracted from input audio $x_{r}^{u}$ to a latent vector $z_{r}$ with a distribution $p(z_{r}|h_{r})$. Here, $z_{r}=\mu_{r}+\sigma_{r} \odot \upvarepsilon$, where $\upvarepsilon\sim N(0,I)$. With the given latent vector $z_{r}$, the feature decoder restores $h'_{r}$ via a distribution $q(h'_{r}|z_{r})$. The restored features $h'_{r}$ are sent as input to the bottleneck and decoded into a frequency-domain FIR filter. 
Lastly, the filtering module converts the FIR filter to the time domain and then performs time-varying filtering on the input audio. 

\noindent\textbf{LFAP Generator.} 
$G_{l}$ has a multi-layer Deep Neural Network (DNN) consisting of a stack of fully connected (FC) layers with ReLU activation except for the last layer with Tanh activation. Upon receiving a random latent vector $z_{c}$, the DNN produces a perturbation vector of length 512. 
$G_{l}$ then proceeds with post-processing techniques to suppress the high-frequency components and the perturbation audibility. Specifically, an audio fading~\cite{borowski2018experimental} ensures smooth transitions when connecting perturbation vectors to match the length of $G_{f}(x_{r}^{u})$. We then apply biquad low-pass filter to suppress the high-frequency components with the cut-off frequency of 500~Hz. Lastly, we adopt a normalization process to limit the amplitude of the LFAP. Here, we set the Signal-to-Noise Ratio (SNR) as our normalization, which closely approximates the masking effect in the human auditory system~\cite{brungart2001informational}. We empirically find an optimal SNR $\rho$ such that the source signal thoroughly dominates the perception of the perturbing signal while preventing SSEAs.

\noindent\textbf{Training Loss.} 
We denote the processes that generate adversarial audio as $\bar{x}_{r}^{u}=G_{r}(x_{r}^{u},z_{c}, \rho)=G_{f}(x_{r}^{u}) + G_{l}(z_{c}, \rho)$.
According to the optimization goal in Eq.~\ref{eq:general1}, we formulate the following objectives to train $G_{r}$ and $D^{p}_{r}$:
\begin{equation} \label{eq:general6}
\begin{aligned}
\operatorname*{max}_{D^{p}_{r}} \operatorname*{min}_{G_{r}} \quad & \mathcal{L}_{adv}(G_{r},D^{p}_{r}) + \lambda_{kl} \mathcal{L}_{kl}(G_{r})   \\
& + \lambda_{ens}\mathcal{L}_{ens}(G_{r}) + \lambda_{rec} \mathcal{L}_{rec}(G_{r}), 
\end{aligned}
\end{equation}
where $\mathcal{L}_{adv}$, $\mathcal{L}_{kl}$, $\mathcal{L}_{ens}$, and $\mathcal{L}_{rec}$ are the GAN loss, the KL loss, the ensemble loss, and the reconstruction loss. $\lambda_{kl}$, $\lambda_{ens}$, and $\lambda_{rec}$ are weight parameters. With this loss function, we iteratively train $G_{r}$ and $D^{p}_{r}$ until reaching equilibrium. We define each loss function as:

\noindent $\bullet$ \textbf{Adversarial Loss.} We define the adversarial loss as: 
    \begin{equation} \label{eq:general7}
    \begin{aligned}
    \centering
    \mathcal{L}_{adv} = \underset{x_{r}}{\mathbb{E}}[\log D^{p}_{r}(x_{r})] + \underset{\bar{x}_{r}}{\mathbb{E}}[\log (1-D^{p}_{r}(\bar{x}_{r}))].
    \end{aligned}
    \end{equation}

\noindent $\bullet$ \textbf{KL Loss.} We employ the KL loss to diminish the gap between the posterior distribution $p(z_{r}|h_{r})$ and the prior distribution as:
    \begin{equation} \label{eq:general8}
    \begin{aligned}
    \centering
    \mathcal{L}_{kl} = \underset{h_{r}}{\mathbb{E}}[\mathbb{D}_{KL}(p(z_{r}|h_{r}) || N(0,I))],
    \end{aligned}
    \end{equation}
    where $\mathbb{D}_{KL}$ means the KL divergence. The prior is assumed to follow a multivariate normal distribution.  

\noindent $\bullet$ \textbf{Ensemble Loss.} To subvert an eavesdropper's ML model, we build an ensemble of surrogate models with $K$ configurations of SSEA for speech recognition and audio classifier. Then, the few-shot translator $T_{r,s}$ bridges the gap between the PGM and surrogate models to calculate $\mathcal{L}_{ens}$ as:
    \begin{equation} \label{eq:general9}
    \begin{aligned}
    \centering
    \mathcal{L}_{ens} = & \underset{\bar{x}_{r},\tau_{s}}{\mathbb{E}} [\sum_{k=1}^{K} (\log \PP(y_{sr} | \tilde{x}_{r,s}) + Y^{k}_{s}(y_{ac} | \tilde{x}_{r,s}))], \\
    \end{aligned}
    \end{equation}
    where $\{\PP(y_{sr} | \tilde{x}_{r,s})\}^{K}_{k=1}$ is a set of predicted probability that will be transcribed into $y_{sr}$ following a Connectionist Temporal Classification (CTC) loss~\cite{graves2006connectionist}. $\{Y^{k}_{s}(y_{ac} | \tilde{x}_{r,s})\}^{K}_{k=1}$ is a set of the probability belonging to $y_{ac}$. $\tilde{x}_{r,s}$ is obtained from Eq.~\ref{eq:general_perturbed}. Note that the model parameters for speech recognition and audio classifier are different depending on sensor type $s$ and ensemble index $k$.

\noindent $\bullet$ \textbf{Reconstruction Loss.} To ensure that our perturbations are undetectable to the human ear but cause failure of SSEA's audio construction, we calculate $\mathcal{L}_{rec}$ as:
    \begin{equation} \label{eq:general10}
    \resizebox{0.87\columnwidth}{!}{$
    \begin{aligned}
    \centering
    \mathcal{L}_{rec} = \underset{x_{r},\bar{x}_{r},\tau_{s}}{\mathbb{E}} [\norm{(\bar{x}_{r} - x_{r})}^{1}_{1} -\mathcal{L}_{stft}(\tilde{x}_{r,s}, T_{r,s}(x_{r}, \tau_{s}))],  
    \end{aligned}$}
    \end{equation}
    where $\mathcal{L}_{stft}$ is the multi-resolution STFT loss~\cite{yamamoto2020parallel}. By adopting $\mathcal{L}_{stft}$, we maximize the spectral difference between the eavesdropping results for the original and perturbed audio. To make our perturbation inaudible, we use the mean absolute error to restrict the difference between original and perturbed audio.
\presec
\section{Evaluation}
\label{sec:eval}



\subsection{SSEA Setup and Implementation}
\label{sec:eval_setup}
\begin{figure}[t]
     \centering
     \begin{tabular}{@{}cc@{}}
     \centering
         \includegraphics[width=0.42\linewidth]{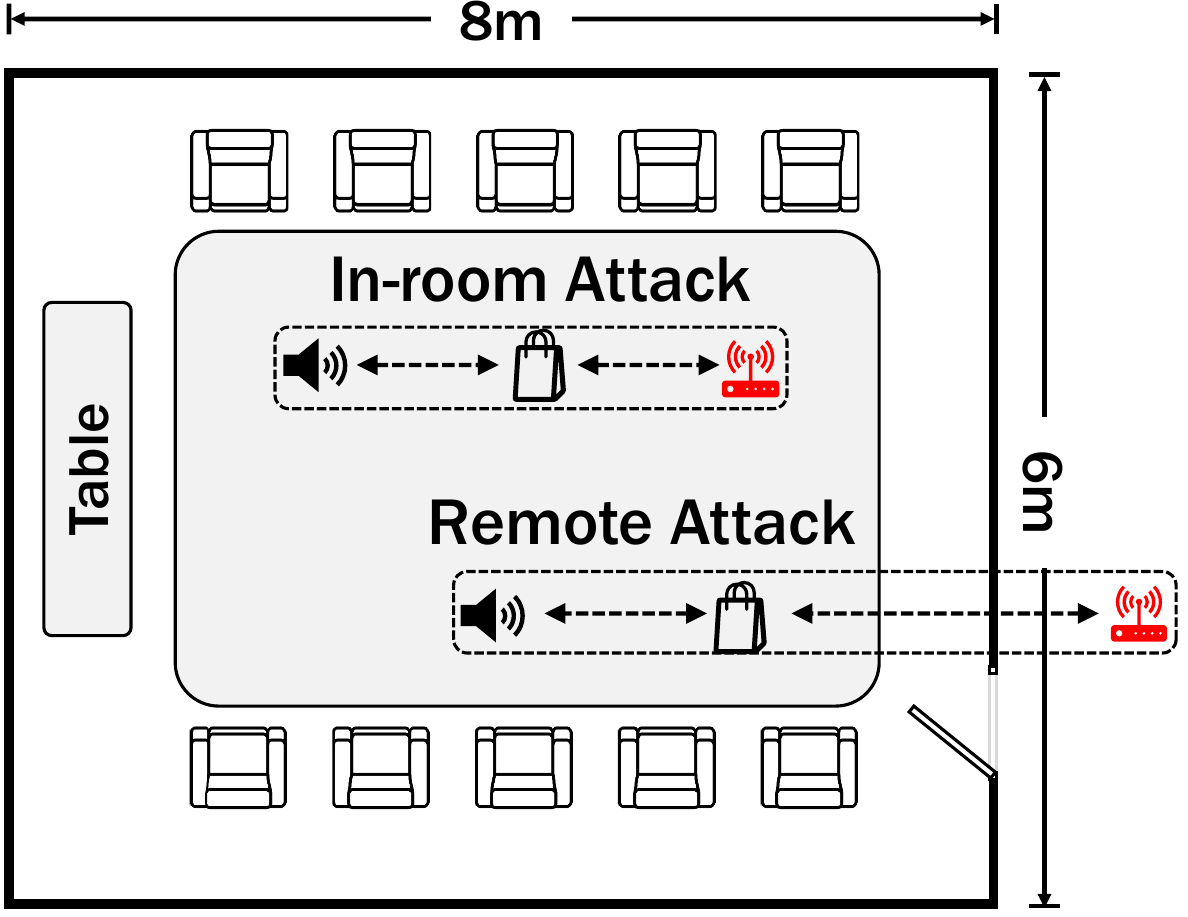}  &
         \includegraphics[width=0.42\linewidth]{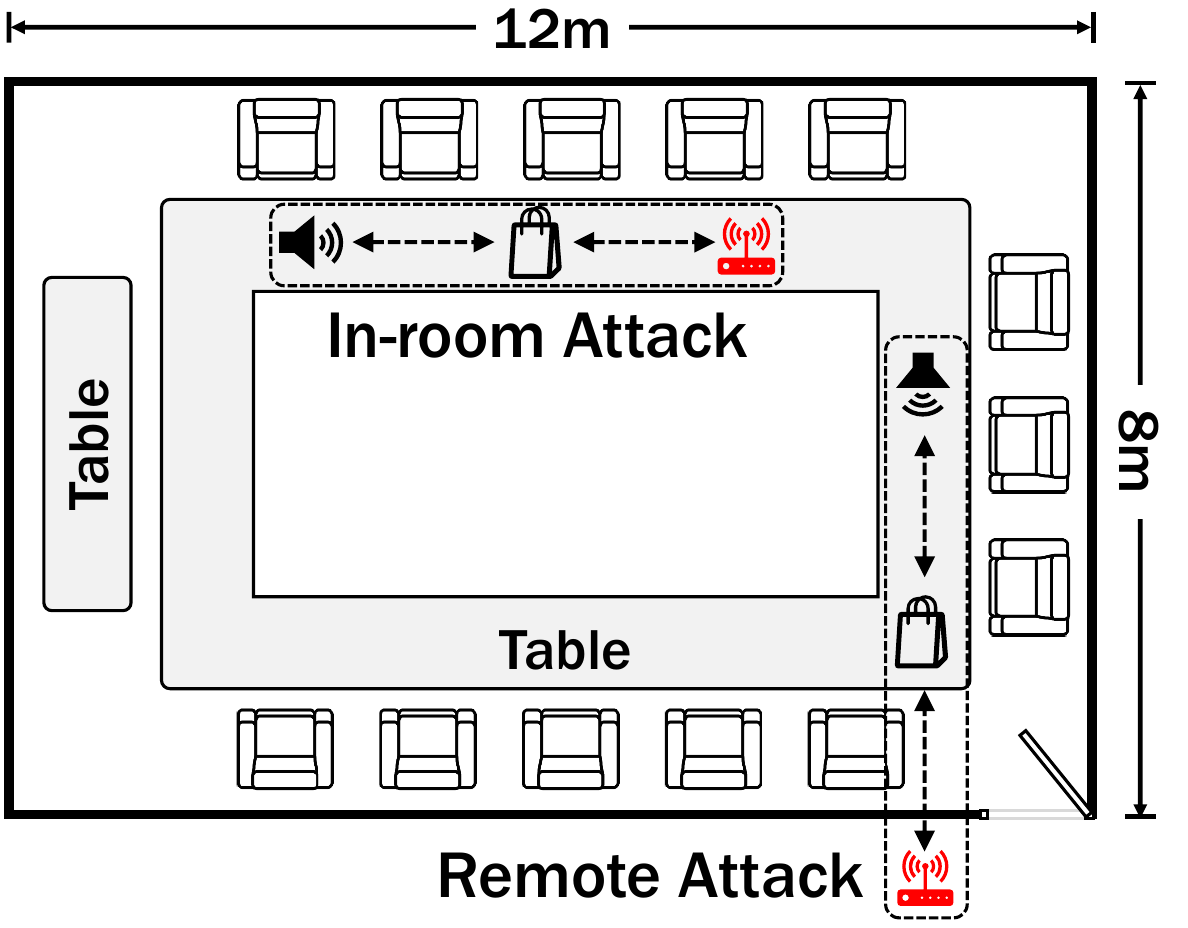} \\
         (a)  &
         (b) \\
     \end{tabular}
     \caption{Layout of the experimental environment. (a) The room where the \sys is trained. (b) The room where the \sys evaluation is conducted.}
     \label{fig:layout}
 \end{figure}

\noindent\textbf{SSEA Scenario Setup.} 
We re-implement vibration-based SSEAs using three representative side channels: mmWave radar, accelerometer and optical sensor.
Figure~\ref{fig:layout} shows the layout of our experimental environment. We use the conference room depicted in Figure~\ref{fig:layout}(a) to train EvGuard and the room in Figure~\ref{fig:layout}(b) for its evaluation.
Appendix Table~\ref{tab:data3} provides the detailed experimental setup.

We establish basic attack scenarios as discussed in Sec. \ref{sec::preli:understand} and Figure \ref{fig:feasibility}.
We further assess \sys's performance not only in the basic attack scenarios but also in unseen scenarios (Sec. \ref{sec:eval_mmwave} and Sec. \ref{sec:eval_motion}).

\noindent\textbf{SSEA Implementation.} 
Unlike previous SSEAs~\cite{hu2023mmecho, shi2023privacy, hu2022milliear, hu2022accear, sun2023stealthyimu} that limit the attacker's capabilities, we consider a stronger threat model, assuming that the eavesdropper is powerful enough to collect SSEA data from the victim's room using the same equipment as the victim, \jungwoo{as shown in Appendix Table~\ref{tab:data3}}. This allows us to assess \sys in the worst-case scenario, while still integrating sophisticated attack techniques established in SOTA SSEAs. Our implementation of the SOTA SSEAs \cite{hu2023mmecho, shi2023privacy, hu2022milliear, hu2022accear, sun2023stealthyimu} comprises three parts.
\begin{enumerate}[leftmargin=*]
\item \textbf{Signal Processing} (SP). We follow the steps discussed in Sec. \ref{sec::preli:understand} and~\cite{hu2023mmecho} to pre-process the raw eavesdropping signals and improve reconstruction quality.  Note that SP only applies to mmWave radar.

\item \textbf{Machine Learning} (ML). To enhance the probability of successful eavesdropping, we further implement an ML-based speech enhancement model by following well-established cGAN models~\cite{hu2022milliear, hu2022accear}. 

\item \textbf{Speech Recognition} (SR). The final step involves training dedicated speech recognition or audio classification models using the processed signals from SP and ML. It aims to extract explicit private information. The results of SP and ML in the evaluation are derived through SR.
\end{enumerate}

%
%
\jungwoo{The ML model architecture and implementation details are documented in Appendix~\ref{sec:append_SSEA}.}

\noindent\textbf{Datasets.} 
\jungwoo{Appendix~\ref{sec:append_dataset}} describes the speech datasets utilized for training each ML model for SSEA. We leverage MILLIEAR~\cite{hu2022milliear}, LJSpeech~\cite{ljspeech}, AudioMNIST~\cite{becker2018interpreting}, which have been employed in SOTA SSEA works~\cite{hu2022milliear, hu2022accear, shi2023privacy, sun2023stealthyimu}. These datasets are used for both SSEA training and \sys evaluation purposes. 
The datasets are split into non-overlapping training and testing sets with an 8:2 ratio.

\noindent\textbf{Defense Comparison.} We compare \sys against two baselines: (a) Gaussian noise, and (b) vanilla audio perturbations (VAP)~\cite{xie2021enabling}. Gaussian noise introduces randomly sampled noise into input audio, and VAP is designed to subvert speech recognition~\cite{warden2018speech}. To ensure fairness, each method is evaluated with the same budget (i.e., SNR) for the magnitude of perturbations.

\subsection{Defense Implementation} 
As described in Sec.~\ref{sec:design_objective}, \textit{\sys is a black-box defender}, i.e., it lacks knowledge about the attacker's ML model, attack scenario, and the devices utilized by the attacker. In line with this black-box assumption, we i). collect data in an entirely distinct environment from the SSEA attack implementation (\jungwoo{Figure~\ref{fig:layout}}); ii). employ surrogate SSEA models distinct from those used by the attacker (\jungwoo{Appendix~\ref{sec:append_surrogate}}); iii).  utilize different speech datasets to train \sys models than that used by the attacker's SSEA training (\jungwoo{Appendix~\ref{sec:append_dataset}}).

We implement the \sys using Pytorch 
and perform training with Adam optimizer~\cite{kingma2014adam} at a learning rate of 0.001. More details including hyperparameters can be found in \jungwoo{Appendix~\ref{sec:append_training}}. Our PGM $G_{r}$ is trained on the audio datasets with sampling rates of 16kHz and 48kHz (i.e., $r \in \{16,48\}$). This is because 16kHz audio is widely utilized in speech recognition and VoIP for its efficiency, while 48kHz audio is primarily used in video streaming platforms to ensure high-quality sound~\cite{Picovoice2024}. \sys selects either $G_{16}$ or $G_{48}$ based on the audio's sampling rate. \jungwoo{\sys segments audio into 50ms intervals and feeds it into the PGM.}

\begin{figure}[t]
    \centering
    \begin{tabular}{@{}cccc@{}}
    \centering
        \includegraphics[width=0.21\linewidth]{Fig/4_clean_mmwave.png}  &
        \includegraphics[width=0.21\linewidth]{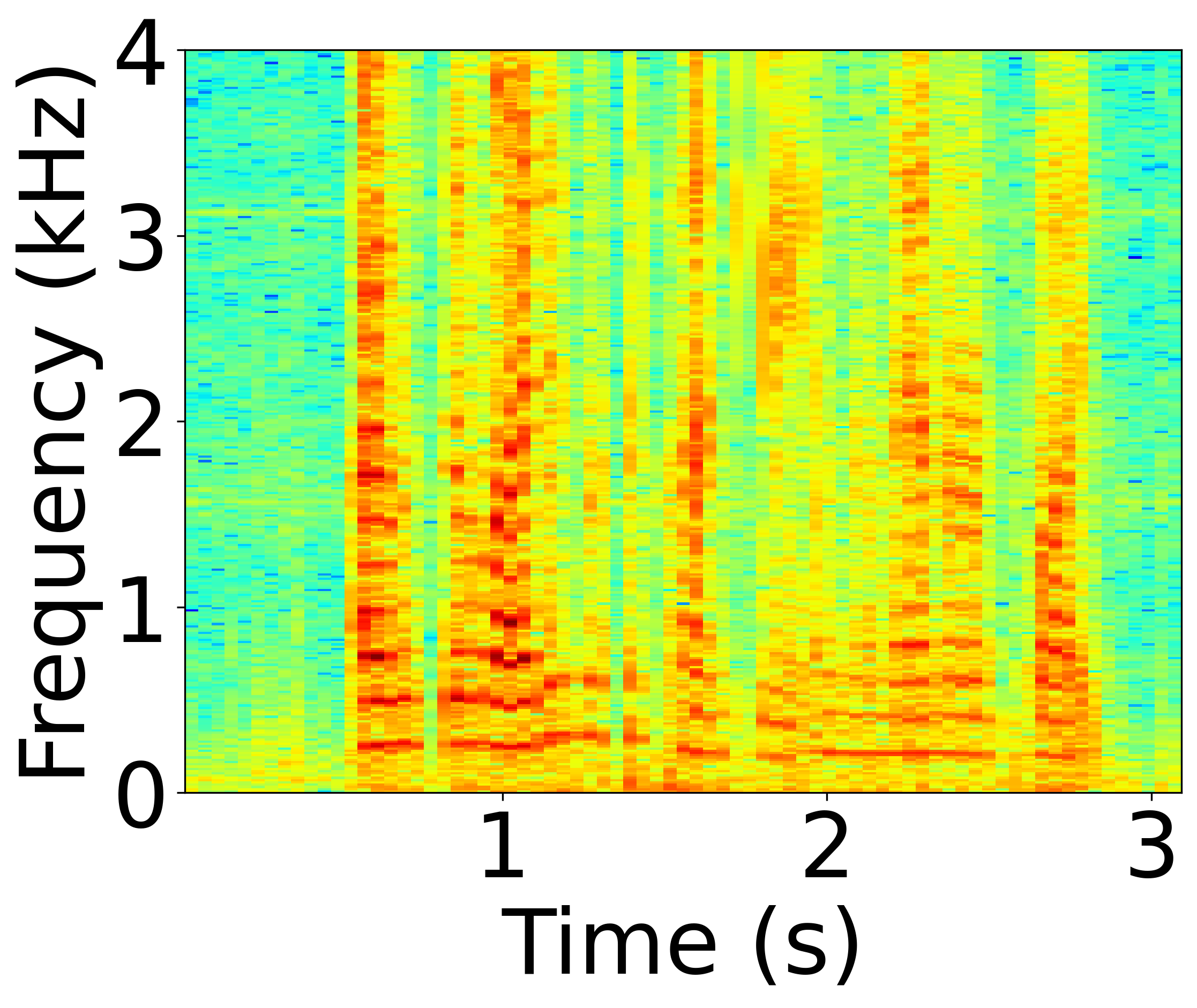} &
        \includegraphics[width=0.21\linewidth]{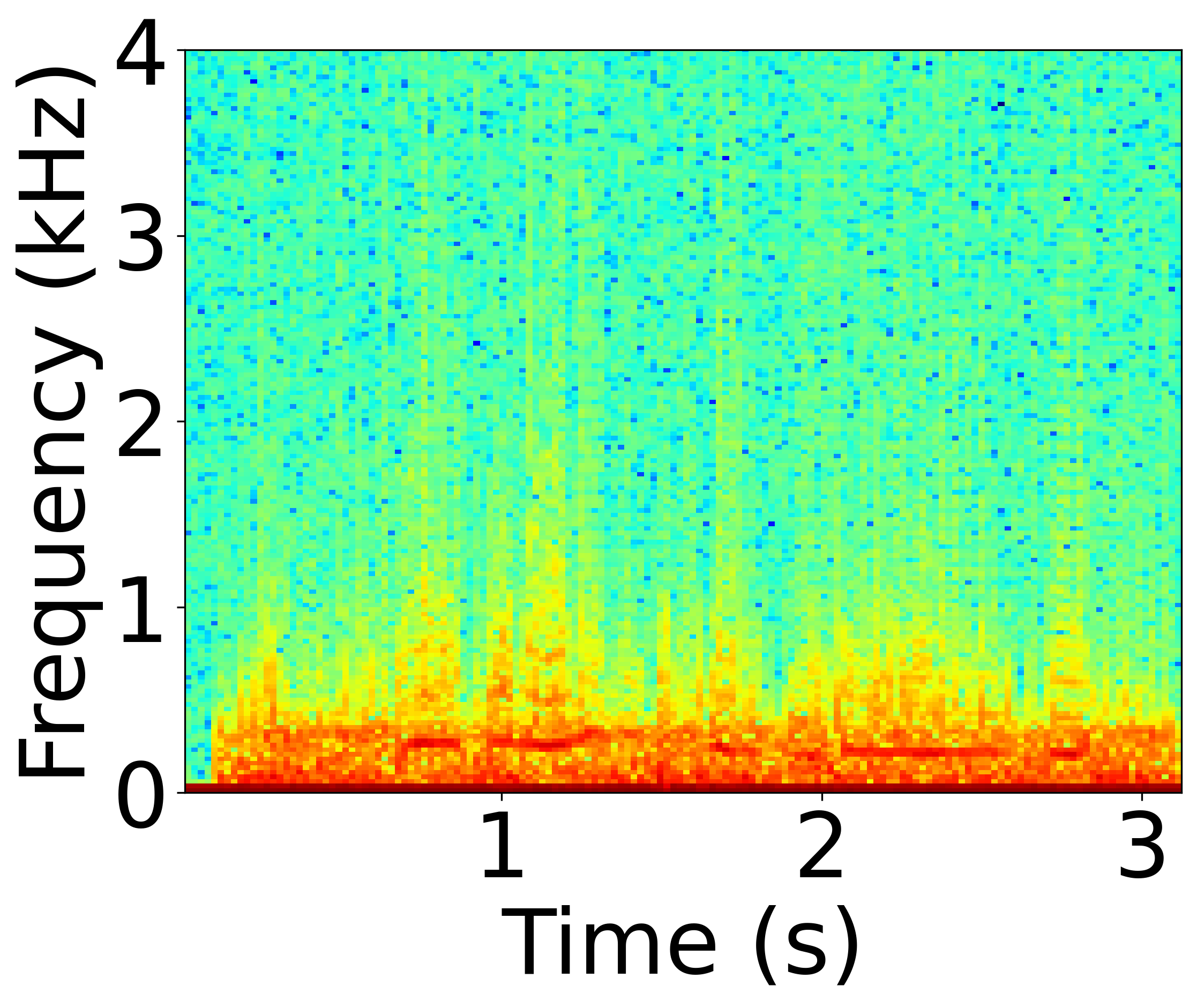}  &
        \includegraphics[width=0.21\linewidth]{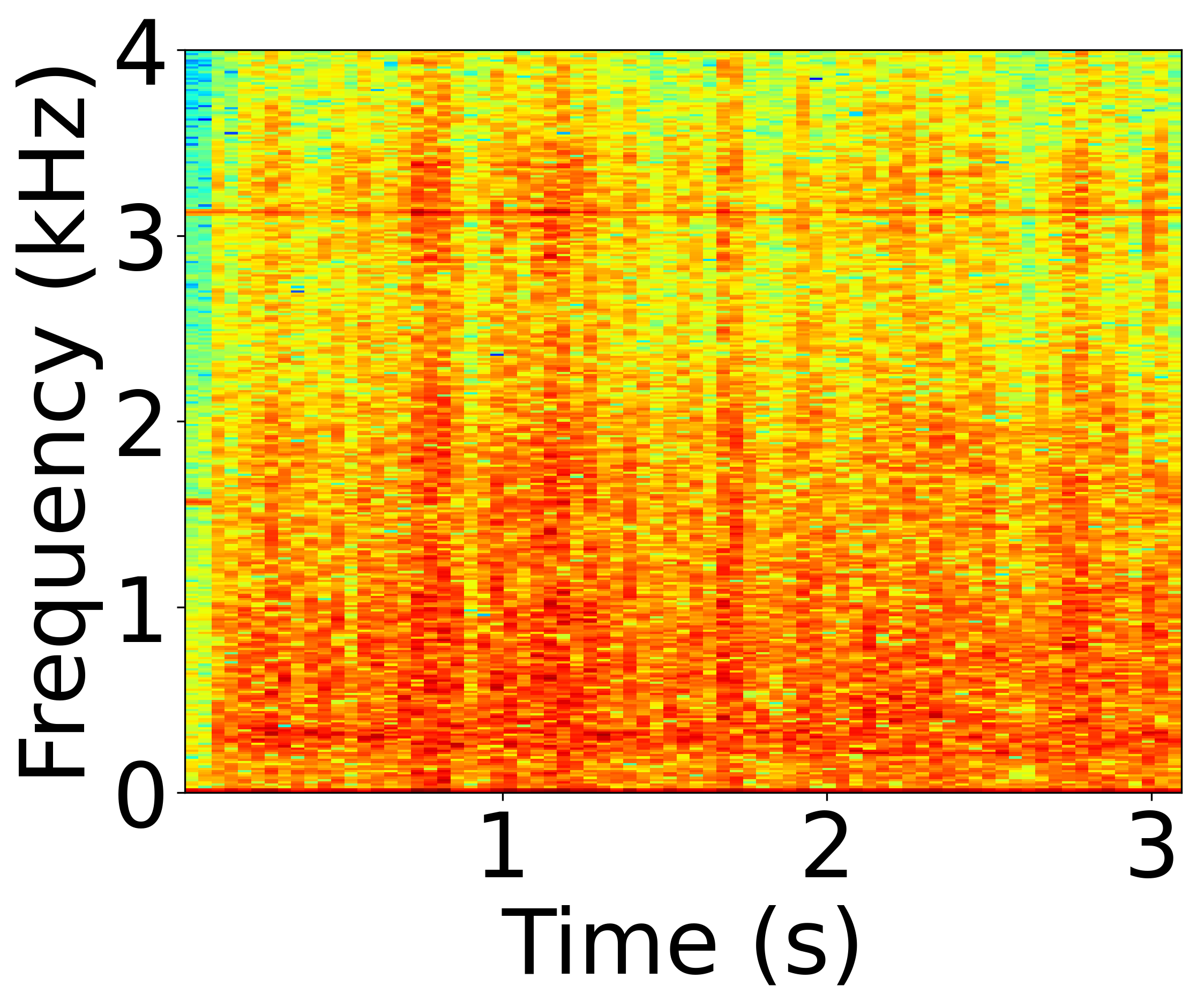} \\
        \multicolumn{2}{c}{\footnotesize (a) Clean mmWave Signal} &
        \multicolumn{2}{c}{\footnotesize (b) Perturbed mmWave Signal} \\
        \includegraphics[width=0.21\linewidth]{Fig/4_clean_imu_sub.png}  &
        \includegraphics[width=0.21\linewidth]{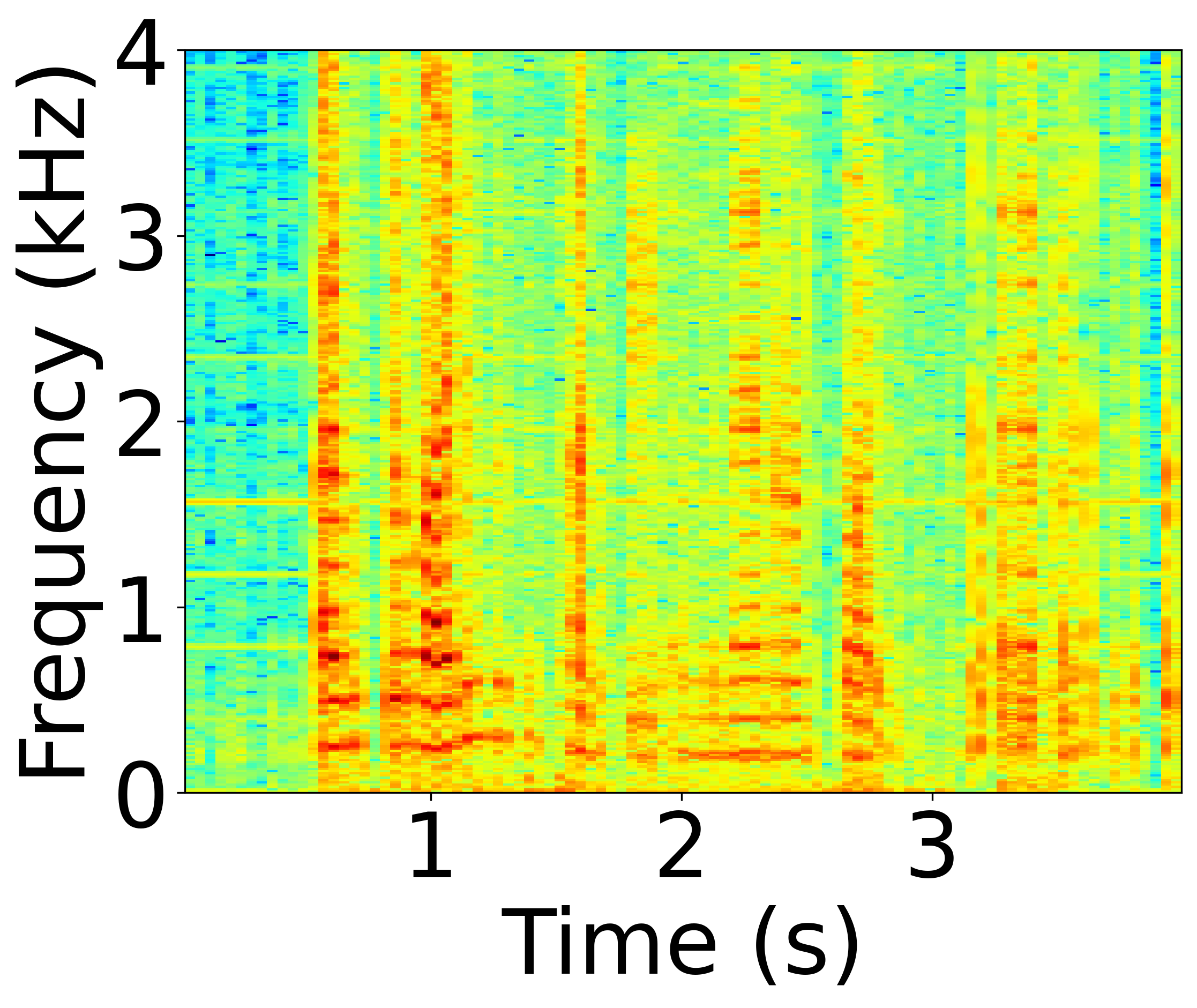} &
        \includegraphics[width=0.21\linewidth]{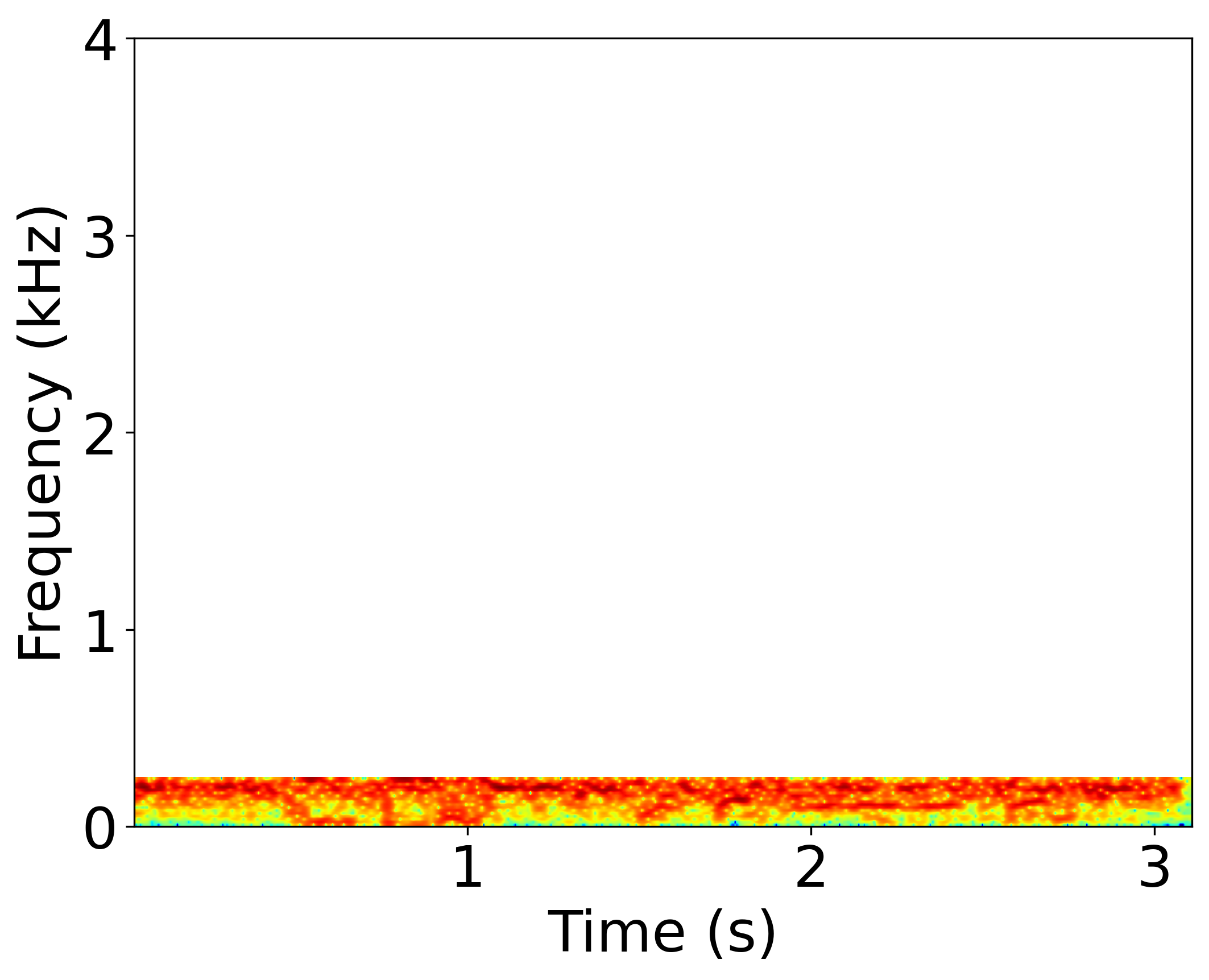}  &
        \includegraphics[width=0.21\linewidth]{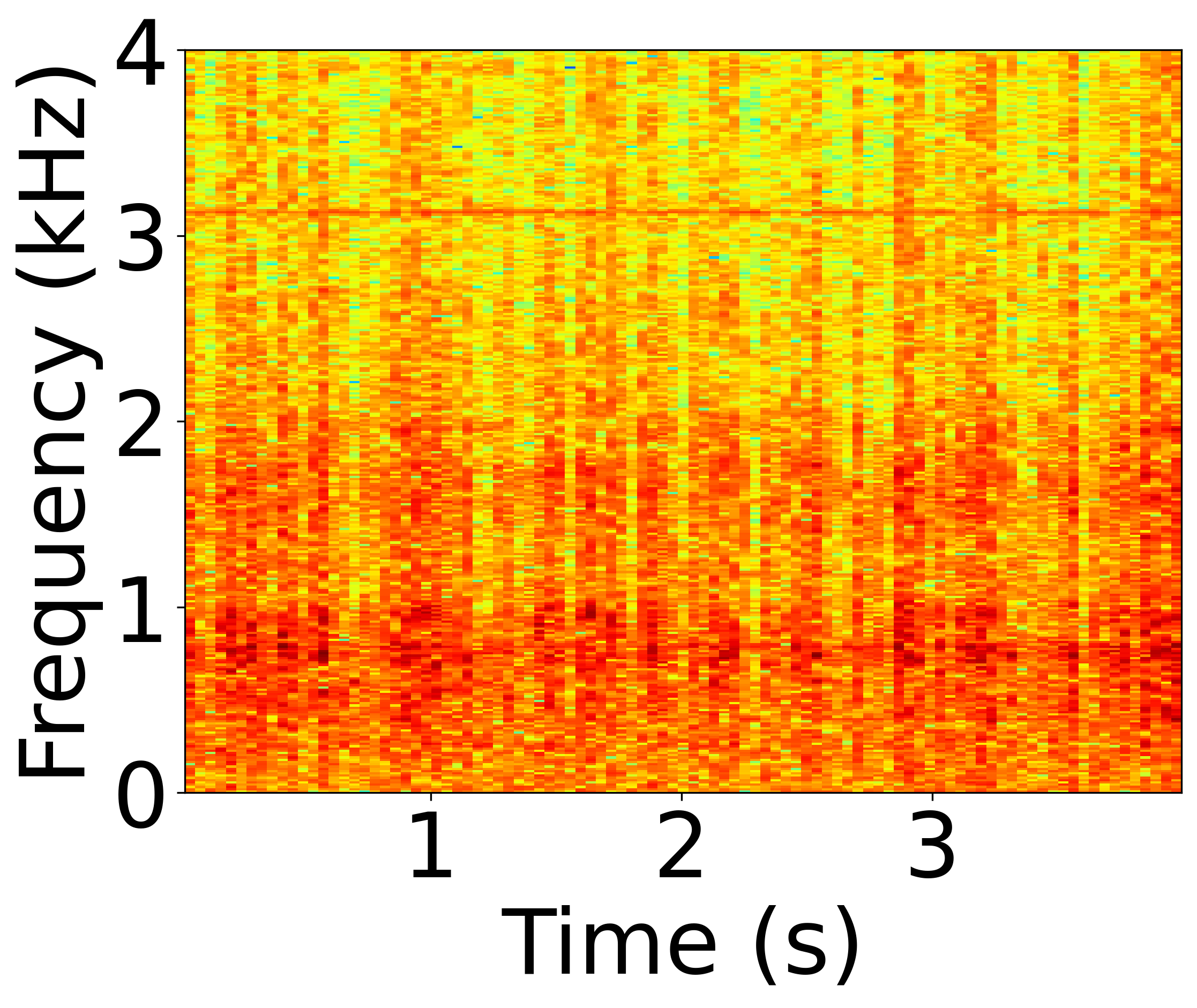} \\
        \multicolumn{2}{c}{\footnotesize (c) Clean IMU Signal} &
        \multicolumn{2}{c}{\footnotesize (d) Perturbed IMU Signal} \\
        \includegraphics[width=0.21\linewidth]{Fig/4_clean_laser.png}  &
        \includegraphics[width=0.21\linewidth]{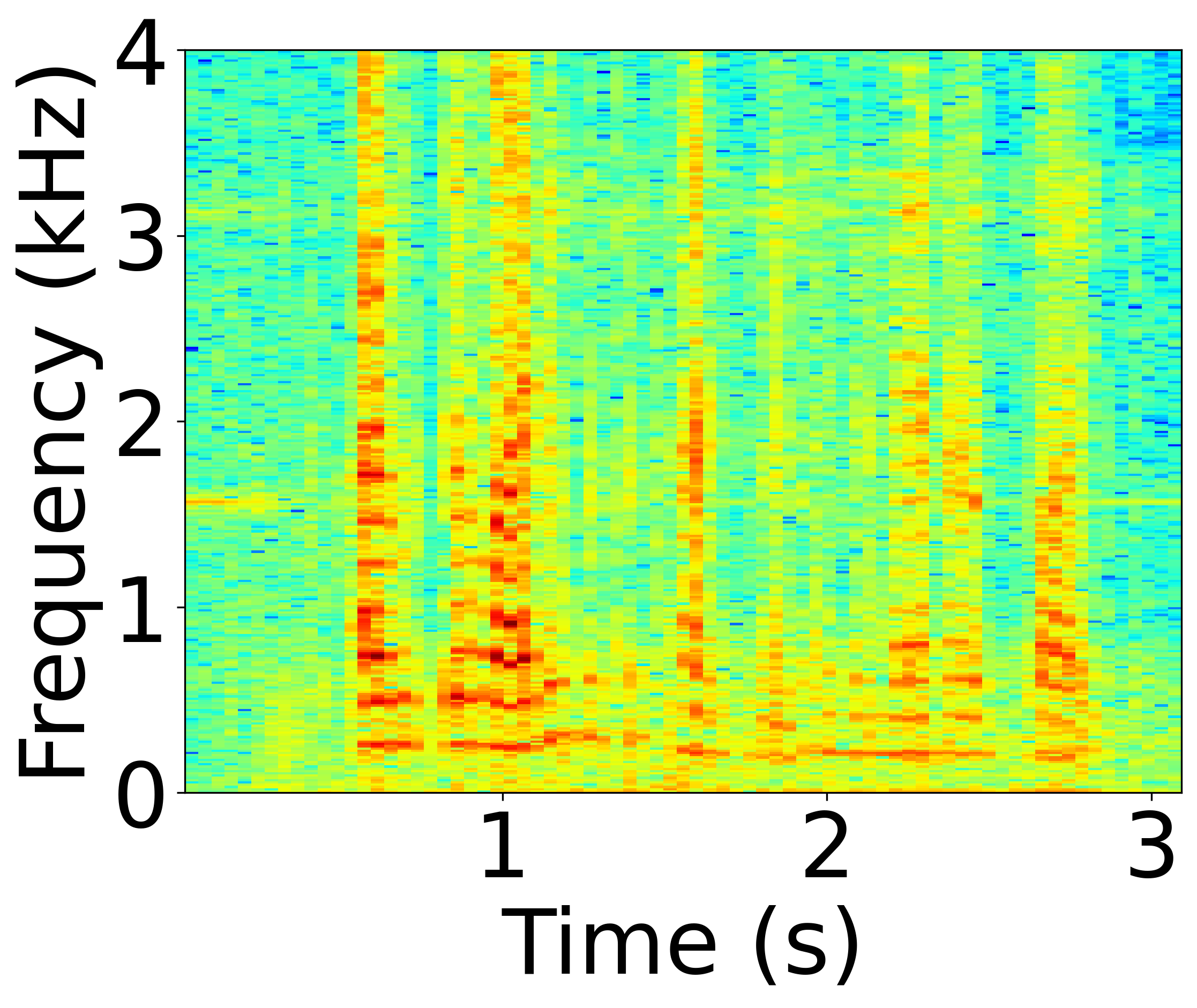} &
        \includegraphics[width=0.21\linewidth]{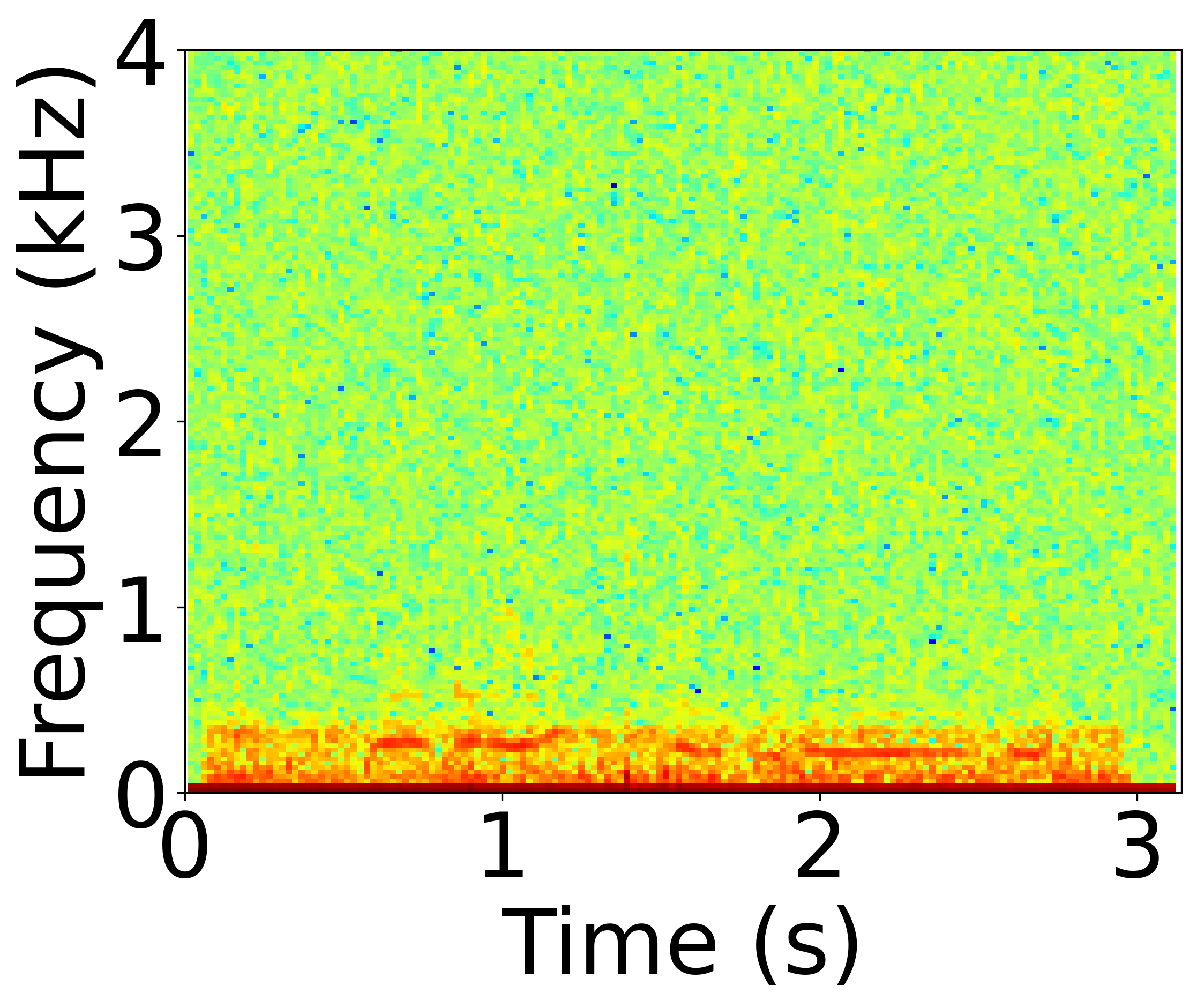}  &
        \includegraphics[width=0.21\linewidth]{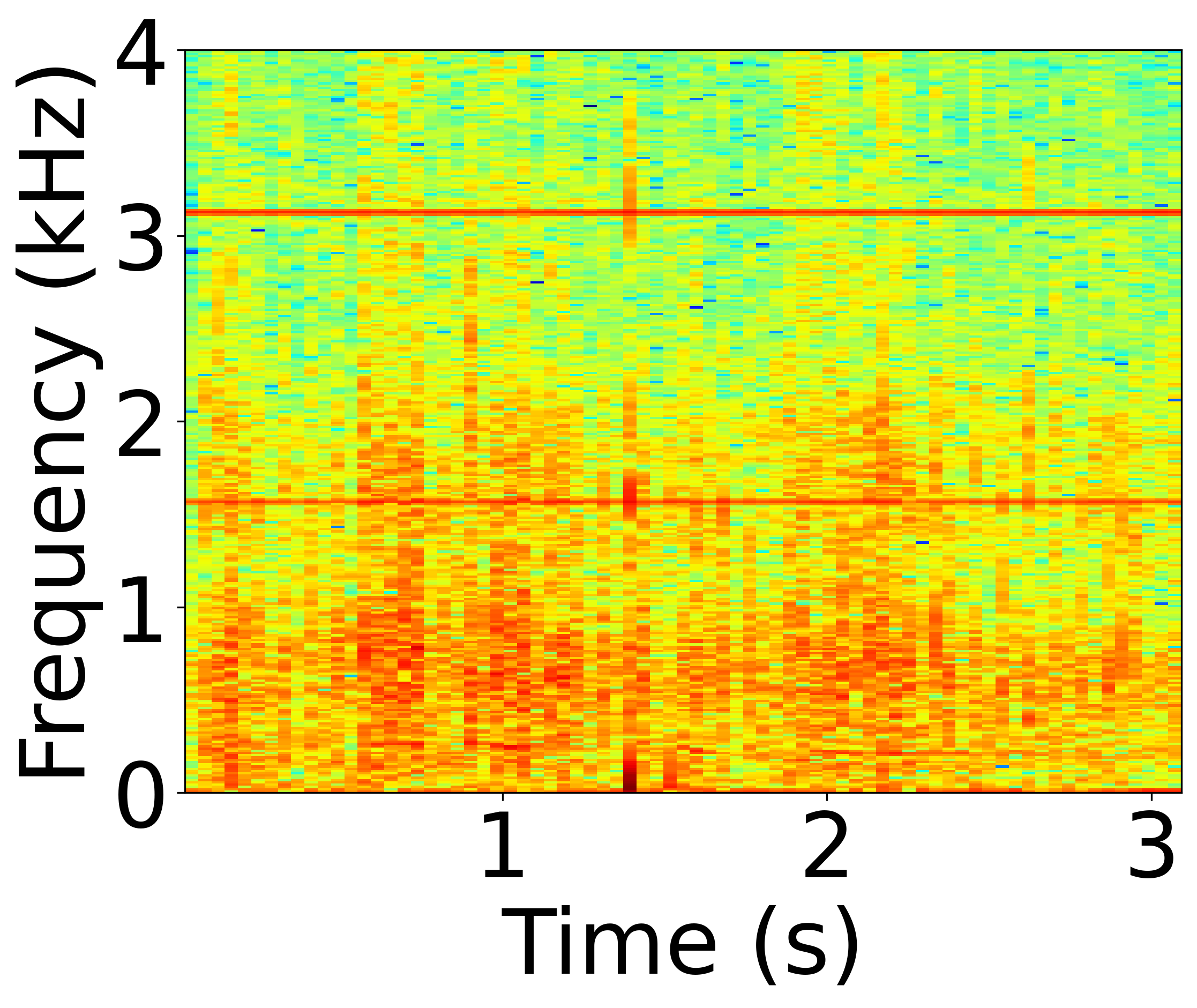} \\
        \multicolumn{2}{c}{\footnotesize (e) Clean Laser Signal} &
        \multicolumn{2}{c}{\footnotesize (f) Perturbed Laser Signal} \\
    \end{tabular}
    \caption{Spectrograms measured by different side channels (e.g., mmWave, IMU, and Laser). The left figures in (a)-(f) show raw-recovered audio. The right figures in (a)-(f) show reconstructed audio from the left figures through ML-based audio enhancement.}
    \label{fig:result_spec}
\end{figure}

\begin{table}[b]
\vspace{+0.1in}
\caption{\sys against baseline mmWave attacks.}
\centering
\label{tab:eval1}
\resizebox{0.95\columnwidth}{!}{
\midsepremove
\begin{tabular}{ ccccc|cccc }
\toprule
\multirow{2}{*}{Defense} & \multicolumn{4}{c|}{ML-SSEA} & \multicolumn{4}{c}{SP-SSEA}\\
\cline{2-5} \cline{6-9}
& MCD & WER & DDR & PESQ & MCD &WER & DDR & PESQ \\
\hline
OFF & $3.3 $ & $8.5\%$ & $98\%$ & - & $3.4$ & $9.2\%$ & $96\%$ & - \\
Gaussian & $7.7 $ & $12.8\%$ &  $94\%$ &  2.54 &  $8.5$ &  $18.5\%$ &  $88\%$ &  2.44 \\
VAP & $7.4 $ & $15.5\%$ &  $78\%$ &  2.63 &  $8.1$ &  $20.6\%$ &  $73\%$ &  2.63 \\
\sys & $13.4 $ & $68.2\%$ &  $3\%$ &  3.42 &  $13.6$ &  $70.1\%$ &  $2\%$ &  3.42 \\
\bottomrule
\end{tabular}}
\end{table}

\begin{figure*}[t]
    \centering

    \resizebox{0.7\textwidth}{!}{
    \begin{tabular}{@{}c@{}}
    \includegraphics[width=\linewidth]{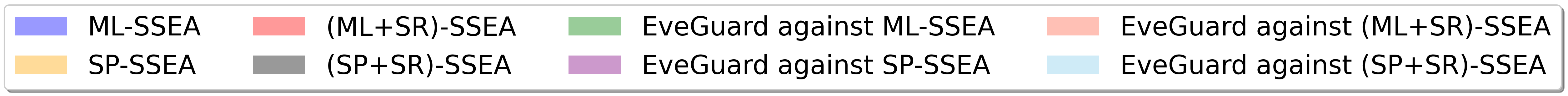} \\  
    \end{tabular}}

    \begin{tabular}{@{}cccc@{}}
        \includegraphics[width=0.21\linewidth]{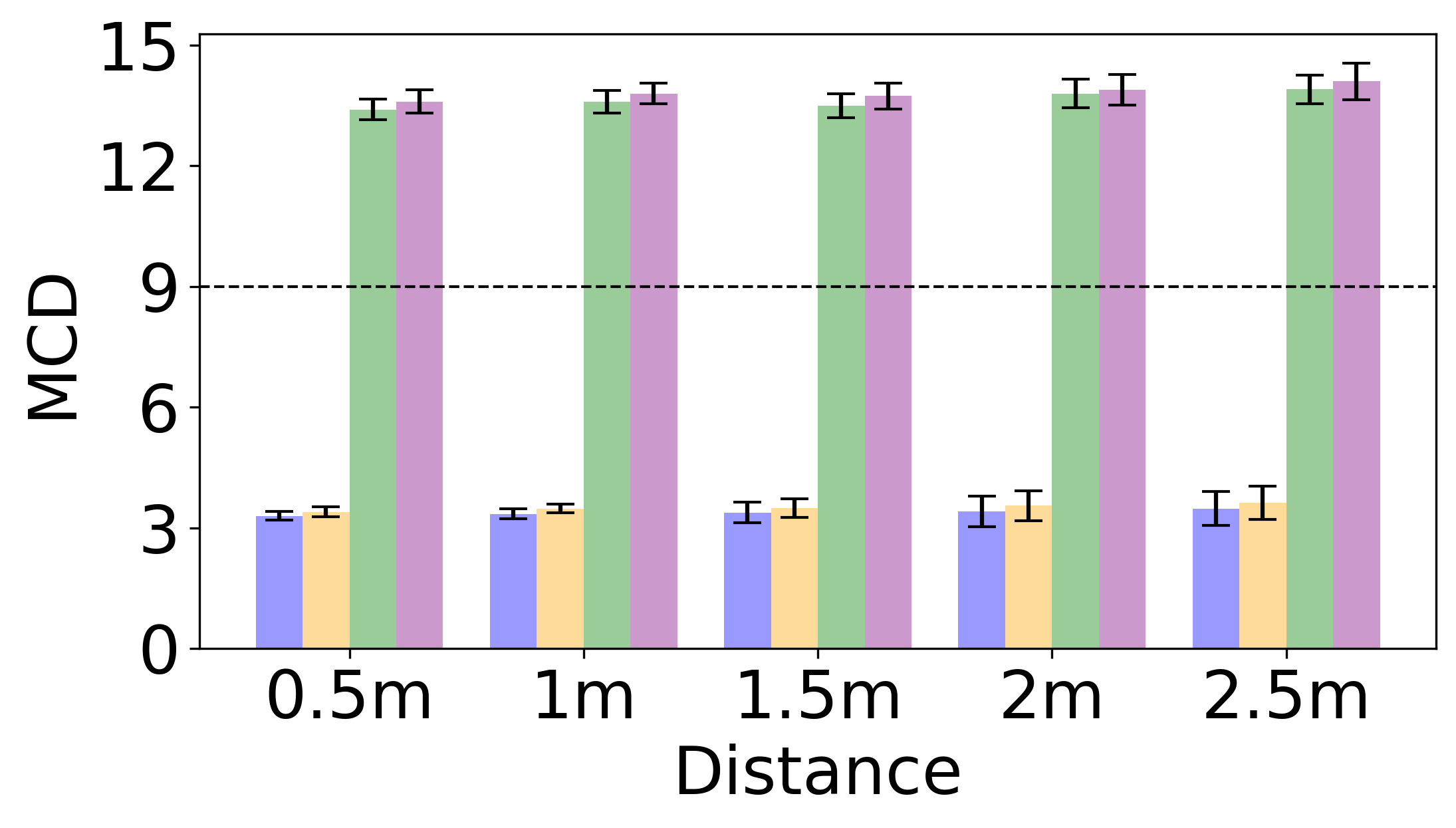} &
         \includegraphics[width=0.21\linewidth]{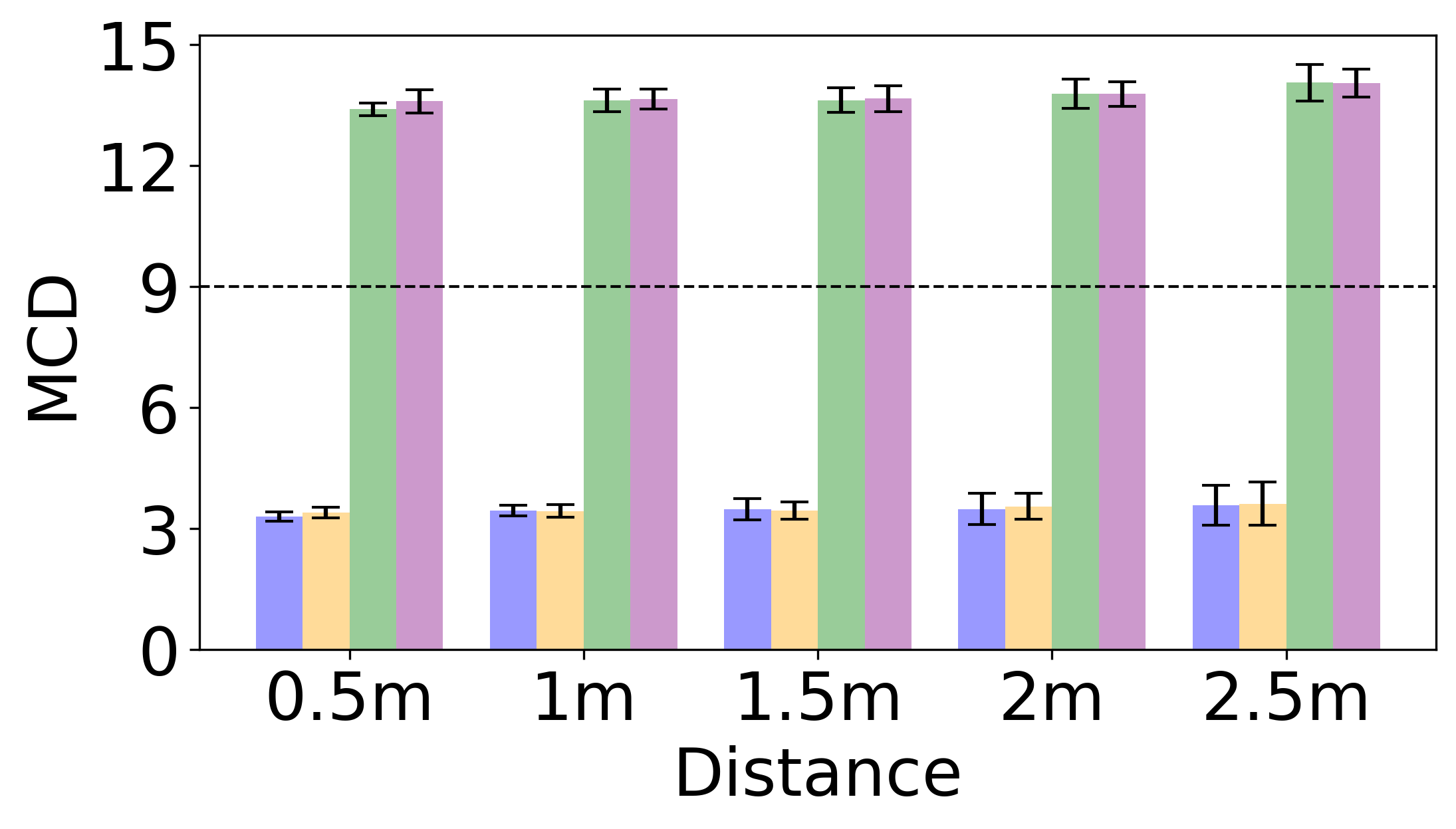} &
         \includegraphics[width=0.21\linewidth]{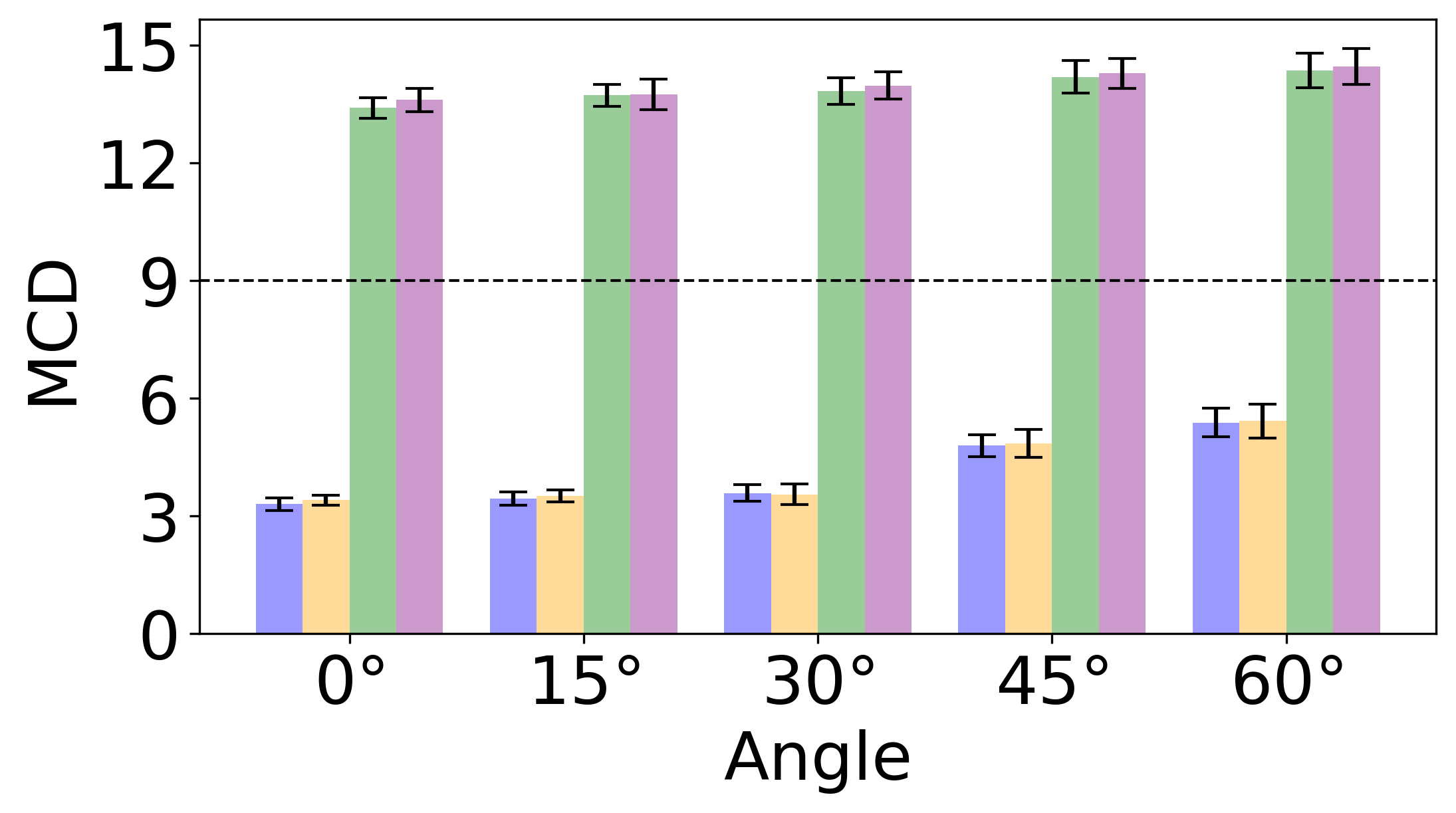} &
        \includegraphics[width=0.21\linewidth]{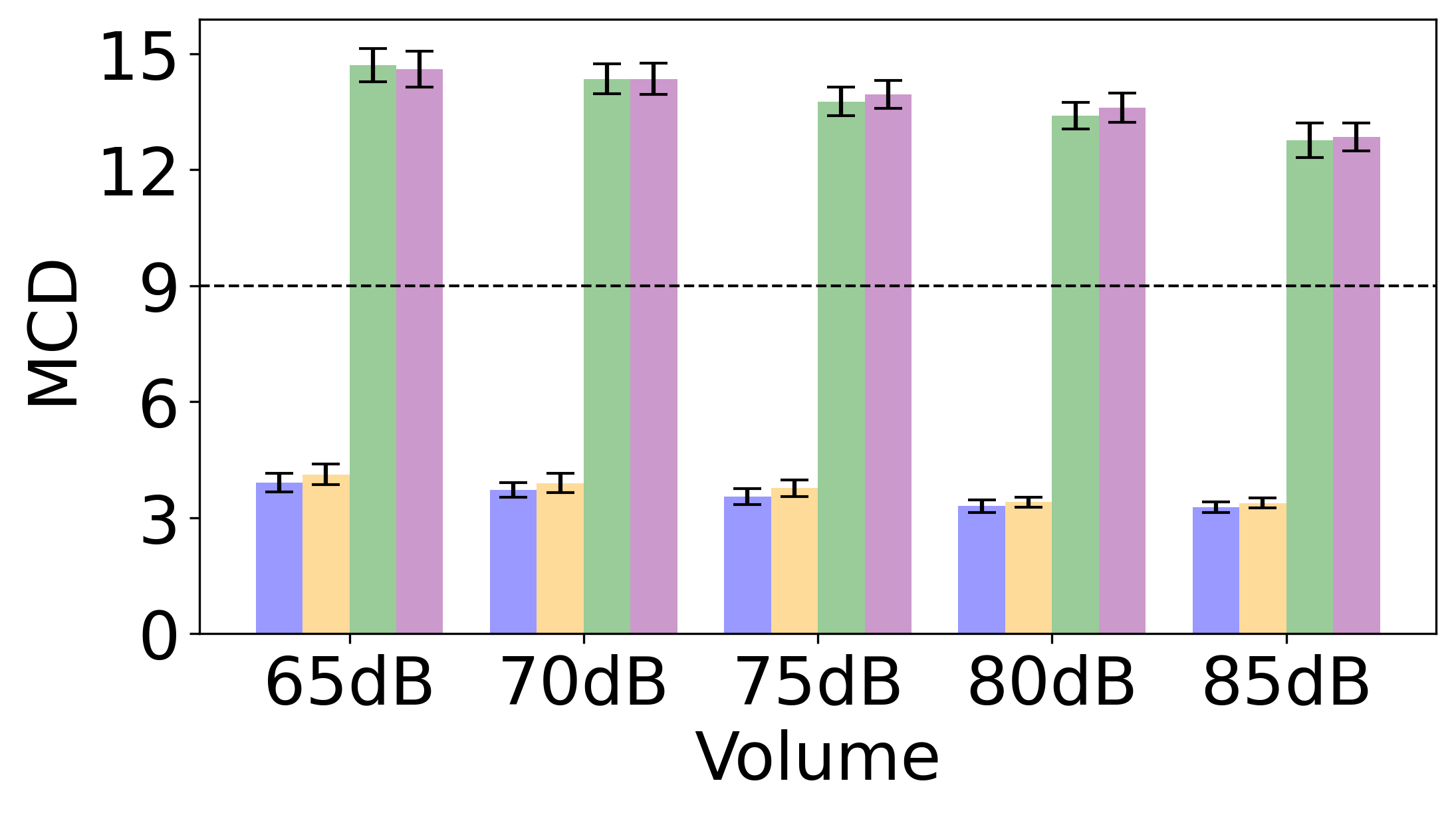} \\
        \includegraphics[width=0.21\linewidth]{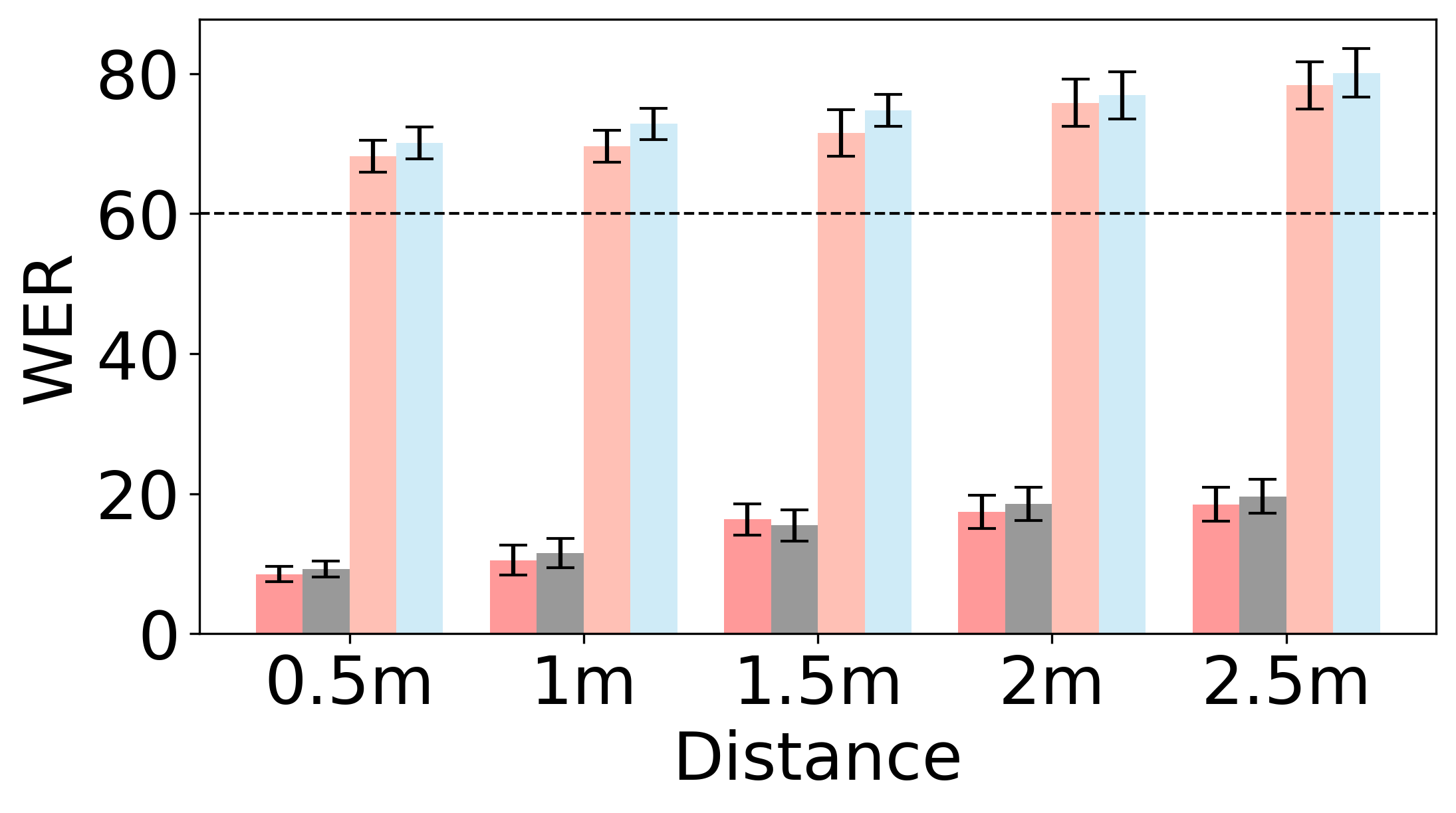} &
         \includegraphics[width=0.21\linewidth]{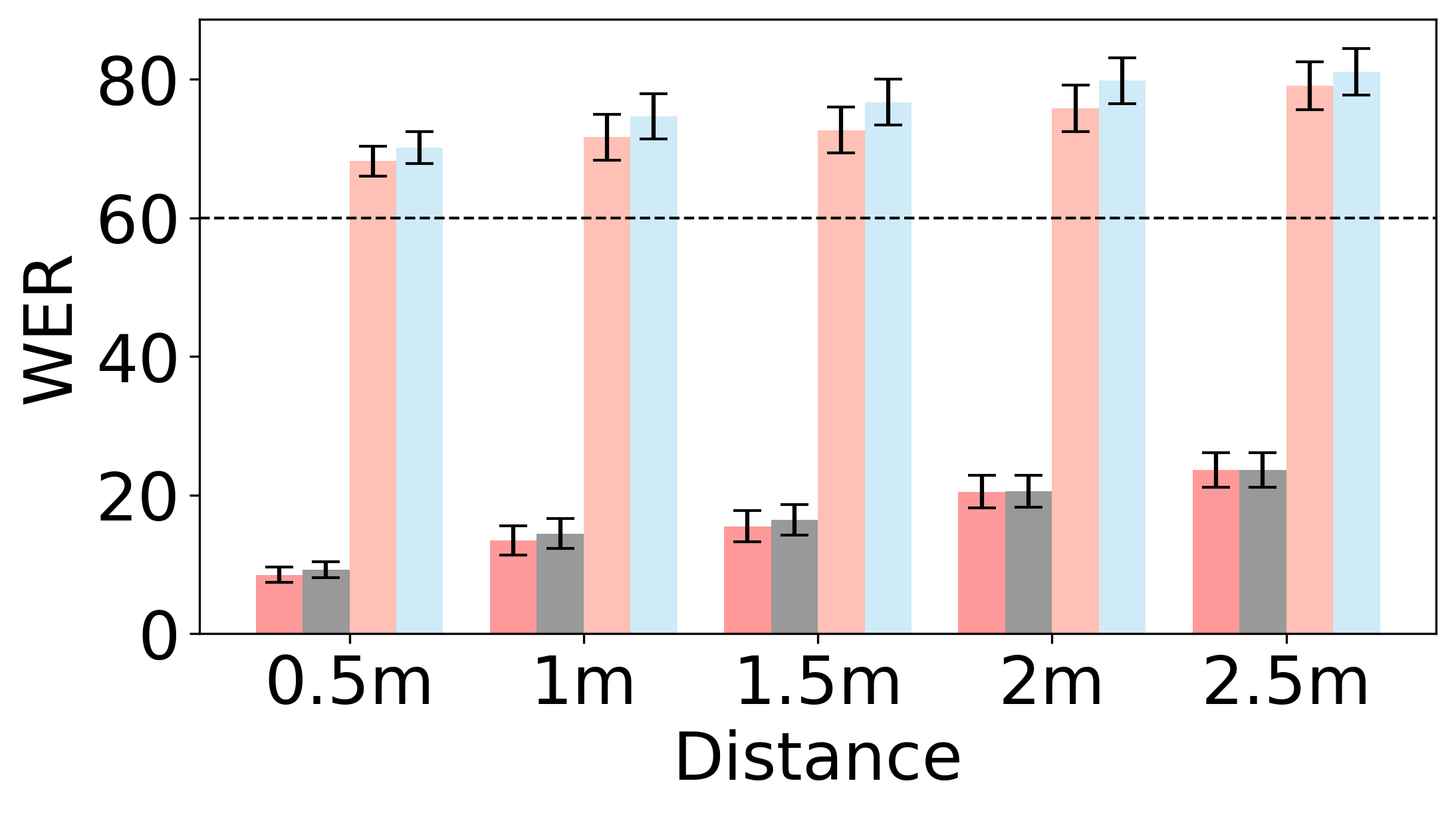} &
         \includegraphics[width=0.21\linewidth]{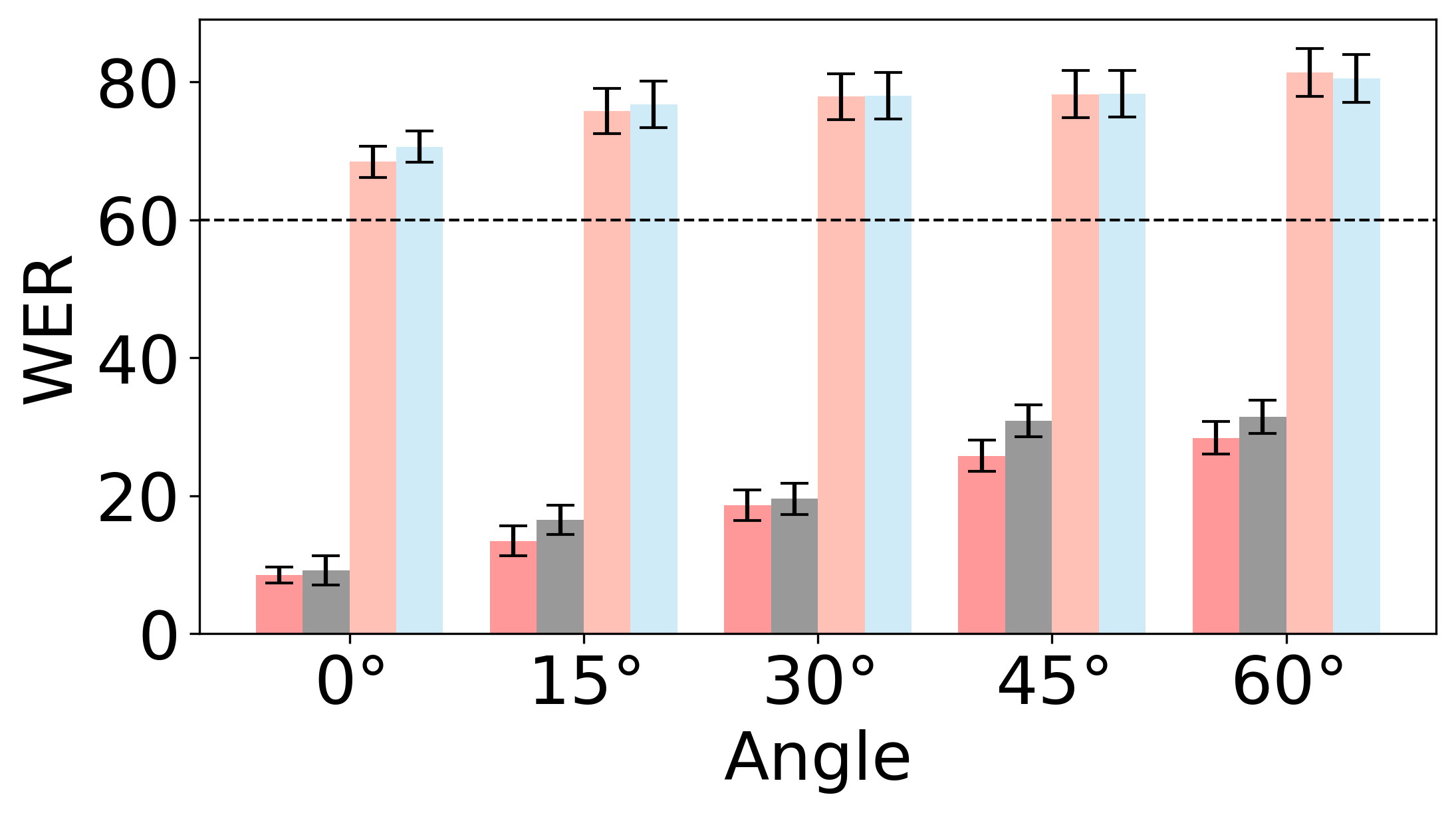} &
        \includegraphics[width=0.21\linewidth]{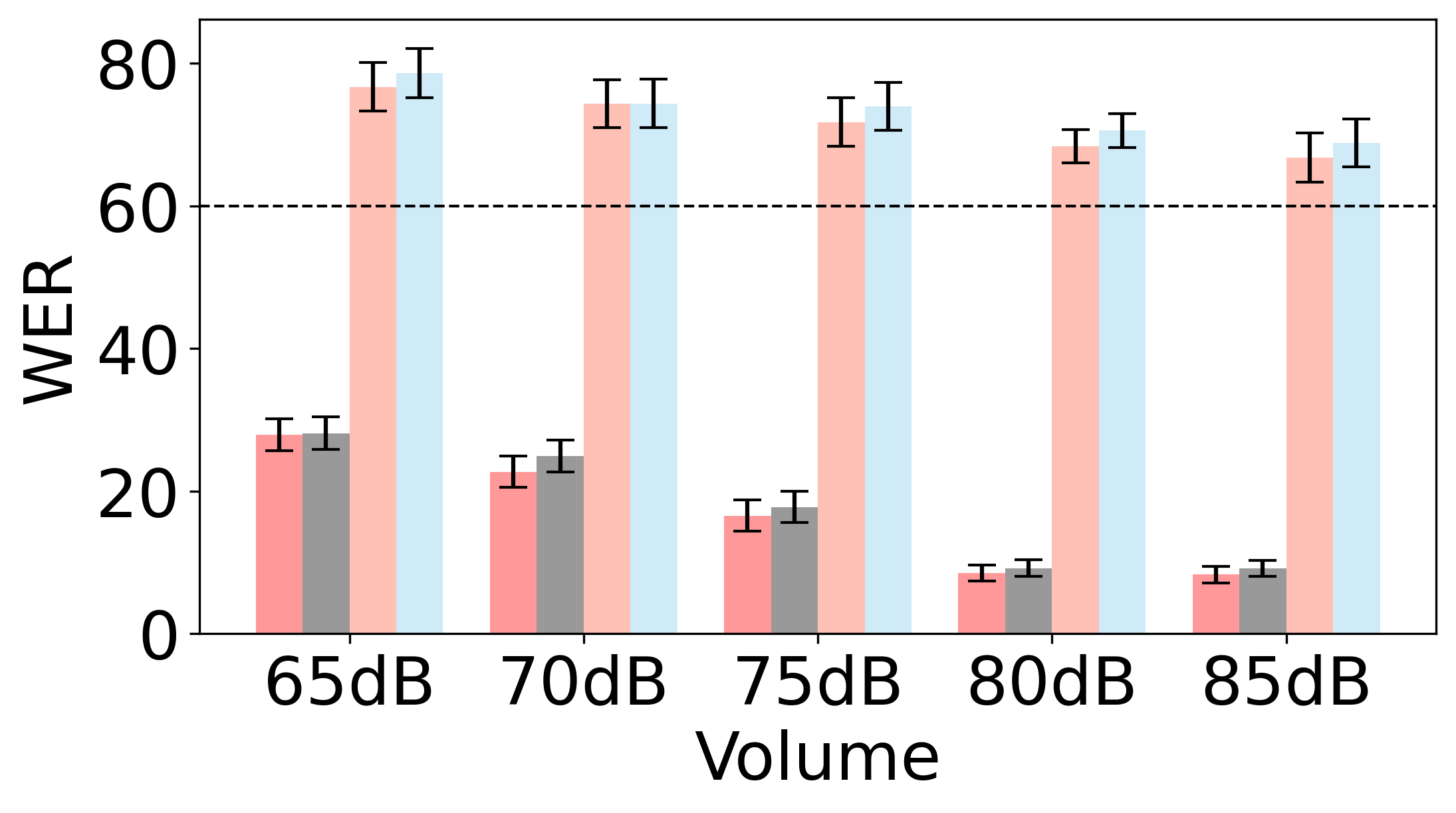} \\
        \includegraphics[width=0.21\linewidth]{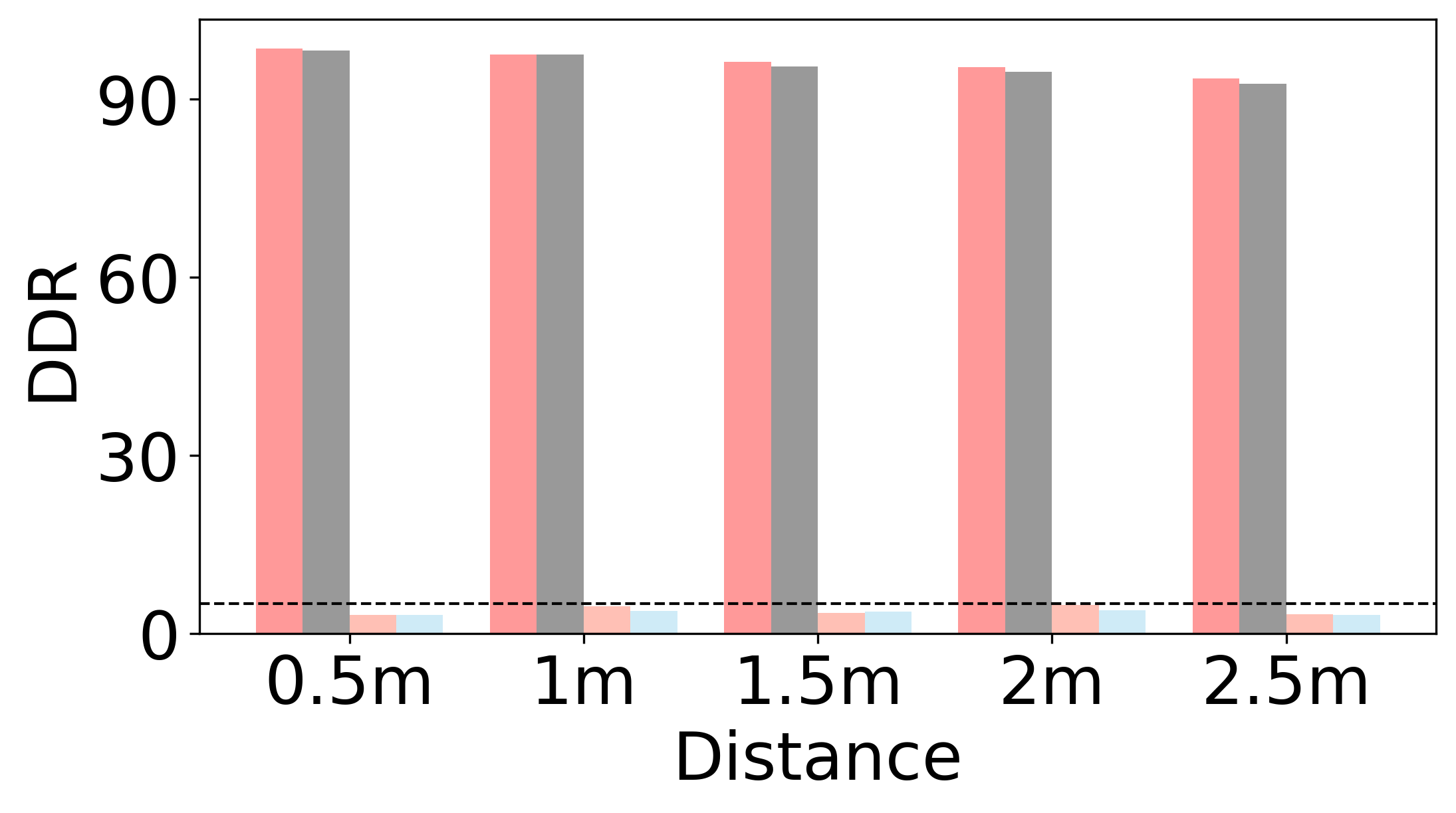} &
         \includegraphics[width=0.21\linewidth]{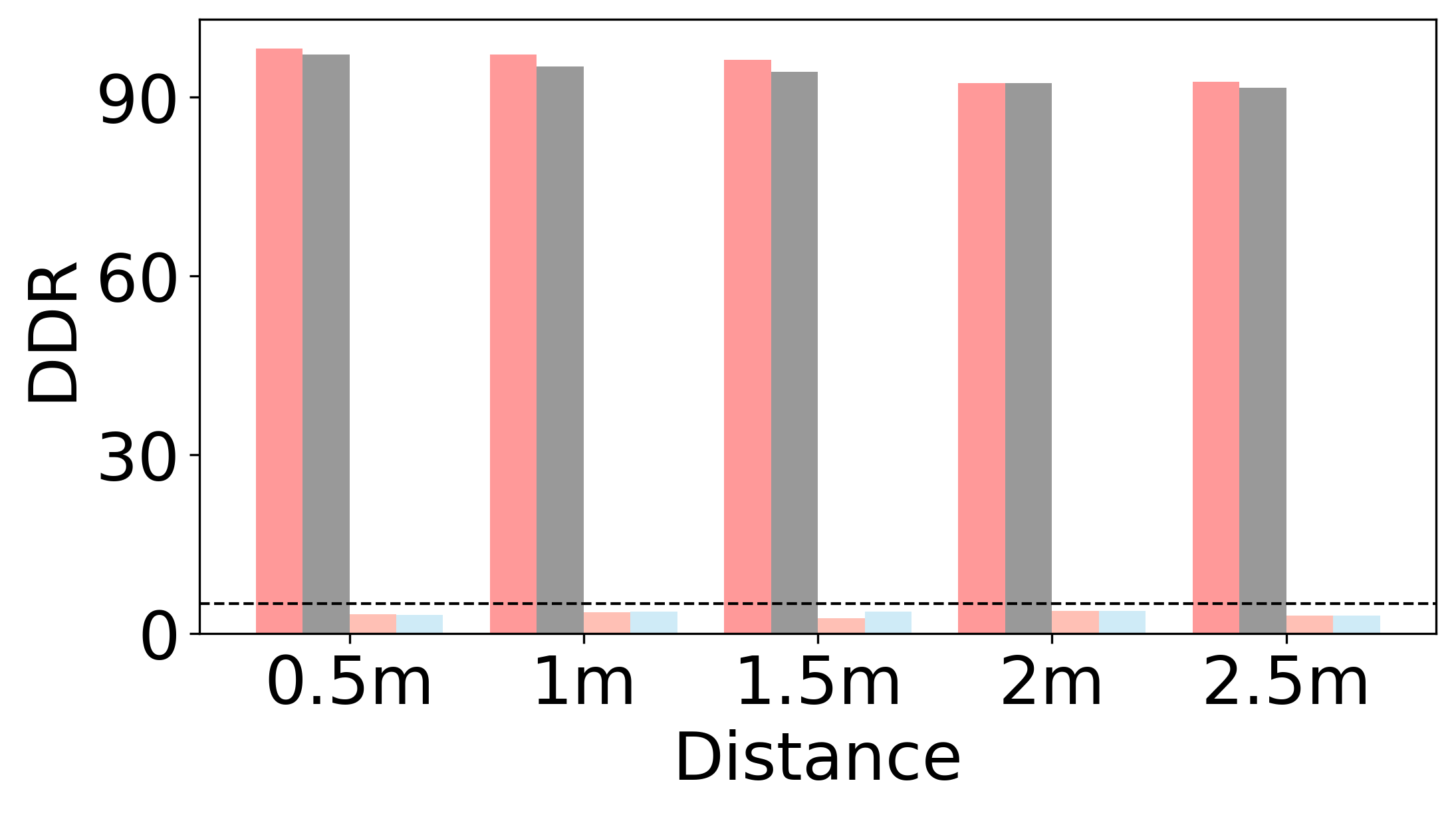} &
         \includegraphics[width=0.21\linewidth]{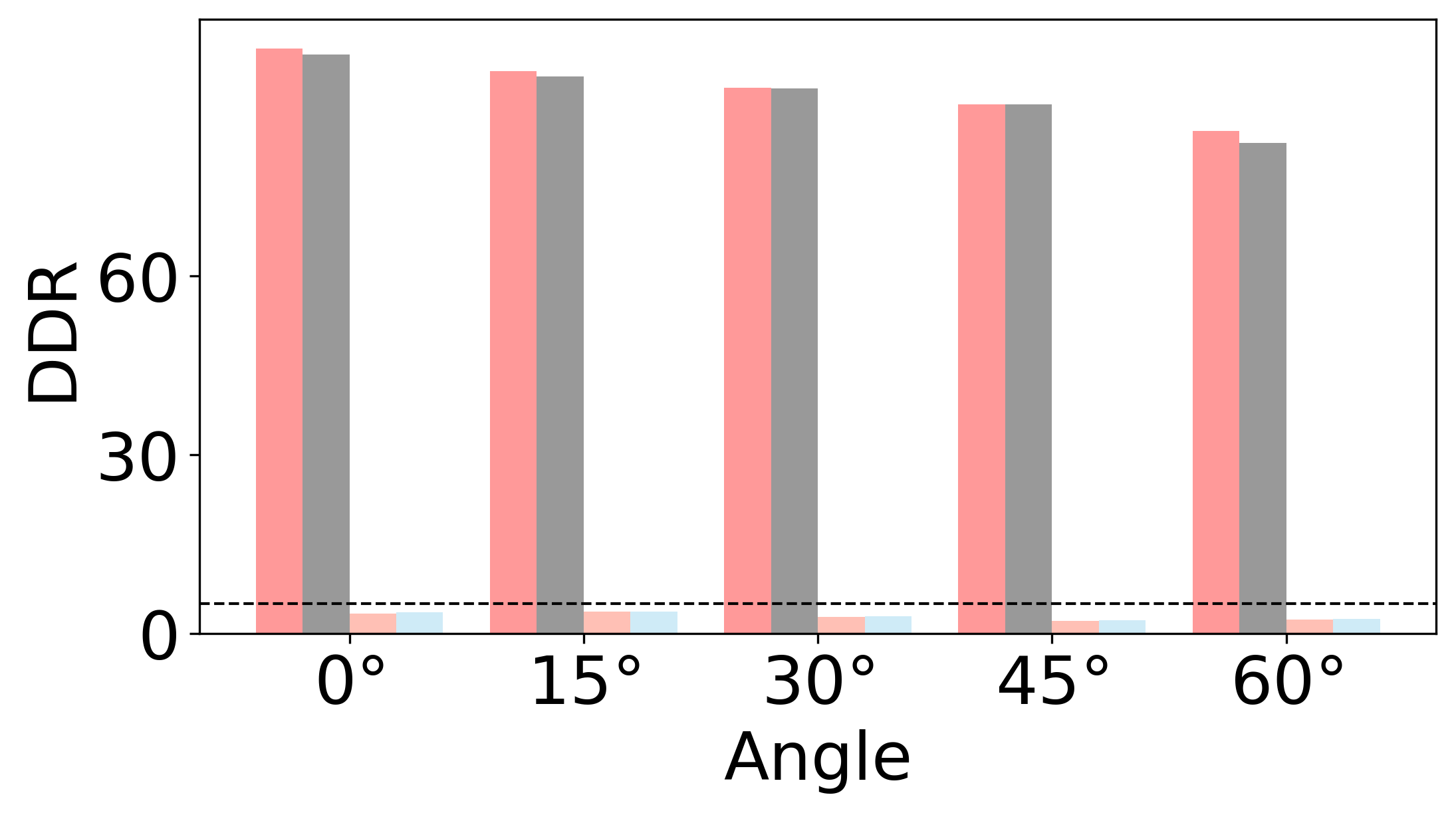} &
        \includegraphics[width=0.21\linewidth]{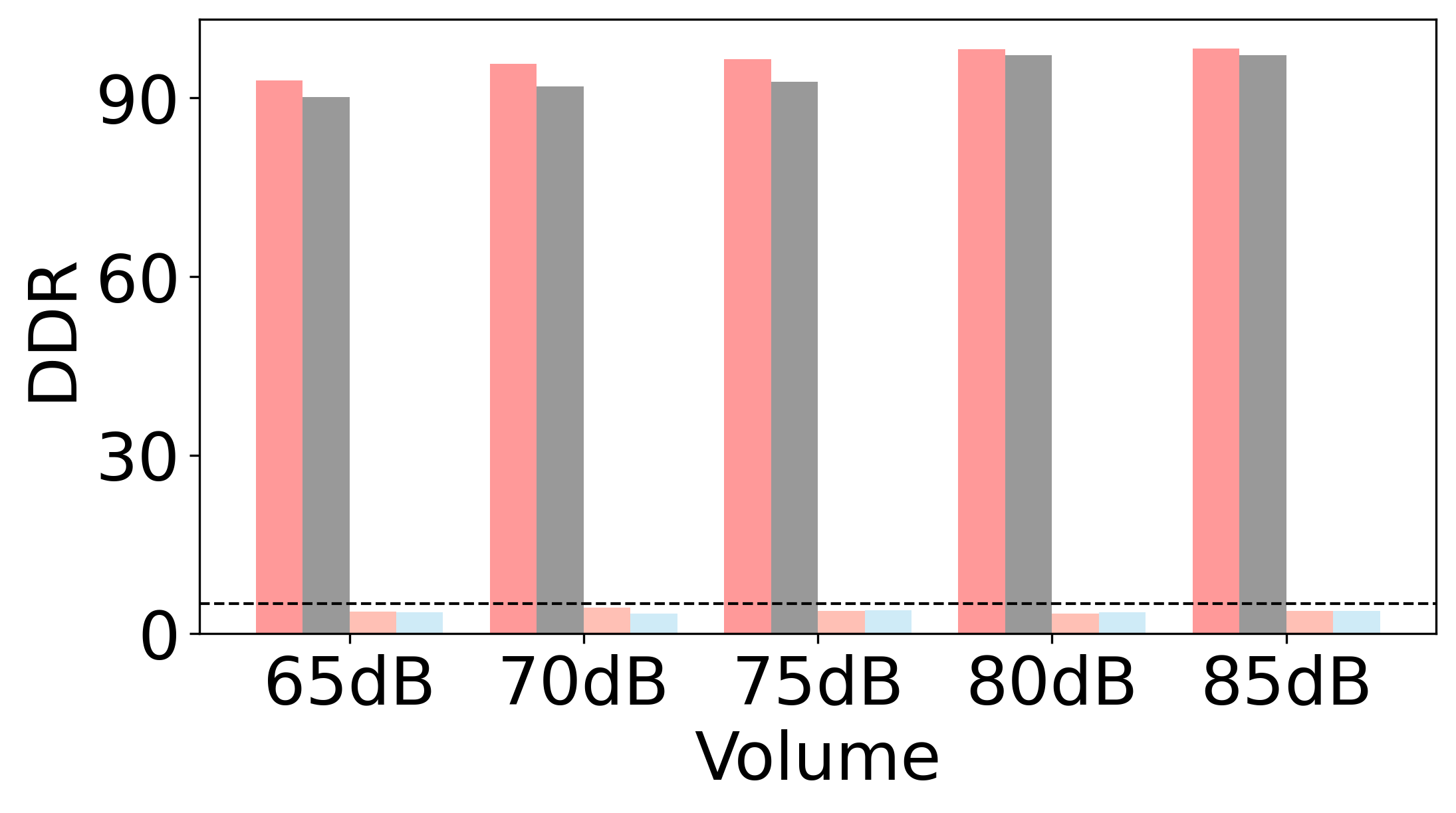} \\
        \footnotesize (a) R-to-O Distance &
        \footnotesize (b) V-to-O Distance &
        \footnotesize (c) Various Angles  &
        \footnotesize (d) Various Volume  \\
    \end{tabular}
    \caption{Micro benchmark. V, O, and R denote voice source, reverberating object, and radar, respectively.}
    \label{fig:exp1}
\end{figure*}

\presec
\subsection{Evaluation Metrics} \label{sec:metrics}
We use the following evaluation metrics to quantify the effectiveness of \sys: 

\begin{enumerate}[leftmargin=*]
    
\item \textbf{Mel-Cepstral Distortion} (MCD)~\cite{kominek2008synthesizer} quantifies the difference between the original speech and the attacker's reconstruction.
Audio with an MCD below 8 is typically recognizable by speech recognition models~\cite{yan2019feasibility}.

\item \textbf{Word Error Rate} (WER)~\cite{wang2003word} measures the fraction of wrong words produced by a speech recognition model.

\item \textbf{Digit Detection Rate} (DDR) is an objective metric to measure the performance of the audio classifier.

\item \textbf{Perceptual Evaluation of Speech Quality} (PESQ)~\cite{recommendation2001perceptual} is a standardized speech quality metric. A score above 3.0 is required for good-quality voice communication.
\end{enumerate}
\emph{To optimize \sys performance, we target higher MCD and WER along with a lower DDR, which reduces the chance of leaking privacy information. Additionally, we aim for a higher PESQ score.}

\begin{table}[b]
\vspace{+0.1in}
\caption{Different operating frequencies on \sys.}
\centering
\scriptsize
\label{tab:eval_freq}
\makebox[\linewidth][c]{
\midsepremove
\begin{tabular}{ cccc|ccc }
\toprule
\multirow{2}{*}{SSEA} & \multicolumn{3}{c|}{TI IWR6843 (60GHz)} & \multicolumn{3}{c}{TI IWR1843 (77GHz)}  \\
\cline{2-7}
& \scriptsize  MCD & \scriptsize WER & \scriptsize DDR & \scriptsize MCD &\scriptsize  WER &\scriptsize  DDR  \\
\hline
ML & 13.7 & $70.3\%$ & $2\%$ & 13.4 & $68.2\%$ & $3\%$  \\
SP & 13.9 & $71.4\%$ & $2\%$ & 13.6 & $70.1\%$ & $2\%$  \\
\bottomrule
\end{tabular}}
\end{table}

\begin{table}[b]
\vspace{+0.1in}
\caption{Different sampling rates on \sys.}
\centering
\scriptsize
\label{tab:eval_sampling}
\makebox[\linewidth][c]{
\midsepremove
\begin{tabular}{ cccc|ccc }
\toprule
\multirow{2}{*}{SSEA} & \multicolumn{3}{c|}{8kHz} & \multicolumn{3}{c}{12kHz} \\
\cline{2-7}
& \scriptsize  MCD & \scriptsize WER & \scriptsize DDR & 
\scriptsize MCD &\scriptsize  WER &\scriptsize  DDR \\
\hline
ML & 13.5 & $68.6\%$ & $3\%$ & 13.4 & $68.5\%$ & $3\%$ \\
SP & 13.6 & $69.8\%$& $2\%$ & 13.5 & $68.8\%$ & $2\%$ \\
\bottomrule
\end{tabular}}
\end{table}

\begin{table}[b]
\vspace{+0.1in}
\caption{Different antenna configurations on \sys.}
\centering
\scriptsize
\label{tab:eval_antenna}
\makebox[\linewidth][c]{
\midsepremove
\begin{tabular}{ cccc|ccc }
\toprule
\multirow{2}{*}{SSEA} & \multicolumn{3}{c|}{1 Tx + 4 Rxs} & 
\multicolumn{3}{c}{MIMO (3 Txs + 4 Rxs)}\\
\cline{2-7}
& \scriptsize  MCD & \scriptsize WER & \scriptsize DDR & 
\scriptsize  MCD & \scriptsize  WER &\scriptsize  DDR \\
\hline
ML & 13.4 & $68.3\%$ & $3\%$ & 13.9 & $72.5\%$ & $1\%$ \\
SP & 13.5 & $69.2\%$ & $3\%$ & 14.1 & $75.1\%$ & $1\%$ \\
\bottomrule
\end{tabular}}
\end{table}

\presec
\subsection{Results of mmWave Radar-based SSEA}
\label{sec:eval_mmwave}
\noindent\textbf{Overall Performance.} Table~\ref{tab:eval1} shows defense performance against the mmWave radar in the baseline attack scenario. With \sys activated, the MCD significantly increases from $3.4$ to $13.6$, underscoring its effectiveness.
Specifically, sentences translated by SR exhibit a WER exceeding $68\%$, and the accuracy of the audio classifier drops to below $5\%$. Furthermore, \sys achieves a PESQ of $3.42\pm0.25$, indicating that the perturbed audio remains perceptually similar to the original. 
Gaussian noise and VAP are not as effective. \sys outperforms the baselines by 5.3$\times$ and 4.4$\times$ respectively on the WER metric. We confirm that the effects of VAP do not transfer to vibration-based SSEA recognition models, as VAP is designed to subvert microphone-based audio recognition models. This result underscores the importance of designing specialized perturbations to prevent SSEA effectively.
\jungwoo{To analyze the impact of our perturbations, we visualize the spectrograms of mmWave signals in Figure~\ref{fig:result_spec}(a) and \ref{fig:result_spec}(b). From Figure~\ref{fig:result_spec}(a), we see that the raw-recovered audio can be restored to high-quality audio similar to the original audio (Figure~\ref{fig:feasibility}(a)) using SSEA's audio enhancement model. However, SSEA fails to improve perturbed mmWave signals, as shown in Figure~\ref{fig:result_spec}(b).} Specifically, the restored audio becomes dominated by noise and unintelligible to humans.

Next, we evaluate a comprehensive set of attack scenarios by altering one of the environmental factors involved in the baseline attack setup.

\noindent\textbf{Impact of Distance and Direction.} We vary distances between the radar and the tinfoil, as well as between the loudspeaker and the tinfoil, from 0.5 to 2.5m at a fixed angle of $0^{\circ}$. As shown in Figure~\ref{fig:exp1}(a) and \ref{fig:exp1}(b), \sys maintains high performance even when the attacker is close to the vibrating sound source. 
Figure~\ref{fig:exp1}(c) further shows the results when varying the angles between the attacking radar and tinfoil at a fixed 0.5m. Again, \sys achieves high performance regardless of the relative angle since the perturbation propagates uniformly across different directions.

\noindent\textbf{Impact of Sound Volume.} The audio source's volume directly influences the reverberator vibration intensity. We evaluate the \sys by adjusting the sound volume at the baseline attack setting. 
As shown in Figure~\ref{fig:exp1}(d), although SSEA can achieve better performance at higher volumes, 
\sys is still able to achieve consistent defense performance. Specifically, at a volume of 85dB, \sys achieves an MCD of 12.7, a WER of 66.8\%, and a DDR of 3\% on average.

\begin{figure}[t]
    \centering
    \resizebox{0.95\linewidth}{!}{
    \begin{tabular}{@{}c@{}}
        \includegraphics[width=\linewidth]{Fig/fig_exp_legend.png} \\  
    \end{tabular}}
    \begin{tabular}{@{}cc@{}}
    \centering
        \includegraphics[width=0.43\linewidth]{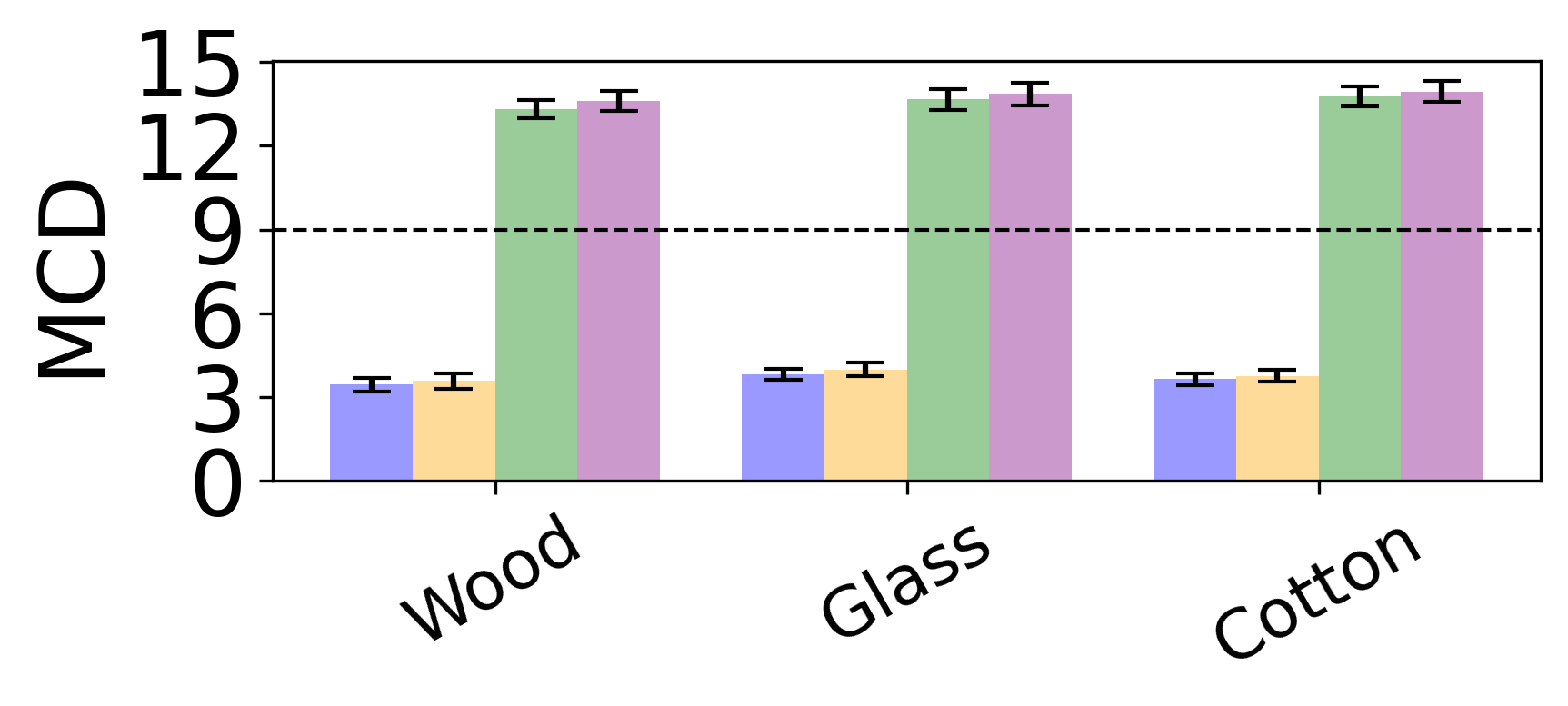}  &
        \includegraphics[width=0.43\linewidth]{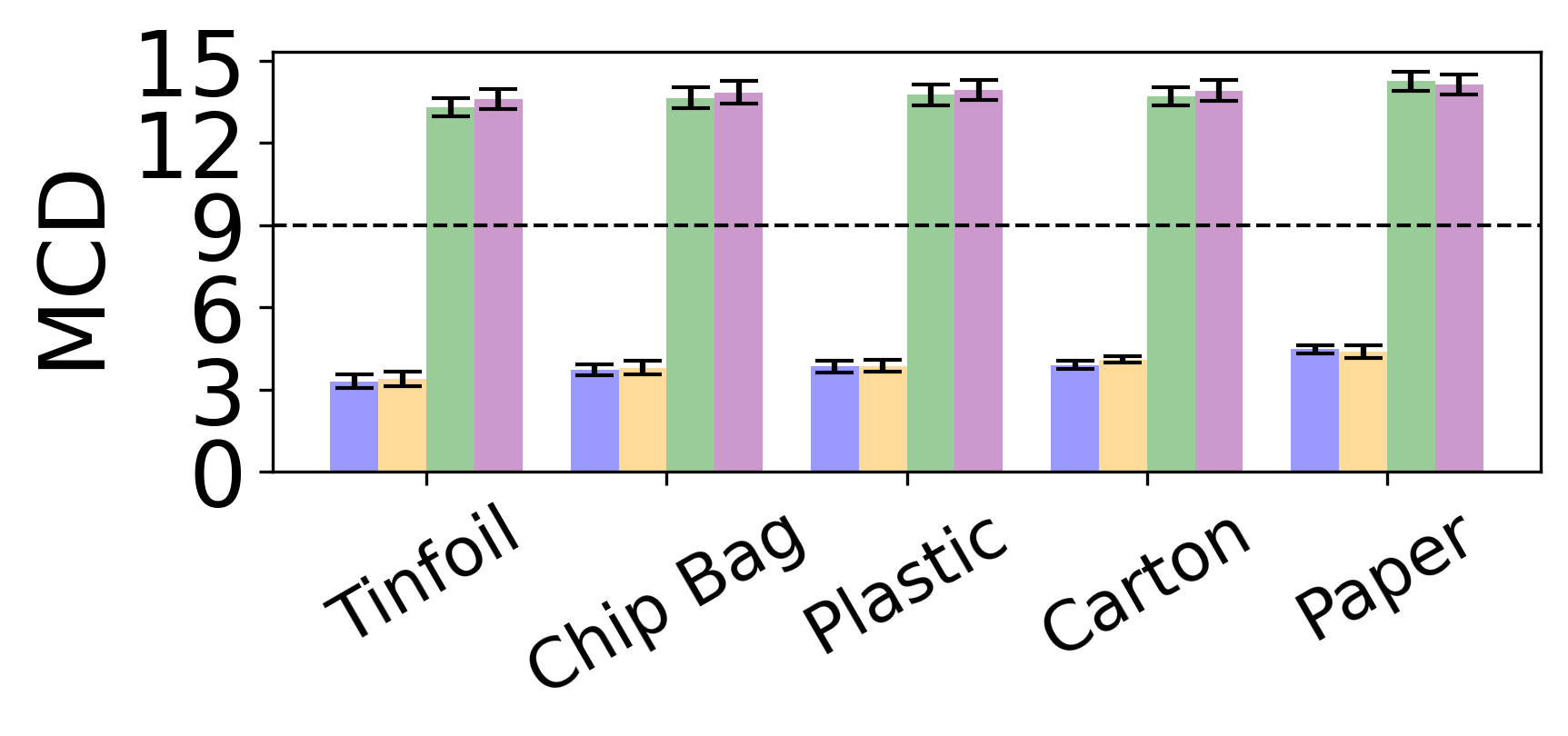} \\
        \includegraphics[width=0.43\linewidth]{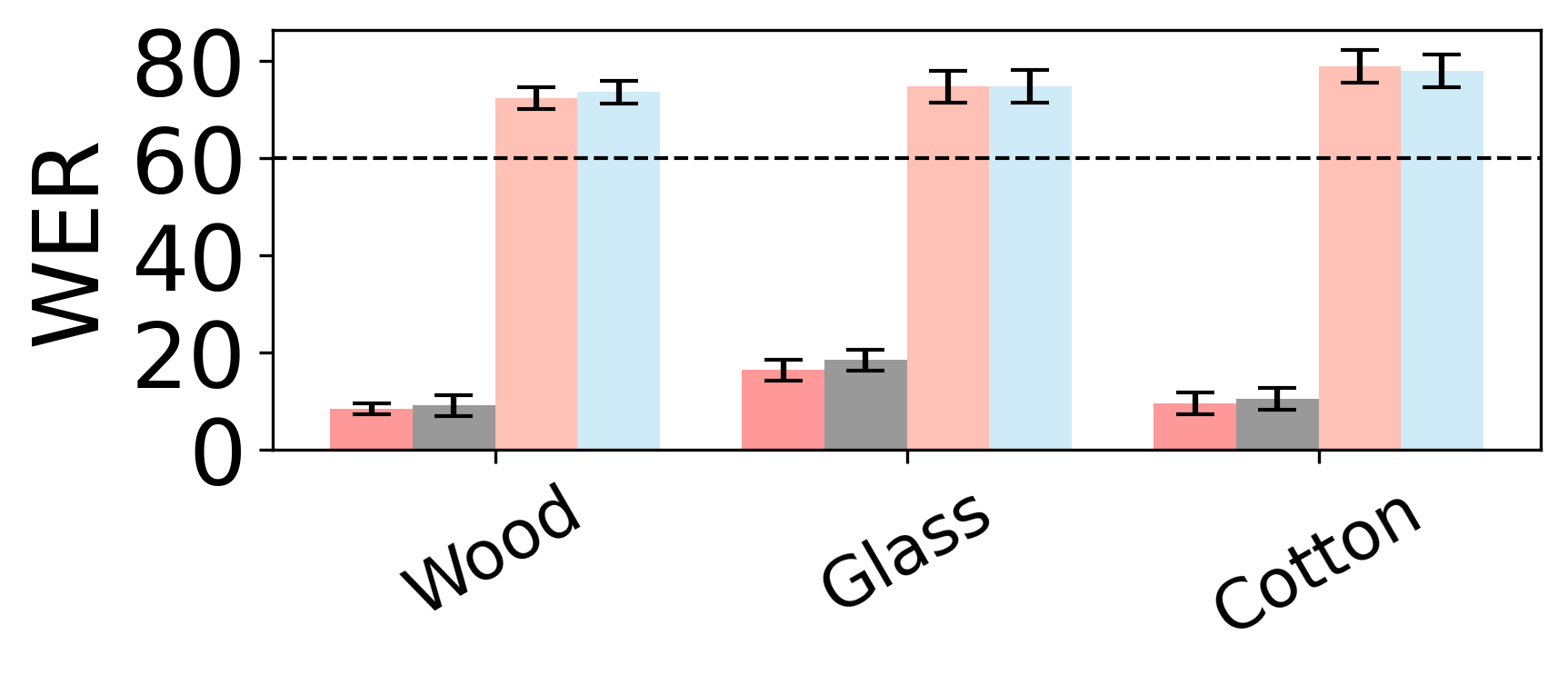}  &
        \includegraphics[width=0.43\linewidth]{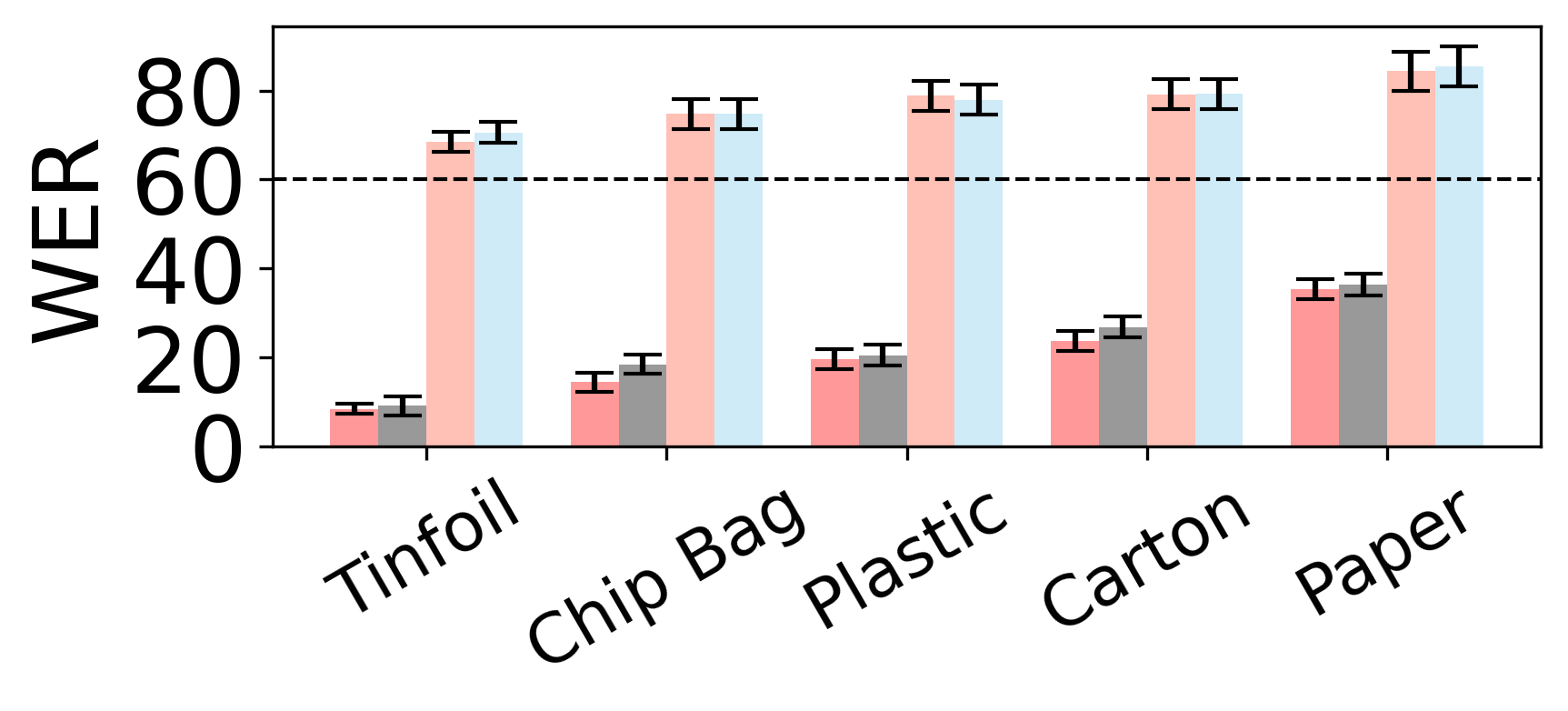} \\
        \includegraphics[width=0.43\linewidth]{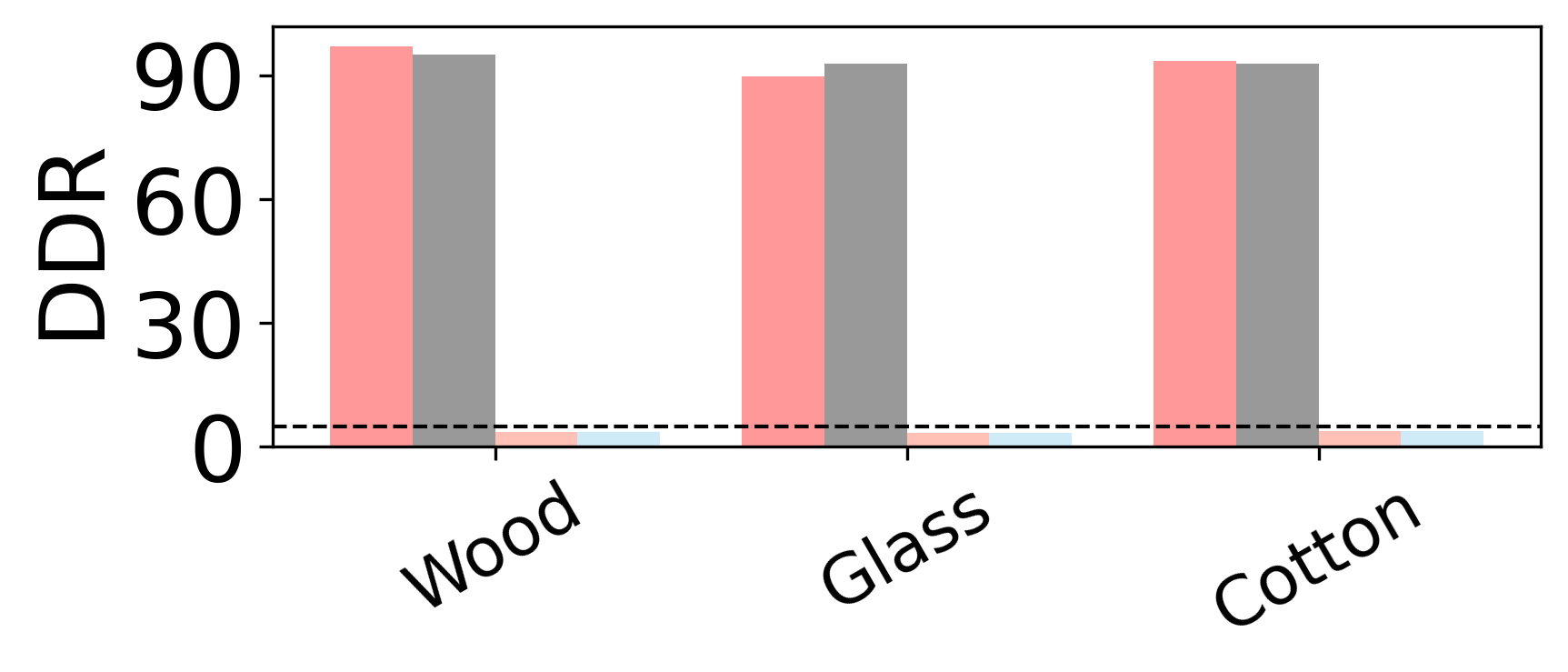}  &
        \includegraphics[width=0.43\linewidth]{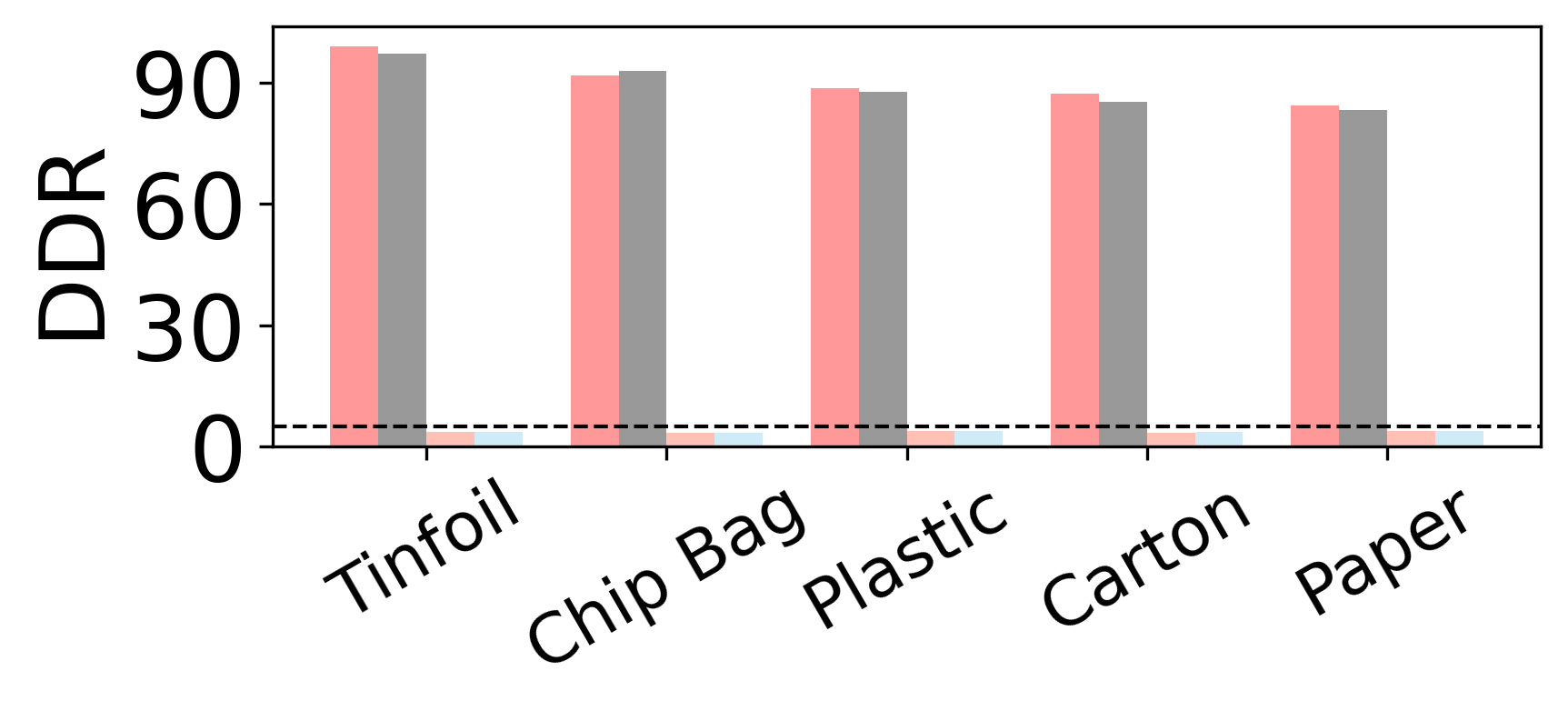} \\
        \footnotesize (a) &
        \footnotesize (b) \\
    \end{tabular}
    \caption{MCD, WER, and DDR of different (a) insulators and (b) reverberators.}
    \label{fig:exp2}
\end{figure}

\noindent\textbf{Impact of Acoustic Insulators.} We install various insulating materials between the mmWave radar and tinfoil in the baseline attack setup and then evaluate the performance. 
In Figure~\ref{fig:exp2}(a), we can observe that regardless of the insulator, \sys maintains a high MCD and WER and low DDR. The acoustic insulators do not affect mmWave radar's ability to capture the vibration of sound sources or reverberating materials. Yet \sys precedes the audio emission and thus remains as effective as the case with insulators. 

\noindent\textbf{Impact of Different Reverberating Materials.} The same audio can induce vibrations of varying intensity depending on the reverberating materials~\cite{hu2023mmecho, shi2023privacy}. We evaluate \sys by replacing the tinfoil with other materials. As shown in Figure~\ref{fig:exp2}(b), \sys has an MCD of up to 14.2, a WER of up to 85.5, and a DDR of at least 3$\%$. These results highlight that our perturbations, trained on an ensemble of SSEA samples, effectively adapt to significant deviations in the acoustic properties of various reverberating materials.

\noindent\textbf{Impact of Radar Frequency.}
We evaluate the transferability of the \sys to an unseen radar frequency. We use the TI 60~GHz radar (IWR6843-Boost~\cite{iwr6843}) to measure the sound-induced vibration. As shown in Table~\ref{tab:eval_freq}, \sys performs even better on the 60~GHz radar. Since the 60~GHz radar has a lower vibration resolution than the default 77~GHz radar, it is less capable of detecting high-frequency bands. Thus, LFAPs are more prominent.

\begin{table}[b]
\vspace{+0.1in}
\caption{\sys against baseline attack setups of optical sensor and accelerometer.}
\centering
\label{tab:imu_laser}
\resizebox{0.95\columnwidth}{!}{
\midsepremove
\begin{tabular}{ ccccc|cccc }
\toprule
\multirow{2}{*}{Defense} & \multicolumn{4}{c|}{Optical Sensor} & \multicolumn{4}{c}{Accelerometer}\\
\cline{2-5} \cline{6-9}
&  MCD & WER &  DDR & PESQ &  MCD & WER & DDR & PESQ \\
\hline
OFF & $5.8 $ & $12.2\%$ & $92\%$ & - & $6.5$ & $15.4\%$ & $88\%$ & - \\
Gaussian & $9.5 $ & $20.5\%$ & $71\%$ & 2.54 & $10.4$ & $24.4\%$ & $65\%$ & 2.54 \\
VAP & $9.2 $ & $22.6\%$ & $65\%$ & 2.63 & $9.6$ & $27.5\%$ & $62\%$ & 2.63 \\
\sys & $14.5 $ & $73.2\%$ & $3\%$ & 3.42 & $14.7$ & $88.6\%$ & $1\%$ & 3.42 \\
\bottomrule
\end{tabular}}
\end{table}

\noindent\textbf{Impact of Sampling Rate.}
We adjust the chirp rate of the mmWave radar to capture vibrations at sampling rates different from those in the training set. Table~\ref{tab:eval_sampling} shows that \sys's defense performance is consistent regardless of the sampling rate. As mentioned in Sec.~\ref{sec::preli:understand}, SNR tends to decrease to almost 0dB for frequencies above 2kHz, so even when the radar's sampling rate is increased to 12~kHz, it still cannot capture high-frequency vibrations.

\noindent\textbf{Impact of Antenna Configurations.}
We consider two types of widely-used multi-antenna setups in SSEAs~\cite{hu2023mmecho, shi2023privacy}. As shown in Table~\ref{tab:eval_antenna}, although \sys is trained using 1 Tx and 1 Rx, its performance is invariant across antenna settings. Multi-antenna can enhance sensing by focusing a directional beam toward the reverberator, but it only improves the low- and mid-frequency bands, and the SNR in the high-frequency bands remains close to zero~\cite{shi2023privacy}. 
This means that the multi-antenna eavesdropping signals are more strongly biased by our perturbations from the sound sources.

\begin{table}[b]
\vspace{+0.1in}
\caption{Different sampling rates of the accelerometer.}
\centering
\scriptsize
\label{tab:eval2}
\makebox[\linewidth][c]{
\midsepremove
\begin{tabular}{ cccc|ccc }
\toprule
\multirow{2}{*}{\scriptsize \sys} & \multicolumn{3}{c|}{167Hz} & \multicolumn{3}{c}{200Hz} \\
\cline{2-7}
& \scriptsize MCD & \scriptsize WER & \scriptsize DDR & 
\scriptsize  MCD &\scriptsize  WER &\scriptsize  DDR \\
\hline
OFF & 8.7 & $21.5\%$& $81\%$& $7.6 $ & $20.3\% $ & $82\% $ \\
\cellcolor{Gray} ON & \cellcolor{Gray} 14.9 & \cellcolor{Gray} $93.4\%$& \cellcolor{Gray} $0\%$ & \cellcolor{Gray} $14.9 $ & \cellcolor{Gray} $92.2\% $ & \cellcolor{Gray} $1\% $ \\
\bottomrule
\end{tabular}}
\end{table}

\begin{table}[b]
\vspace{+0.1in}
\caption{FIR perturbation and LFAP on \sys.}
\centering
\label{tab:eval3}
\resizebox{0.95\columnwidth}{!}{
\midsepremove
\begin{tabular}{ cccc|ccc|ccc }
\toprule
\multirow{2}{*}{SSEA} & \multicolumn{3}{c|}{Two-Stage PGM} & \multicolumn{3}{c|}{FIR Perturbation} & \multicolumn{3}{c}{LFAP}\\
\cline{2-10}
&  MCD & WER & DDR & MCD & WER & DDR & MCD & WER & DDR \\
\hline
ML & 13.4 & $68.2\%$ & $3\%$ & 4.8 & $52.5\%$ & $17\%$ & 13.1 & $59.6\%$ & $55\%$ \\
SP & 13.6 & $70.1\%$& $2\%$ & 5.2 & $56.3\%$ & $15\%$ & 13.3 & $61.5\%$ & $50\%$ \\
\bottomrule
\end{tabular}}
\end{table}

\presec
\subsection{Results of Different Sensor-based SSEAs}
\label{sec:eval_motion}

\noindent\textbf{Overall Performance.} We play perturbed audio from the baseline attack setups of the optical sensor and the accelerometer. Table~\ref{tab:imu_laser} shows the results of the \sys defense. We confirm that adversarial perturbations severely impede audio restoration.
\sys outperforms the defense baselines (i.e., Gaussian noise and VAP) by a large margin, increasing the WER by up to 3.6$\times$. This result highlights the necessity of designing specialized perturbations to prevent SSEA, as shown in Figure~\ref{fig:overview}.
\jungwoo{As shown in Figure~\ref{fig:result_spec}(c)-(f), we observe that \sys defeats the attacker's cGAN-based audio enhancement. Since the optical sensor has a strong response in the low-frequency range below 500Hz, eavesdroppers are vulnerable to our perturbations. The motion sensor is located on the same surface as the smartphone speaker, making it challenging for the accelerometer to evade our perturbations.} 

\begin{figure}[t]
    \centering
    \resizebox{0.7\linewidth}{!}{
    \begin{tabular}{@{}c@{}}
        \includegraphics[width=\linewidth]{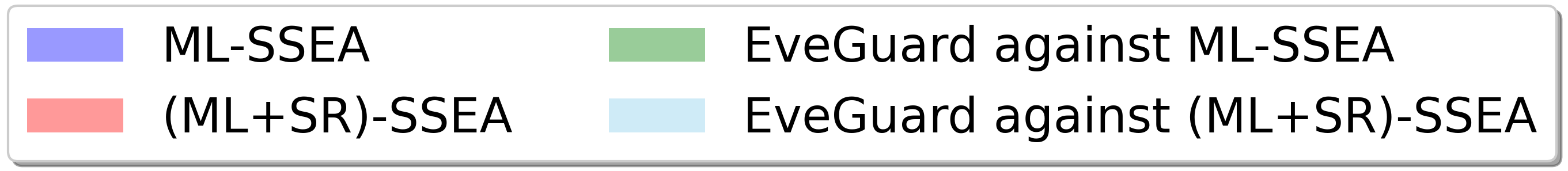} \\  
    \end{tabular}}
    \begin{tabular}{@{}cc@{}}
    \centering
        \includegraphics[width=0.43\linewidth]{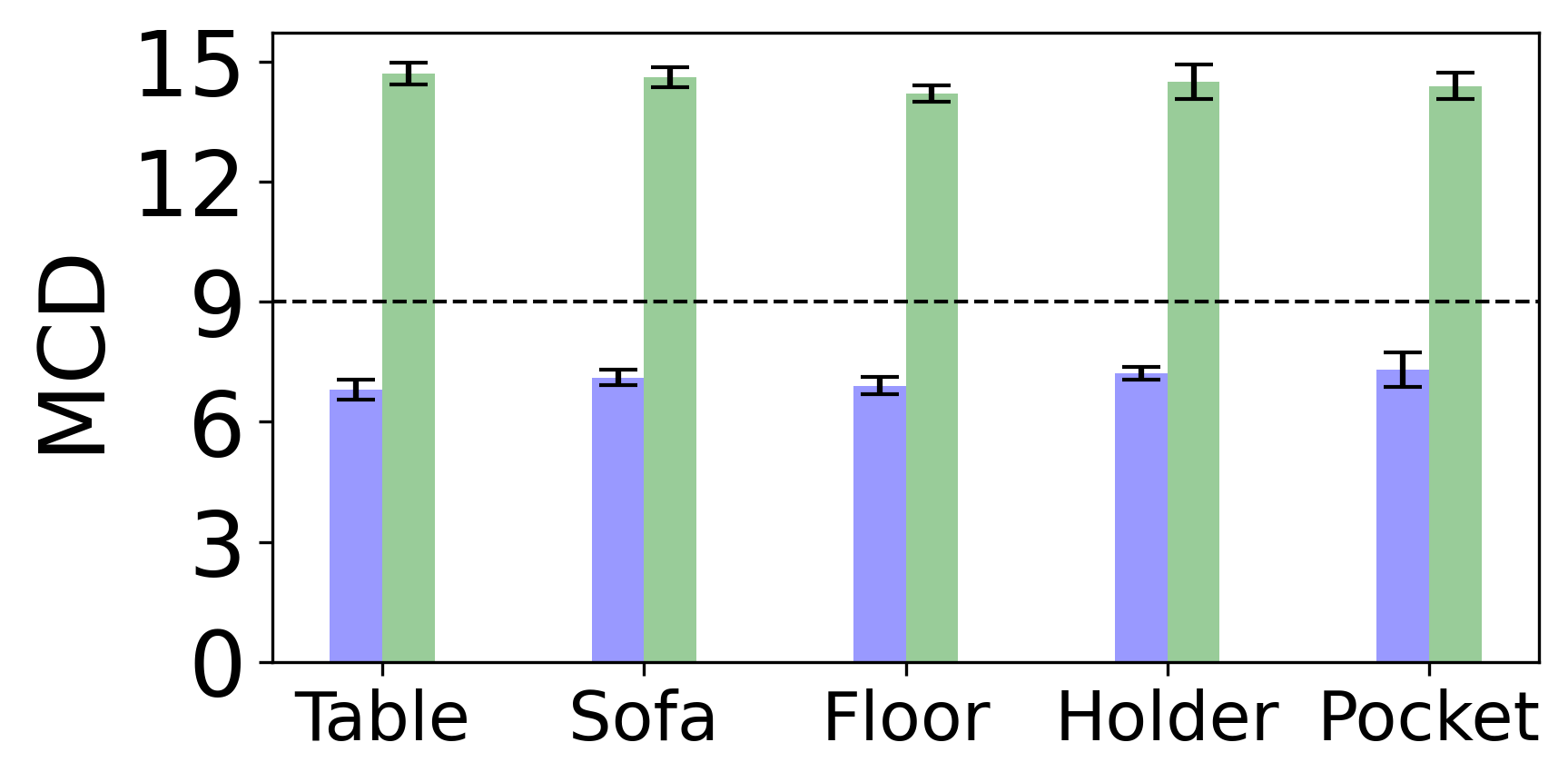}  &
        \includegraphics[width=0.43\linewidth]{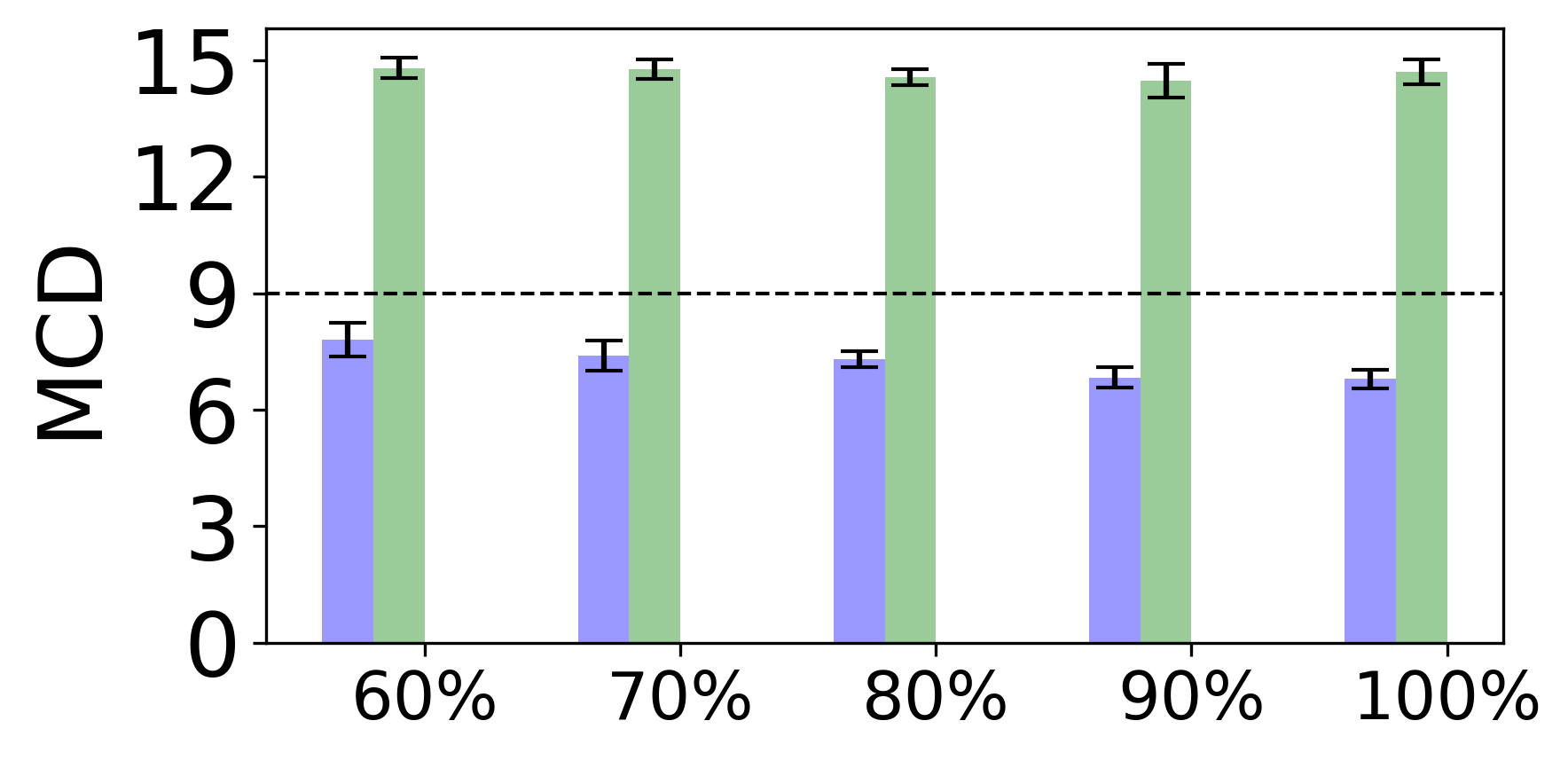} \\
        \includegraphics[width=0.43\linewidth]{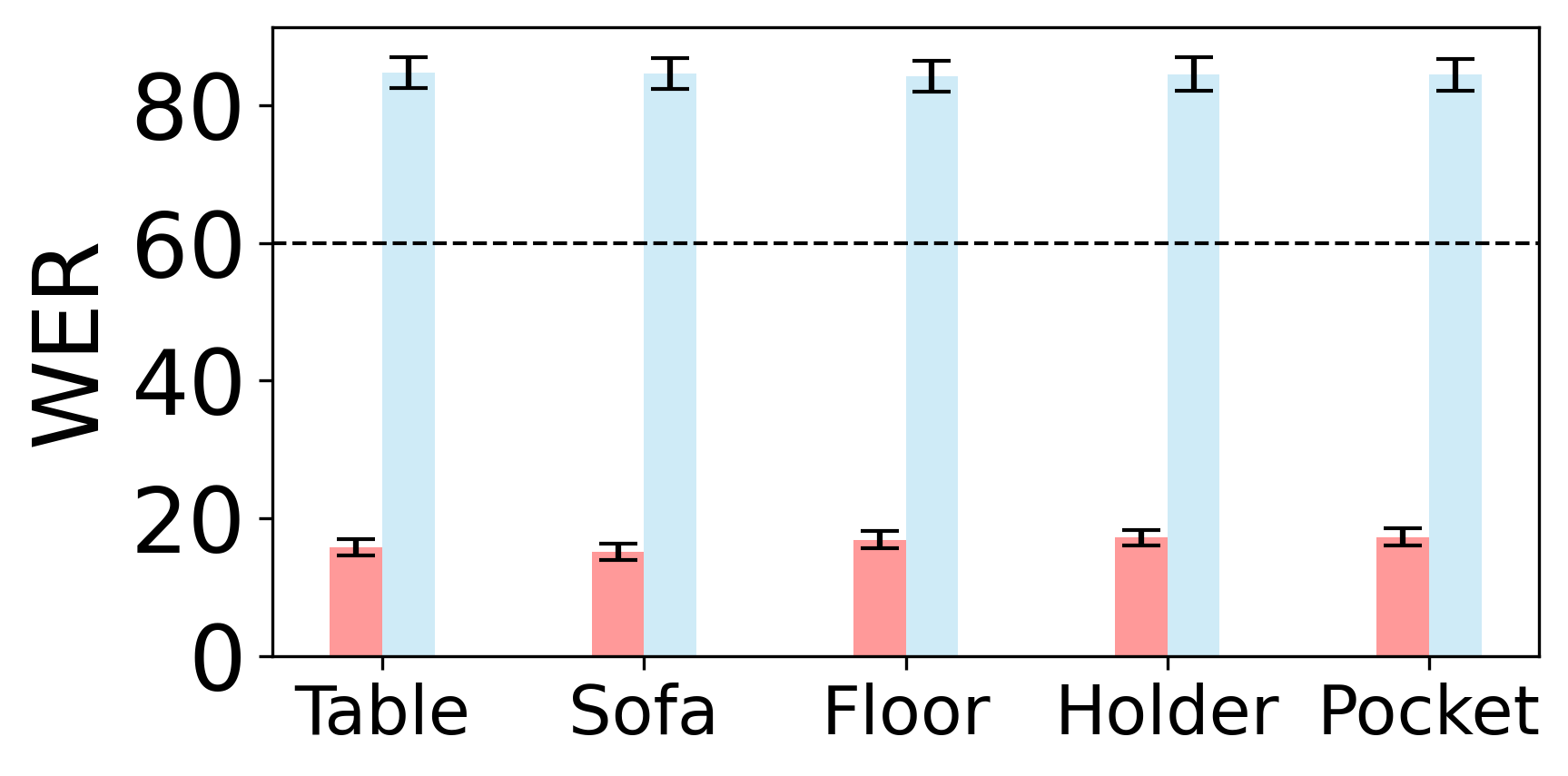}  &
        \includegraphics[width=0.43\linewidth]{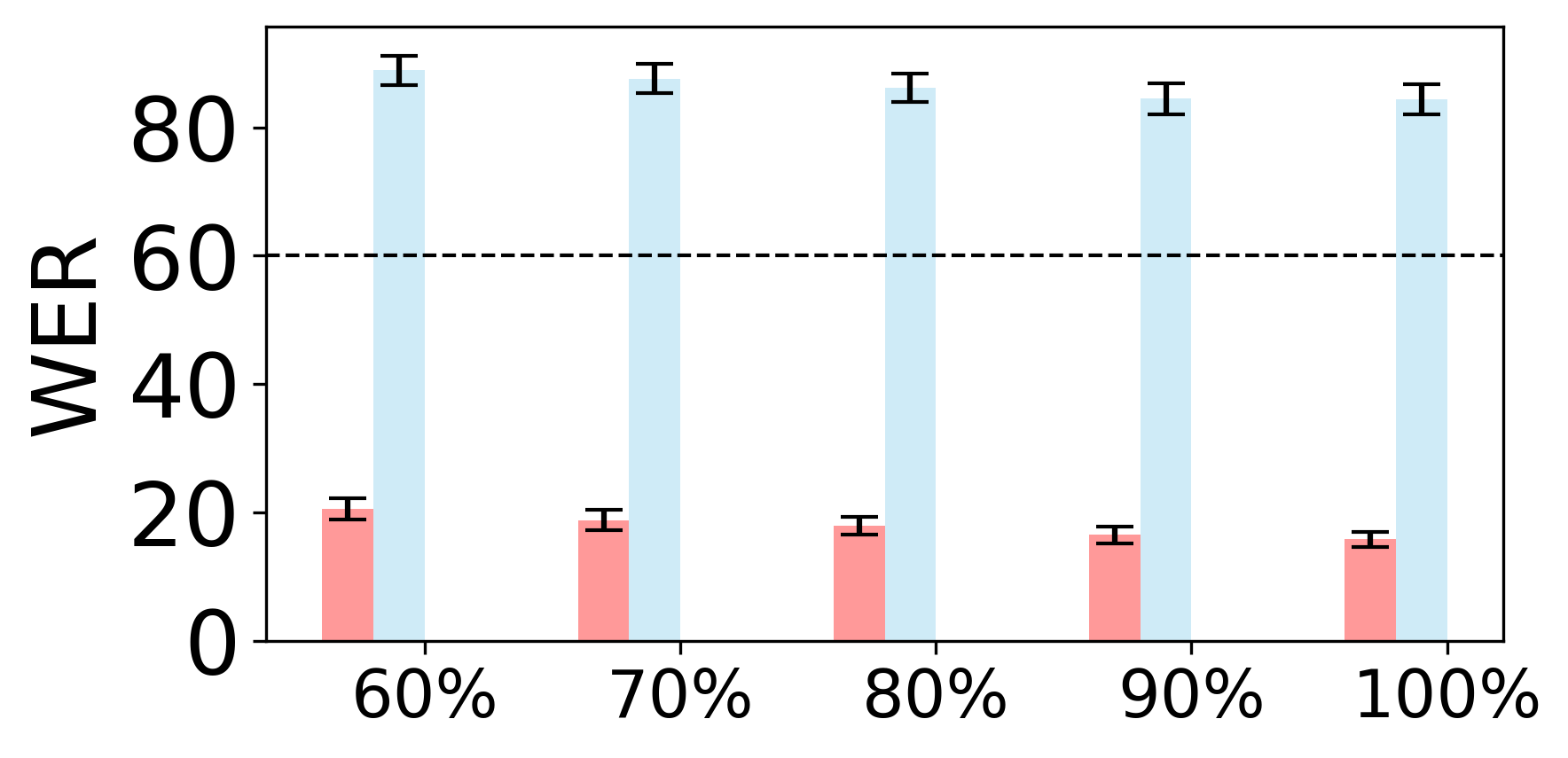} \\
        \includegraphics[width=0.43\linewidth]{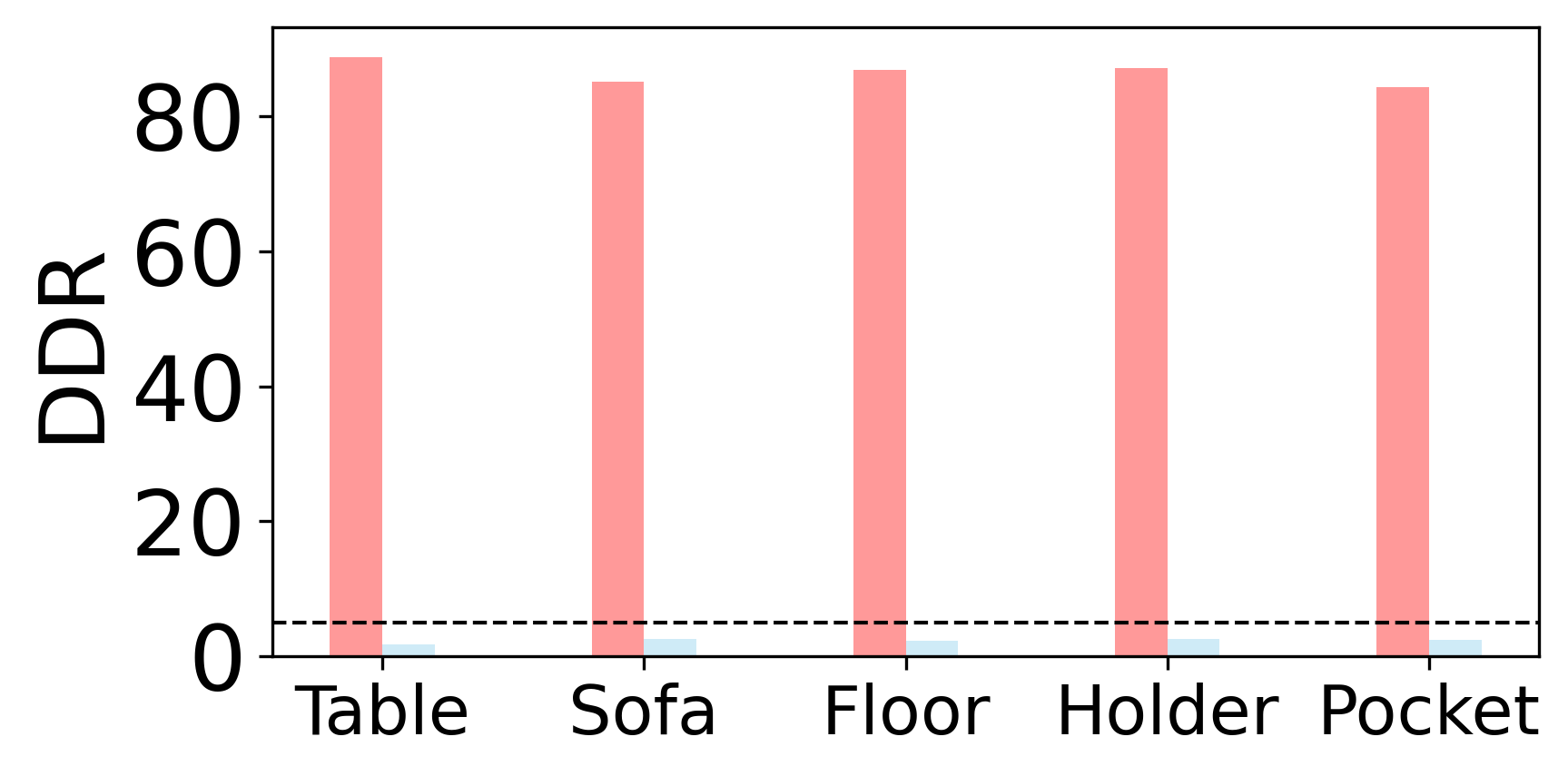}  &
        \includegraphics[width=0.43\linewidth]{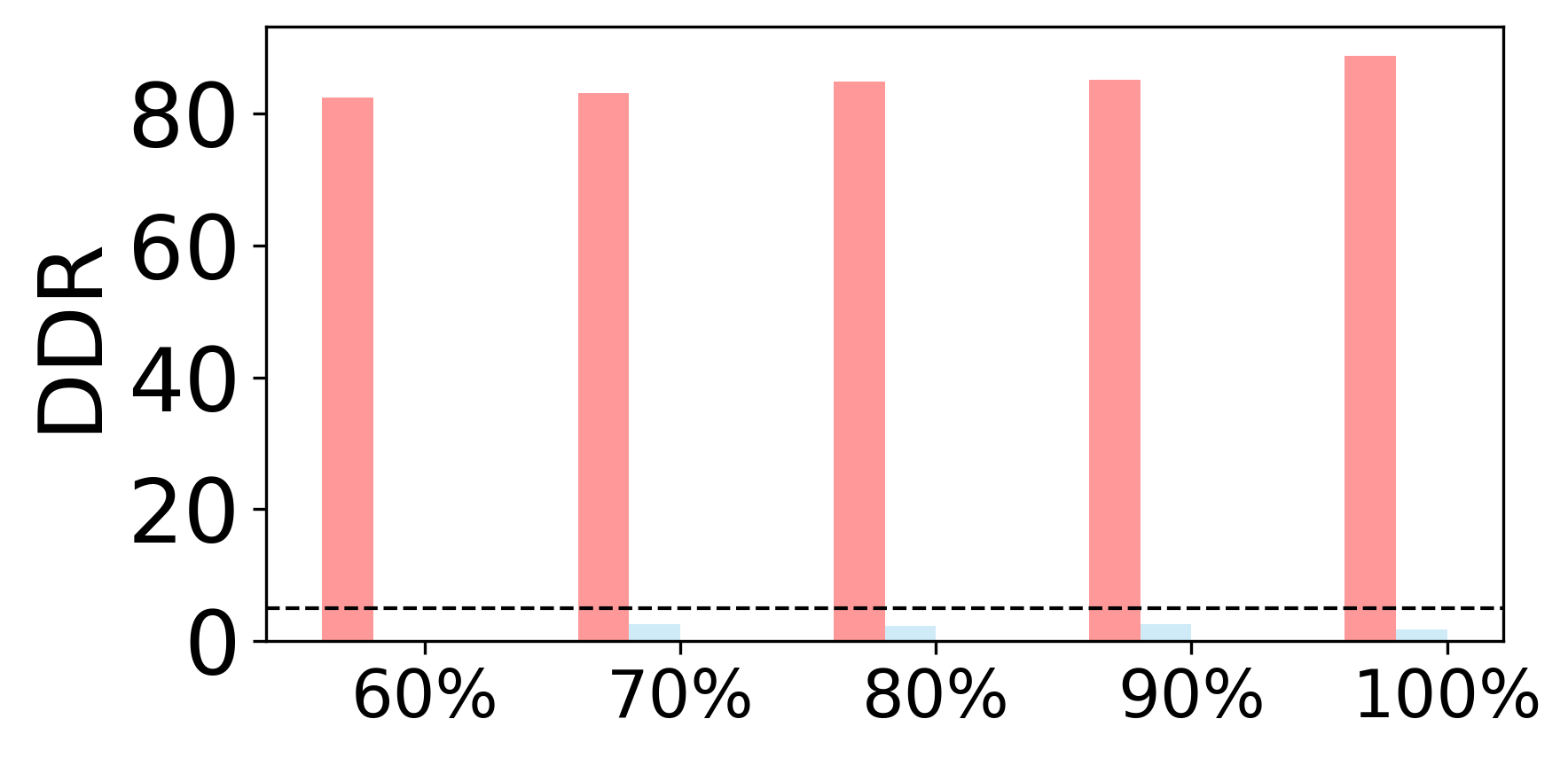} \\
        \footnotesize (a) &
        \footnotesize (b) \\
    \end{tabular}
    \caption{MCD, WER, and DDR of different (a) placements and (b) smartphone volumes.}
    \label{fig:exp3}
\end{figure}

\noindent\textbf{Effectiveness across different attack scenarios.} 
To understand the impact of the sampling rate, we evaluate accelerometer data at 167Hz and 200Hz sampling rates, following \cite{hu2022accear}. As shown in Table~\ref{tab:eval2}, \sys consistently performs well at these rates. Additionally, we vary the surface on which the smartphone sits and the audio volume. As shown in \jungwoo{Figure~\ref{fig:exp3}}, the defense performance of \sys meets all MCD, WER, and DDR thresholds. This robustness is attributed to \sys's design, ensuring low-frequency perturbations remain dominant in the restored audio, making it resilient to various environmental factors.

\presec
\subsection{User Study for Speech Quality}
\label{sec:user}

\revision{In this study, we recruited 24 volunteers to evaluate the naturalness of audio samples. Specifically, we conducted an online survey in which participants listened to 30 pairs of audio samples, comprising a balanced mix of clean speech, perturbation-added speech, and speech eavesdropped from these two types, totaling 120 samples. To mitigate bias, the sample sequence was randomized, and the study’s objective was concealed. Participants were also encouraged to remain engaged throughout the evaluation.
We received an Institutional Review Board (IRB) exemption from our institute (IRB $\#$812263), confirming compliance with ethical guidelines. We collected limited demographic information, including participants' age range (20–40) and professional backgrounds: 58.3$\%$ graduate students, 33.3$\%$ industry professionals, and 8.3$\%$ educators. No personally identifiable information (PII) was collected.}

The participants were requested to rate two criteria based on a Likert scale~\cite{likert1932technique} from 1 to 5: (1) the intelligibility of the eavesdropper's reconstructed audio and (2) perceptual quality of \sys's perturbed audio. In order to avoid bias, participants are not informed that the audio is perturbed.
\jungwoo{The complete survey can be found in Appendix~\ref{sec:apend_user}.}
As shown in Figure~\ref{fig:exp_user}(b), we found that 94.8$\%$ of participants could not discern any information from the SSEA-reconstructed audio, while 5.2$\%$ could vaguely hear a few words. 
Participants rated the naturalness of \sys' perturbed audio highly, with an average score of 4.59, closely approaching the original audio's perceptual quality score of 4.69, indicating that \sys preserves the audio quality.

\begin{table}[b]
\vspace{+0.1in}
\caption{\revision{\sys performance against adversarial training (AT) and perturbation removal (PR).}}
\centering
\label{tab:eval4}
\resizebox{0.96\columnwidth}{!}{
\midsepremove
\begin{tabular}{ cccc|ccc|ccc }
\toprule
\multirow{2}{*}{\begin{tabular}[c]{@{}c@{}}$\#$ of \\ Samples \end{tabular}} & \multicolumn{3}{c|}{Original} & \multicolumn{3}{c|}{AT} & \multicolumn{3}{c}{PR}\\
\cline{2-10}
& MCD & WER & DDR & MCD & WER & DDR & MCD &  WER & DDR \\
\hline
100 secs & 13.4 & $68.2\%$ & $3\%$ & 12.6 & $63.9\%$ & $6\%$ & 13.5 & $70.4\%$ & $3\%$ \\
30 mins & 13.4 & $68.2\%$ & $3\%$ & 11.8 & $61.5\%$ & $9\%$ & 12.8 & $65.4\%$ & $8\%$ \\
\bottomrule
\end{tabular}}
\end{table}

\presec
\subsection{Ablation Study}
\label{sec:ablation}
\noindent\textbf{Analysis of \gan.} We evaluate the translation performance of \gan, which converts audio into the SSEA samples. To achieve this, we utilize the widely used Structural Similarity Index Measure (SSIM)~\cite{wang2004image} to quantify the spectrogram similarity between the generated and actual SSEA samples. \revision{The test dataset contains 5,600 audio samples.}
We find that \gan achieves SSIM of 94.02$\%$ for mmWave and 96.51$\%$ for accelerometer, respectively. \revision{The standard deviation for radar and accelerometers is 1.54$\%$ and 2.27$\%$.} These high similarity rates show that \gan effectively produces samples that closely mirror real SSEA samples, thereby enabling the PGM to efficiently train on adversarial examples generated by \gan.


\begin{figure}[t]
    \centering
        \resizebox{0.96\linewidth}{!}{
    \begin{tabular}{@{}c@{}}
        \includegraphics[width=\linewidth]{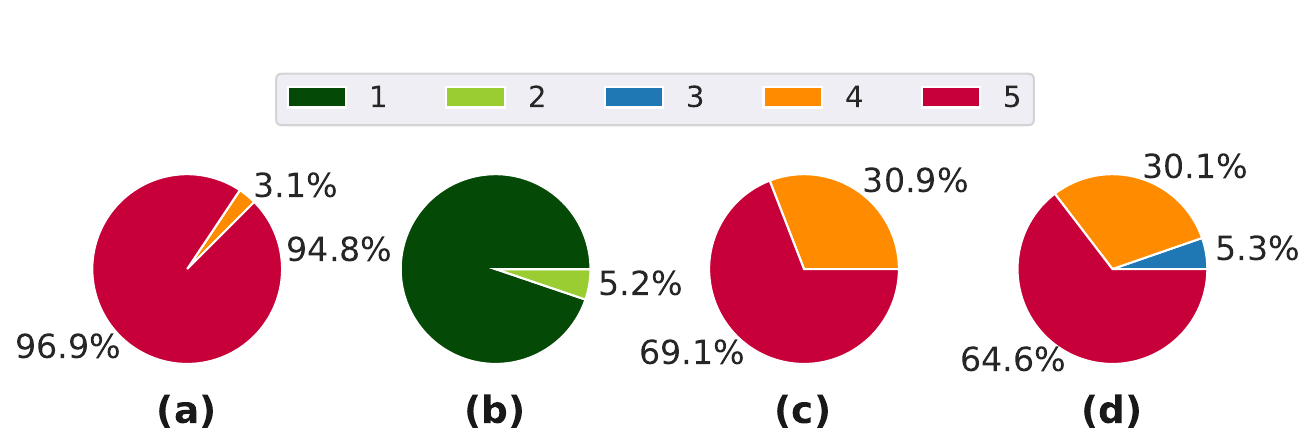} \\  
    \end{tabular}}
    \caption{User study assessing the intelligibility of eavesdropped results for (a) original and (b) perturbed audio, along with the perceptual quality evaluation of (c) original and (d) perturbed audio.}
    \label{fig:exp_user}
\end{figure}

\noindent\textbf{Analysis of Two-Stage PGM.} To understand the impact of FIR perturbation and LFAP, we exclude each module from the PGM and verify the defense for each case under the mmWave radar-based basic attack scenario. As shown in Table~\ref{tab:eval3}, we confirm that FIR filtering has a low impact on the MCD but is effective in perturbing the ML models. This is because FIR perturbations subtly manipulate the frequency spectrum rather than causing a noticeable sound in the perturbed audio. In contrast, LFAP effectively degrades the quality of reconstructed audio by increasing MCD. Thus, integrating the two perturbations can create a synergistic effect.

\begin{table}[b]
\vspace{+0.1in}
\caption{\sys performance against Speech transformation from WaveGuard~\cite{hussain2021waveguard}.}
\centering
\label{tab:eval5}
\resizebox{0.95\columnwidth}{!}{
\midsepremove
\begin{tabular}{ cccc|ccc|ccc }
\toprule
\multirow{2}{*}{SSEA} & \multicolumn{3}{c|}{QT} & \multicolumn{3}{c|}{AR} & \multicolumn{3}{c}{FF}\\
\cline{2-10}
& MCD & WER & DDR & MCD & WER & DDR & MCD & WER & DDR \\
\hline
ML & 13.6 & $66.4\%$ & $4\%$ & 13.6 & $65.7\%$ & $5\%$ & 12.5 & $68.5\%$ & $3\%$ \\
\bottomrule
\end{tabular}}
\scriptsize
     \begin{tablenotes} 
       \item [1] $\bullet$ \textbf{QT}: Quantization; \textbf{AR}: Audio Resampling; \textbf{FF}: Frequency Filtering.
     \end{tablenotes}
\end{table}

\noindent\textbf{Impact of LFAP Power Level.} We also study how different values of $\rho$ in LFAP can balance the trade-off between defense performance and speech quality of perturbed audio, as shown in Figure~\ref{fig:exp_snr}. 
As $\rho$ decreases, MCD increases because LFAP occupies a relatively higher proportion of the eavesdropped audio. Conversely, PESQ, which indicates the quality of perturbed audio, tends to decrease. We find that setting $\rho=16$ satisfies both the MCD and PESQ criteria.


\presec
\subsection{Runtime System Overhead of \sys}
\label{sec:runtime}

\begin{figure}[t]
    \centering
    \begin{minipage}[b]{0.25\textwidth}
        \includegraphics[width=\linewidth]{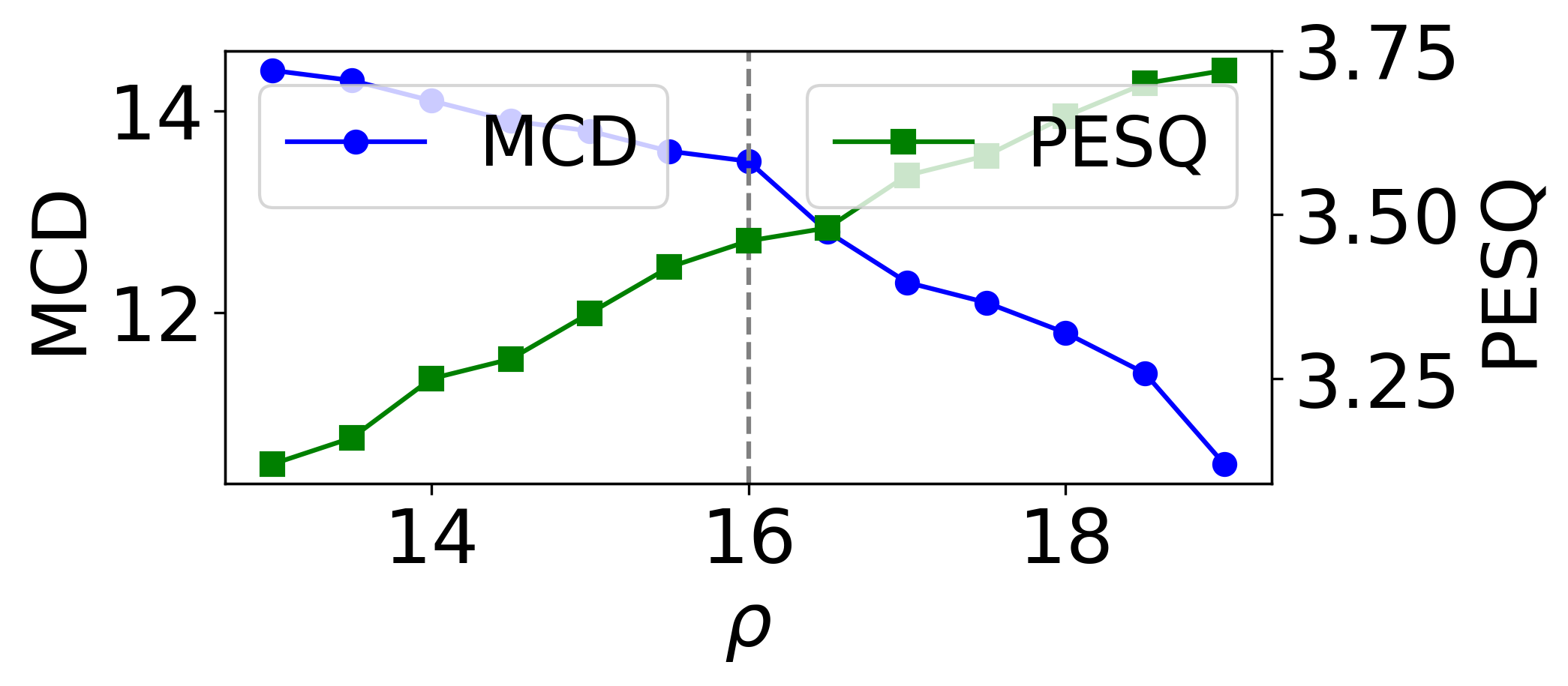}  \\
        \vspace{-0.2in}
        \caption{Impact of LFAP's power.}
        \label{fig:exp_snr}
    \end{minipage}
    \hfill
    \begin{minipage}[b]{0.21\textwidth}
        \hspace{-0.3cm} \includegraphics[width=\linewidth]{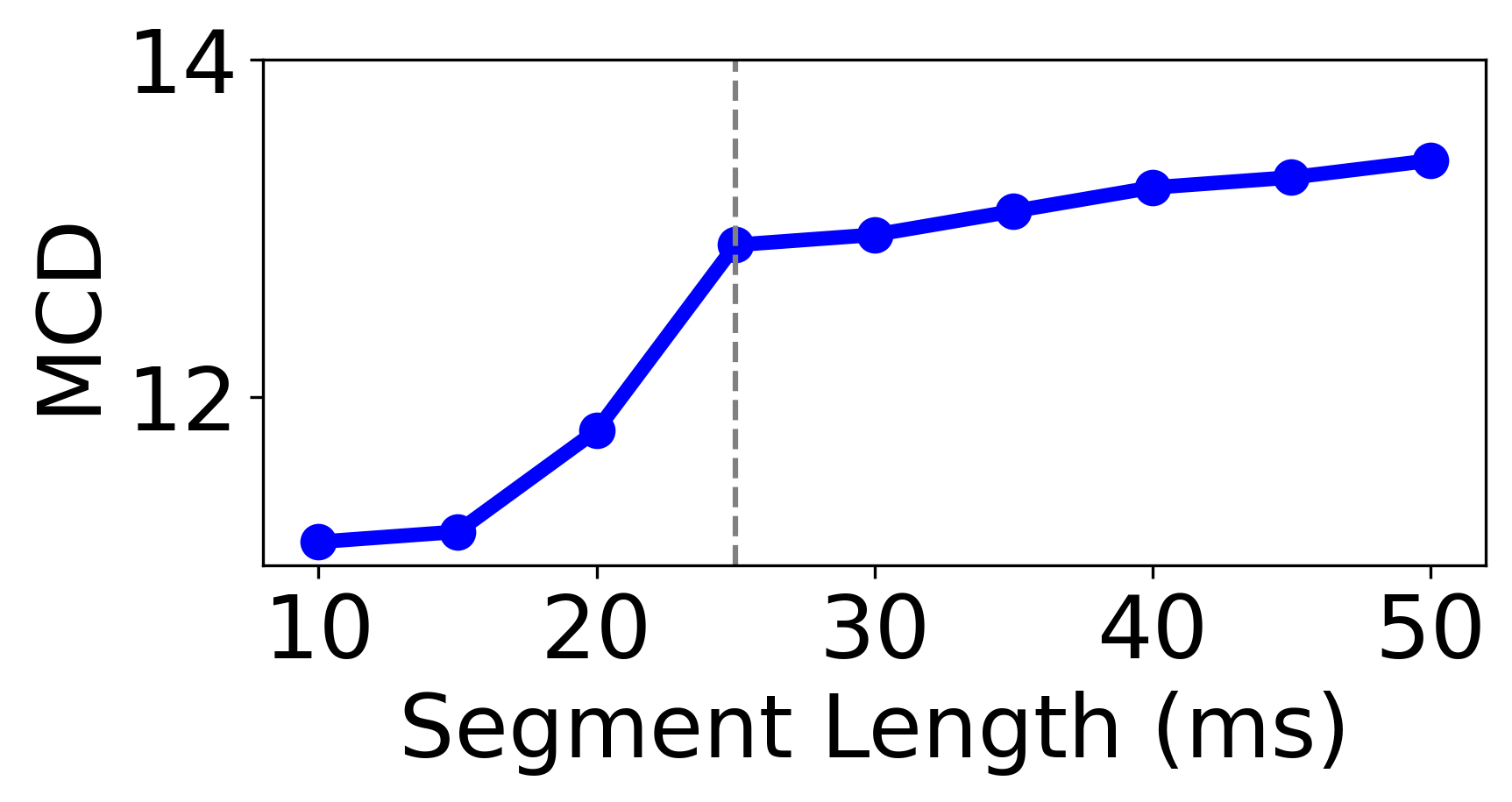}  \\
        \vspace{-0.2in}
        \caption{Impact of PGM's input length.}
        \label{fig:exp_runtime}
    \end{minipage}
\end{figure}

We evaluated the run-time latency of \sys on two hardware platforms, including a workstation with NVIDIA RTX A6000 and a low-end desktop with RTX 2060. \sys converts audio with sampling rates of 16kHz and 48kHz into adversarial audio at 50ms granularity, respectively. 
Experimental results show that the high-end desktop experiences latency of 2.7ms and 6.5ms when the sampling rate is 16kHz and 48kHz, respectively. Additionally, the low-end desktop requires 4.6ms and 11.7ms to process 16kHz and 48kHz audio, respectively.
Since the latency threshold for high-quality VoIP communications is set at 150ms~\cite{series2003transmission}, \sys is feasible for deployment in both cloud and on-device platforms. Furthermore, we investigate \sys performance based on the segment length of the audio fed to the PGM. From Figure~\ref{fig:exp_runtime}, we see that MCD remains nearly constant for segment lengths above 25ms, indicating potential for improving latency overhead. \revision{As a further improvement, we will consider offloading computation from the loudspeaker to a cloud for collaborative defense.}

\presec
\subsection{Robustness to Adaptive Attackers}
\label{sec:adaptive}

We evaluate \sys against adaptive attackers who seek to further enhance ML-SSEA and SR-SSEA based on knowledge of \sys. We define an expert attacker who is aware that the loudspeaker is protected by \sys and knows the defense methodology. However, the attacker does not know the exact ML model of the \gan and PGM. \revision{We assume that the attacker has access to multiple eavesdropping results from original-perturbed sample pairs. To observe the impact of sample quantity on defense performance, we consider two scenarios where the length of sample pairs corresponds to 100 seconds and 30 minutes.} The attacker then trains a substitute \sys with a different architecture (two more layers, different number of neurons) with his/her training data. Additionally, we explore an attack strategy in which the attacker may apply transformation operations to mitigate the perturbation effects. Overall, we consider three types of strategies:



\noindent $\bullet$ \textbf{Adversarial Training.} We aim to robustly train SSEA by allowing attackers to expand the training data using substitute PGM. Specifically, the attacker crafts his/her adversarial audio, performs SSEA to obtain the reconstructed audio, and aggregates it to form a new dataset.

\noindent $\bullet$ \textbf{Perturbation Removal.}
The attacker has learned the perturbation estimates through the substitution PGM. Thus, the attacker attempts to eliminate the disruptive effects of our perturbations within the raw-recovered audio. 

\noindent $\bullet$ \textbf{Speech Transformation.} We verify the robustness of \sys by applying the signal processing techniques~\cite{hussain2021waveguard} that have been used to protect audio systems. Specifically, the attacker can apply audio transformations to the eavesdropped audio obtained from ML-SSEA: (1) quantization-dequantization, (2) down-sampling and up-sampling, and (3) frequency filtering. \jungwoo{The description of each transformation is summarized in Appendix~\ref{sec:append_transformation}}.  

\noindent\textbf{Evaluation Results.} We report experimental results in the baseline attack scenario of mmWave radar, and summarize the results for the adversarial training and perturbation removal in Table~\ref{tab:eval4}. We see that these attacks fail to mitigate the effects of our perturbations because the estimated perturbation used by the attacker has a different distribution from the actual perturbation. Even if the attacker has acquired some eavesdropping results for the actual perturbations, PGM enhances the diversity of perturbations, making it infeasible for attackers to train ML-SSEA and SR-SSEA that are robust to all perturbations. 
\revision{Table~\ref{tab:eval4} presents a performance comparison between two scenarios in which the attacker has access to original-perturbed sample pairs of different lengths. The results indicate a slight improvement in the attacker's eavesdropping when given access to a large volume of \sys's samples; however, the overall eavesdropping capability remains highly limited. This is because the attacker lacks full knowledge of \sys, and the PGM-generated perturbations are highly diverse with random triggers, making adaptive attacks challenging. This observation aligns with the analysis in \cite{chang2023magmaw, bahramali2021robust}.}

Table~\ref{tab:eval5} shows the defense performance against speech transformation-based approaches. We observe that quantization and audio resampling introduce signal distortions in the eavesdropped audio, making ML-SSEA reconstruction worse. Frequency filtering can remove perturbations to some extent, but comes with the trade-off of degrading the original eavesdropping speech quality, resulting in high reconstruction errors.
\section{\revision{Discussion}}

\noindent \revision{\textbf{Electromagnetic (EM) Side-Channel Eavesdropping.}
%
Several studies~\cite{wakabayashi2018feasibility, wang2024wireless, choi2020tempest, yang2024rf, liao2022magear} have explored audio eavesdropping via EM side channels, which can be categorized into two strategies: (a) passive EM sensing and (b) active EM sensing attacks.
Specifically, passive EM sensing~\cite{choi2020tempest, liao2022magear} exploits unintended EM emissions generated when playing audio. Since EM leakage is inherently weak and difficult to detect remotely, passive attacks require close proximity (around 50 cm) to the target.
Active EM sensing attacks~\cite{wakabayashi2018feasibility, wang2024wireless, yang2024rf} involve embedding a retroreflector inside the target device, which reflects the attacker's RF signal back to a receiver, enhancing the leakage for analysis. 
However, this approach relies on stronger assumptions, as it requires prior physical access to the target.}

\revision{Overall, EM side-channel eavesdropping poses weaker security risks compared to vibration-based SSEAs. As part of our future work, we aim to extend \sys to defend against EM side-channel eavesdropping by designing universal perturbations. This approach leverages the frequency response differences between EM side channels and traditional microphones.}

\noindent \revision{\textbf{Eavesdropping on Human Speech.} Our focus is on protecting audio played through loudspeakers. Prior research~\cite{walker2021sok} has experimentally demonstrated that eavesdropping on live human speech is unlikely in real-world scenarios. However, reactive defense mechanisms, such as playing audio perturbations upon detecting speech, could be explored as a potential countermeasure, which we leave for future work.}

\presec
\section{Conclusion}
\label{sec:conclusion}
In this work, we propose \sys, a software-driven defense framework. By utilizing a two-stage PGM and a novel domain translation task called \gan, EveGuard effectively suppresses sensor-based eavesdropping while preserving audio quality. We demonstrate the effectiveness of \sys with state-of-the-art SSEAs. We further validate the \sys with a user study.

\section*{Acknowledgement}
We thank the anonymous reviewers and our shepherd for
their invaluable comments and feedback that helped us improve our manuscript. This work is partially supported by the U.S. Army/Department of Defense award number W911NF2020267 and Google Ph.D. Fellowship.

\bibliographystyle{plain}
\bibliography{main}

\begin{appendices}
\section{Additional Frequency Responses}
\label{sec:append_frequency}
\begin{figure}[h]
    \centering
    \includegraphics[width=0.65\linewidth]{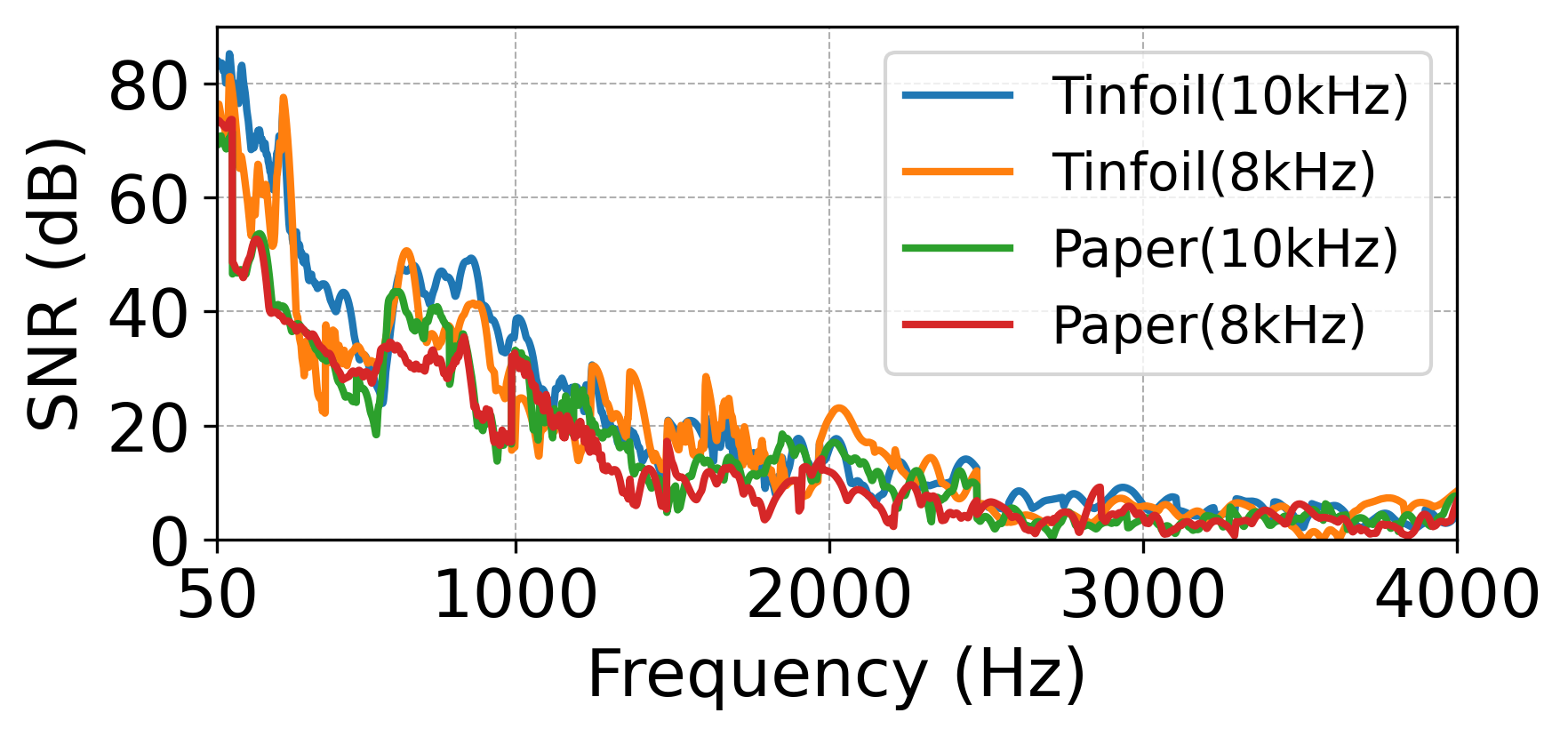} 
    \caption{Comparison of measured frequency responses from different reverberators.}
    \label{fig:append_snr}
\end{figure}

We additionally measure sound-induced vibrations from different reverberators with mmWave radar. Figure~\ref{fig:append_snr} shows a comparison of the measured frequency responses from different reverberators. As seen, the SNR tends to decrease significantly beyond low- and mid-frequencies, reaching nearly 0 dB for frequencies above 2 kHz.

\section{ML Model for SSEA}
\label{sec:append_SSEA}
The attacker adopts ML models introduced in the state-of-the-art literature to achieve high eavesdropping performance. The attacker seeks to enhance the eavesdropped audio with the cGAN model used in Milliear~\cite{hu2022milliear}. As a speech recognition model, the attacker uses the transformer-based speech-to-text model proposed in Radio2Text~\cite{zhao2023radio2text}. Finally, the attacker selects the audio classifier model architecture for digit recognition proposed in ~\cite{shi2023privacy}. The attacker uses the same ML architecture for both mmWave radar and accelerometer, but the learned parameters are different.

\section{Details of Speech Dataset}
\label{sec:append_dataset}
As depicted in Table~\ref{tab:data4}, SSEA and \sys employ distinct sets of speech samples to train their respective ML models. Specifically, SSEA trains cGAN-based audio enhancement models, speech recognition, and audio classifiers. Meanwhile, \sys initiates by establishing a surrogate audio enhancement model to gather SSEA samples, followed by training surrogate speech recognition and surrogate audio classifier, \gan, and two-stage PGM.

\begin{table}[b]
\vspace{+0.1in}
\caption{Experimental setup for victim user, SSEA, and \sys. As seen, SSEA knows the victim's devices (loudspeaker, mmWave radar, and accelerometer) and can collect data from these devices in the victim's room.}
\centering
\label{tab:data3}
\resizebox{0.96\columnwidth}{!}{
\midsepremove
\begin{tabular}{ ccccc }
\toprule
 & \begin{tabular}[c]{@{}c@{}}Loudspeaker\end{tabular} & \begin{tabular}[c]{@{}c@{}} IoT Device \\ (mmWave Radar)\end{tabular} & \begin{tabular}[c]{@{}c@{}}Smartphone \\ (Accelerometer)\end{tabular} & \begin{tabular}[c]{@{}c@{}}Location\end{tabular} \\ 
\hline
\begin{tabular}[c]{@{}c@{}} Victim \\SSEA \end{tabular}  & \begin{tabular}[c]{@{}c@{}} Edifier \\R1700BTs \end{tabular} & \begin{tabular}[c]{@{}c@{}}  1. IWR1843-Boost \\ (with 76-81GHz) \\ 2. IWR6843-Boost \\ (with 60-64GHz) \\ \end{tabular} & \begin{tabular}[c]{@{}c@{}} LG V50 \\ (with 500Hz)  \end{tabular} & \begin{tabular}[c]{@{}c@{}} Figure~\ref{fig:layout}(b)  \end{tabular} \\ 
\hline
\sys  & \begin{tabular}[c]{@{}c@{}}  Logitech \\Z313 \end{tabular} & \begin{tabular}[c]{@{}c@{}} IWR1642-Boost \\ (with 76-81GHz) \end{tabular} & \begin{tabular}[c]{@{}c@{}} Samsung S20 \\ (with 500Hz) \end{tabular} & \begin{tabular}[c]{@{}c@{}} Figure~\ref{fig:layout}(a)  \end{tabular} \\ 
\bottomrule
\end{tabular}}
\end{table}

\begin{table}[t]
\caption{Training dataset used for SSEA and \sys. AE, STT, and AC stand for audio enhancement, speech-to-text model, and audio classification, respectively. F-T denotes fine-tuning. $N_{sc}$ is the number of attack scenarios.} 
\centering
\label{tab:data4}
\resizebox{0.96\columnwidth}{!}{
\midsepremove
\begin{tabular}{ cccccc }
\toprule
 & \begin{tabular}[c]{@{}c@{}}ML\\ Model\end{tabular} & \begin{tabular}[c]{@{}c@{}}Audio\\ Datasets\end{tabular} & \begin{tabular}[c]{@{}c@{}}Sampling\\ Rate\end{tabular} & \begin{tabular}[c]{@{}c@{}}Train\\ Samples\end{tabular} & \begin{tabular}[c]{@{}c@{}}F-T\\ Samples\end{tabular} \\ 
\hline
\multirow{3}{*}{SSEA}  & \multirow{2}{*}{AE} & MILLIEAR~\cite{hu2022milliear} & 48kHz & 8k &  8k $\times N_{sc}$ \\
& & LJSpeech~\cite{ljspeech} & 16kHz & 10k & 10k $\times N_{sc}$ \\
\cline{2-6}
& STT & LJSpeech~\cite{ljspeech} & 16kHz & 10k & 10k $\times N_{sc}$ \\ 
\cline{2-6}
& AC & AudioMNIST~\cite{becker2018interpreting} & 16kHz & 10k & 10k $\times N_{sc}$ \\
\hline
\multirow{5}{*}{\sys}   & PGM & \begin{tabular}[c]{@{}c@{}} VCTK~\cite{yamagishi2019cstr} \\ TIMIT~\cite{garofolo1993darpa} \\ Commands~\cite{warden2018speech} \end{tabular} & \begin{tabular}[c]{@{}c@{}} 48kHz \\ 16kHz \\ 16kHz  \end{tabular} & \begin{tabular}[c]{@{}c@{}} 18k \\ 18k \\ 16k \end{tabular} & \begin{tabular}[c]{@{}c@{}} - \\ - \\ -  \end{tabular} \\
\cline{2-6}
& \begin{tabular}[c]{@{}c@{}}\scriptsize Few-Shot \\\scriptsize \gan \end{tabular} & \begin{tabular}[c]{@{}c@{}} VCTK~\cite{yamagishi2019cstr} \\ TIMIT~\cite{garofolo1993darpa} \end{tabular} & \begin{tabular}[c]{@{}c@{}} 48kHz \\ 16kHz \end{tabular} & \begin{tabular}[c]{@{}c@{}} 18k \\ 18k \end{tabular} & \begin{tabular}[c]{@{}c@{}} - \\ - \end{tabular} \\ 
\cline{2-6}
& \multirow{2}{*}{AE} & VCTK~\cite{yamagishi2019cstr} & 48kHz & 18k & - \\
& & TIMIT~\cite{garofolo1993darpa} & 16kHz & 18k & - \\
\cline{2-6}
& STT & TIMIT~\cite{garofolo1993darpa} & 16kHz & 18k & - \\ 
\cline{2-6}
&  AC & Commands~\cite{warden2018speech} & 16kHz & 16k & - \\ 
\bottomrule
\end{tabular}}
\end{table}

\section{ML Model for \sys}
\label{sec:append_surrogate}
\sys operates under the black-box assumption, employing a surrogate model rather than the target model utilized by SSEA. Specifically, \sys implements audio enhancement via \cite{kameoka2018stargan}. \sys adopts the LSTM-based speech recognition~\cite{shi2019end} and the audio classifier proposed in mmSpy~\cite{basak2022mmspy} as surrogate ML models. 

\revision{Eve-GAN is composed of five key components: a content encoder, a domain encoder, a bottleneck extractor, a generator, and a discriminator. The content encoder maps human speech into a latent content code, while the domain encoder extracts SSEA-specific features. The bottleneck extractor refines these representations before feeding them into the generator, which reconstructs the audio using AdaIN residual blocks and upscale Conv1d layers. The discriminator, based on a multi-period architecture~\cite{kong2020hifi}, evaluates the realism of the generated audio across multiple frequency scales to enhance robustness against SSEA-based attacks.}

\revision{The PGM architecture consists of three main modules. The FIR generator learns the audio perturbation distribution in the frequency domain. The LFAP generator induces low-frequency perturbations to exploit the unique frequency response characteristics of sensors, preventing attackers from accurately restoring speech. Finally, the discriminator distinguishes between real and generated audio, ensuring that perturbations remain imperceptible while making it difficult for adaptive attackers to learn the patterns of \sys.}


\section{Details of \sys Training}
\label{sec:append_training}
We provide comprehensive parameter settings for \sys training. Specifically, we assign $\beta_{con}=1$ and $\beta_{fm}=1$ in Eq. 2. In two-stage PGM training, we set $\lambda_{kl}=1$, $\lambda_{ens}=1$, $\lambda_{rec}=10$ in Eq. 7. We utilize a total of $K=10$ surrogate models for ensemble training. We set $\rho$ to 16, which determines the signal power of LFAP.



\section{Survey Questions for User Study}
\label{sec:apend_user}
\begin{enumerate}
    \item Please select your age group.
    \begin{itemize}
        \item 18-29
        \item 30-39
        \item 40-49
        \item Over 50
    \end{itemize}
    \item Please rate your intelligibility of the eavesdropped audio quality on a scale from 1 to 5.
    \begin{itemize}
        \item 1 (None of the original speech is recovered)
        \item 2 (Little of the original speech is recovered)
        \item 3 (Half of the original speech is recovered)
        \item 4 (Most of the original speech is recovered)
        \item 5 (All the original speech is recovered)
    \end{itemize}
    \item Please rate your perception of the audio quality on a scale from 1 to 5.
    \begin{itemize}
        \item 1 (Bad)
        \item 2 (Poor)
        \item 3 (Fair)
        \item 4 (Good)
        \item 5 (Excellent)
    \end{itemize}
\end{enumerate}

\section{Details of Speech Transformation}
\label{sec:append_transformation}
WaveGuard~\cite{hussain2021waveguard} leverages audio transformation to mitigate adversarial perturbations. In an eavesdropping attack scenario, the attacker does not have access to the perturbed audio played on the loudspeaker. Instead, he/she can perform a transformation function on the reconstructed audio.

\begin{itemize}[leftmargin=*]
    \item \textbf{Quantization-Dequantization} We quantize the bit-width of the audio signal to 8 bits and restore it back to its original bit precision.
    \item \textbf{Down-sampling and Up-sampling} We downsample the sampling rate of the eavesdropped audio and upsample it to the original sampling rate using bi-linear interpolation technique.
    \item \textbf{Frequency Filtering} We perform frequency filtering on the eavesdropped audio, using high/low shelf filters to attenuate signals above and below a certain threshold.

\end{itemize}

\newpage
\section{Meta-Review}

The following meta-review was prepared by the program committee for the 2025
IEEE Symposium on Security and Privacy (S\&P) as part of the review process as
detailed in the call for papers.

\subsection{Summary}
This paper introduces \sys, a software-based defense mechanism designed to protect against vibration-based side-channel eavesdropping attacks. These attacks exploit sensors such as radars, accelerometers, and optical sensors. The primary contributions include the Perturbation Generator Module (PGM), which generates adversarial perturbations to disrupt side-channel attacks without significantly degrading audio quality for human listeners, and Eve-GAN, a generative adversarial network trained with minimal data to simulate various attack scenarios, thereby enhancing generalizability and practicality.

\subsection{Scientific Contributions}
\begin{itemize}
\item Creates a New Tool to Enable Future Science
\item Provides a Valuable Step Forward in an Established Field
\end{itemize}

\subsection{Reasons for Acceptance}
\begin{enumerate}
\item Creates a New Tool to Enable Future Science: \sys presents a software-based approach for defending against vibration-induced side-channel attacks. The methodology, particularly the use of Eve-GAN and adversarial perturbations, is robust and convincingly validated through comprehensive experimentation.
\item Provides a Valuable Step Forward in an Established Field: \sys effectively addresses a recognized security threat, improving the flexibility and applicability of existing defense solutions. The proposed PGM and Eve-GAN end-to-end design improves both defense effectiveness and usability.
\end{enumerate}


\end{appendices}

\end{document}